# *Circular Dichroism Spectroscopy of Single Objects: Problems, Artifacts, and Corrections*

Stefan Goppelt[1], Lisa M. Günther[2], Jürgen Köhler[1,2,3]

[1] Spectroscopy of Soft Matter, University of Bayreuth, Universitätsstr. 30, D-95440 Bayreuth, Germany

[2] Bayreuth Institute for Macromolecular Research (BIMF), Universitätsstr. 30, D-95440 Bayreuth, Germany

[3] Bavarian Polymer Institute, University of Bayreuth, Universitätsstr. 30, D-95440 Bayreuth, Germany

Corresponding author: Jürgen Köhler (juergen.koehler@uni-bayreuth.de)

**ORCID:**

L. M. Günther: 0000-0002-5626-1694

J. Köhler: 0000-0002-4214-4008

**Table of content**

**Abstract**

Molecular aggregates are at the basis of a broad range of functional systems, including photosynthetic light-harvesting complexes, organic photovoltaic materials, optoelectronic devices, and molecular sensors. Their structural organization, however, is often difficult to resolve because intrinsic heterogeneity complicates conventional high-resolution structural characterization. Single-object spectroscopy provides a means to overcome ensemble averaging and directly probe the structural diversity of individual assemblies. In this context, circular dichroism (CD) spectroscopy is particularly powerful because of its sensitivity to the three-dimensional organization of molecular building blocks. Extending CD measurements from ensembles to individual objects therefore offers a unique approach to probing supramolecular architecture and establishing structure–function relationships in heterogeneous, non-crystallizable systems.

Despite this potential, single-object CD spectroscopy has been barely used to study supramolecular structures due to its substantial experimental challenges. In the absence of spatial and orientational averaging, measurements become highly sensitive to imbalances between left- and right-circularly polarized excitation, deviations from ideal circular polarization, and the coupling of residual polarization imperfections with linear dichroism. These effects can produce spurious CD signals that obscure the intrinsic chiroptical response of individual objects, necessitating rigorous characterization, quantification, and suppression of polarization-related artifacts.

Here, we present a systematic theoretical and experimental framework for understanding and controlling artifacts in single-object CD spectroscopy. We develop a mathematical description of the principal artifact mechanisms, identify their origins in practical experimental implementations - including polarization modulation based on Pockels cells - and establish strategies for their minimization and quantitative assessment. Particular emphasis is placed on the generation, characterization, and control of circularly polarized light with the precision required for measurements of individual anisotropic objects. By integrating concepts from anisotropic thin-film CD spectroscopy, single-molecule optics, and precision polarization engineering, we establish a unified methodological framework for reliable single-object CD measurements.

We validate this framework using an experimental implementation capable of acquiring artifact-controlled CD spectra from individual molecular aggregates over the wavelength range of 690 - 785 nm. The results demonstrate that reliable single-object CD measurements can be achieved when polarization-related artifacts are rigorously characterized and controlled. By consolidating methodologies that are currently distributed across several disciplines, this Tutorial Review provides both a theoretical foundation and a practical guide for the development of single-object CD spectroscopy and its application to emerging areas of single-particle chirality, supramolecular photonics, and biophotonics.

# 1. Introduction

Circular dichroism refers to the difference between two absorption spectra $A_{\mathrm{l}}(\lambda)$ and $A_{\mathrm{r}}(\lambda)$

$$\mathrm{CD}(\lambda) = \mathrm{A_l}(\lambda) - A_{\mathrm{r}}(\lambda) \quad (1)$$

obtained using left circularly polarized (LCP) light and right circularly polarized (RCP) light[1–3]. To understand why this could be different from zero, consider an oscillating light field that travels along the $z$-axis and whose electric field vector rotates by $2\pi$ upon traveling a distance of the wavelength λ, Fig.1.

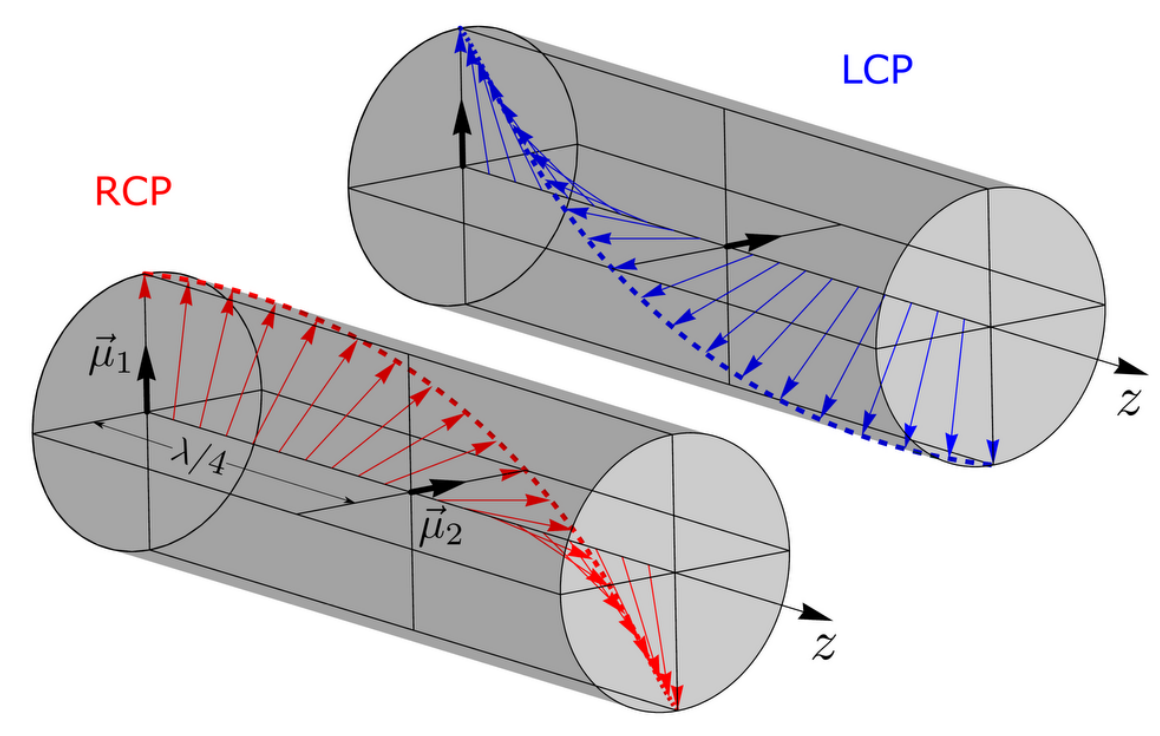

***Fig.1**: Rotation of the electric field vector for left circularly polarized (LCP, blue) and right circularly polarized (RCP, red) light propagating along the $z$-axis. The black arrows indicate two mutually orthogonal transition dipole moments $\vec{\mu}_1$ and $\vec{\mu}_2$ that oscillate with a fixed phase relation with respect to each other, and which are separated by $|\vec{r}_{12}| = \frac{\lambda}{4}$.*

Further, consider two transition dipole moments $\vec{\mu}_{1,2}$ separated by $\vec{r}_{12}$ and which oscillate with a fixed phase relation with respect to each other (coherent coupling). Let the light field initially be polarized exactly along $\vec{\mu}_1$. If $\vec{\mu}_2$ is not oriented coplanar in the $\vec{r}_{12} - \vec{\mu}_1$ plane, then one of the circularly polarized light fields is better phase-matched for the two dipoles than the other, and a difference spectrum $\left(A_{\mathrm{l}}(\lambda) - A_{\mathrm{r}}(\lambda)\right) \neq 0$ will arise. From this simple example, we can draw the following conclusions: (i) the effect will be largest for two dipole moments that are mutually orthogonal and separated by a quarter of the wavelength, (ii) the size of the effect depends on the mutual orientations of the three vectors $\vec{\mu}_1$, $\vec{\mu}_2$, and $\vec{r}_{12}$, (iii) because λ/4 is typically on the order of 100 nm, the effect will become significant for molecular assemblies with extended delocalized electronic wavefunctions (excitons), whereas the expected effect will be very small for molecules, and (iv) given that the CD signal depends on the mutual orientations of coupled transition dipole moments of extended systems, it provides information about the structure of the assembly.

In any case, chirality is a prerequisite for molecules and/or supramolecular structures to exhibit a CD signal; otherwise, the two circularly polarized light fields would give equivalent absorptions[4–6]. For ensembles of proteins or polymeric structures, the relative difference of the absorption spectra is typically on the order of $\frac{\Delta A}{A} \approx 10^{-4}$ to $10^{-3}$. Nowadays, standard CD spectrometers provide sensitivities on the order of $\frac{\Delta A}{A} \approx 10^{-6}$ to $10^{-5}$, and CD spectroscopy became a powerful tool for obtaining structural information of (bio)macromolecules owing to its sensitivity on the mutual orientations of the involved transition dipole moments[2,5,7–10].

Often, the CD effect is characterized by the dissymmetry factor $g$, which is defined as the ratio of the CD effect and the averaged absorbance at a given wavelength

$$g(\lambda) = \frac{A_l(\lambda) - A_r(\lambda)}{\frac{1}{2}\left(A_l(\lambda) + A_r(\lambda)\right)} \tag{2}$$

In conventional absorption spectroscopy on a large ensemble of molecules, the sample is illuminated by a light source, and the signal corresponds to the difference between the incident number of photons and the transmitted number of photons. Although not impossible, performing absorption spectroscopy on single molecules is very challenging because it is rather difficult to discriminate the number of missing (= absorbed) photons against the large background of transmitted photons from the light source[11–14]. This becomes even more challenging when the signal of interest is given by the difference of two absorption spectra, as in CD spectroscopy. The great breakthrough in single-molecule spectroscopy was the application of fluorescence-excitation spectroscopy[15]. Then, the absorption of photons by a single molecule is detected via the induced resonance fluorescence, which occurs against a dark background (at least under ideal conditions) and not against a myriad of photons from the light source. In fluorescence-excitation spectroscopy, the emission $I_{\mathrm{em}}(\lambda_{\mathrm{em}})$ from a sample is integrated over a distinct bandpass as a function of the excitation wavelength $\lambda_{\mathrm{ex}}$.

$$F(\lambda_{\mathrm{ex}}) = \int_{\lambda_1}^{\lambda_2} I_{\mathrm{em}}(\lambda_{\mathrm{em}}) d\lambda_{\mathrm{em}} \tag{3}$$

As long as the fluorescence quantum yield of the sample does not depend on the excitation wavelength, and as long as the excitation intensity is not saturating the optical transition, the fluorescence-excitation spectrum is directly related to the absorption spectrum and therefore

$$A(\lambda_{\mathrm{ex}}) \propto F(\lambda_{\mathrm{ex}}) \tag{4}$$

holds. Accordingly, an obvious approach for developing CD spectroscopy on single objects is to detect the difference in the absorption spectra as the difference of two fluorescence-excitation spectra that are recorded with LCP and RCP light, and the dissymmetry factor is then obtained from

$$g(\lambda_{\mathrm{ex}}) = \frac{A_{\mathrm{l}}(\lambda_{\mathrm{ex}}) - A_{\mathrm{r}}(\lambda_{\mathrm{ex}})}{\frac{1}{2}\left(A_{\mathrm{l}}(\lambda_{\mathrm{ex}}) + A_{\mathrm{r}}(\lambda_{\mathrm{ex}})\right)} = \frac{F_{\mathrm{l}}(\lambda_{\mathrm{ex}}) - F_{\mathrm{r}}(\lambda_{\mathrm{ex}})}{\frac{1}{2}\left(F_{\mathrm{l}}(\lambda_{\mathrm{ex}}) + F_{\mathrm{r}}(\lambda_{\mathrm{ex}})\right)} \tag{5}$$

As pointed out in[16,17], ensemble-averaged dissymmetry factors obtained in bulk measurements differ fundamentally from those accessible at the single-object level. In bulk experiments, orientational and conformational averaging strongly reduces the observable chiroptical response. For individual chiral molecules or proteins, however, $g$-factors on the order of $10^{-3}$ to $10^{-2}$ are theoretically expected, as detailed in[16]. The first report of a dissymmetry factor measured for single molecules appeared in[18]. There, a dissymmetry factor as large as $g \approx 0.5$ was reported at a fixed excitation wavelength. This study triggered an intense debate within the community, and concerns were raised regarding potential experimental artifacts[19,20]. Subsequently, dissymmetry factors of individual nano-objects were reported in several studies[21–23]. These measurements were performed either via differences in fluorescence excitation or via differential extinction, yielding $g$-factors on the order of $10^{-3}$ to $10^{-2}$. In an alternative approach, Orrit and coworkers developed a photothermal detection scheme that was successfully applied for measuring the circular dichroism of gold nanoparticles and for measuring the magnetic circular dichroism of magnetic particles[24–28].

The current work is motivated because it is believed that one of the most efficient light-harvesting systems in photosynthesis, the so-called chlorosome from green sulfur bacteria, features cylindrical symmetry elements for some species and growth conditions[29–33]. In addition, model calculations on cylindrical molecular aggregates have demonstrated that important structural parameters, such as the length and the radius of the tube, can be inferred from circular dichroism (CD) spectroscopy[34]. Owing to the pronounced structural heterogeneity of chlorosomes, our objective was to develop an experimental setup that enables CD spectroscopy on individual chlorosomes. To minimize photobleaching, the samples are kept in a cryostat that served as a vacuum chamber[35,36]. A central challenge in single-object CD spectroscopy is the rigorous suppression of artifacts over the entire spectral range of interest. Any unintended difference in absorbance, $A_{\mathrm{l}}(\lambda) - A_{\mathrm{r}}(\lambda)$, can generate a spurious CD signal[37]. In particular, for individual nano-objects that feature a strong linear dichroism (LD) imperfectly circular polarized light is a major source for polarization artifacts in CD spectra. Although several correction strategies have been developed for CD measurements on anisotropic macroscopic ensembles as for example averaging the signals acquired across different azimuthal sample orientations[38–43], these methods are difficult to implement for individual nanometer-scale objects and become especially impractical in the constrained geometry of a cryogenic environment.

In the following, we systematically analyze possible sources of such artifacts in our setup and describe strategies to reduce them to a tolerable level across a broad spectral range. We demonstrate that our setup enables widefield CD spectroscopy of single nano-objects across the 693 nm to 783 nm wavelength range. Throughout this spectral window, residual artifacts are suppressed to $g < 0.006$, even for the worst-case assumption of maximum sample linear dichroism (LD = 1; see Chapter 3).

This article is organized as follows. After a brief description of the experimental setup, we summarize the aspects of polarized-light formalism relevant to our approach. We then discuss in detail the artifacts introduced by a Pockels cell, followed by an analysis of artifacts arising from the Pockels cell's misalignment. Additional contributions from other optical components, particularly the cryostat windows, are subsequently examined. We next present experimental results that validate the performance and reliability of the setup. Finally, we discuss several alternative approaches that were explored for performing CD spectroscopy on single objects. As none of these strategies achieved the requested accuracy for the $g$-factors, they were ultimately abandoned. Supplementary details and technical clarifications supporting the main text are provided in the Appendix.

The present contribution focuses on the instrumental and methodological aspects of the experiment. The scientific results obtained for the chlorosomes will be reported elsewhere.

## 2. Experimental setup

Circular dichroism spectroscopy on single objects was performed using a home-built setup, schematically shown in Fig.2.

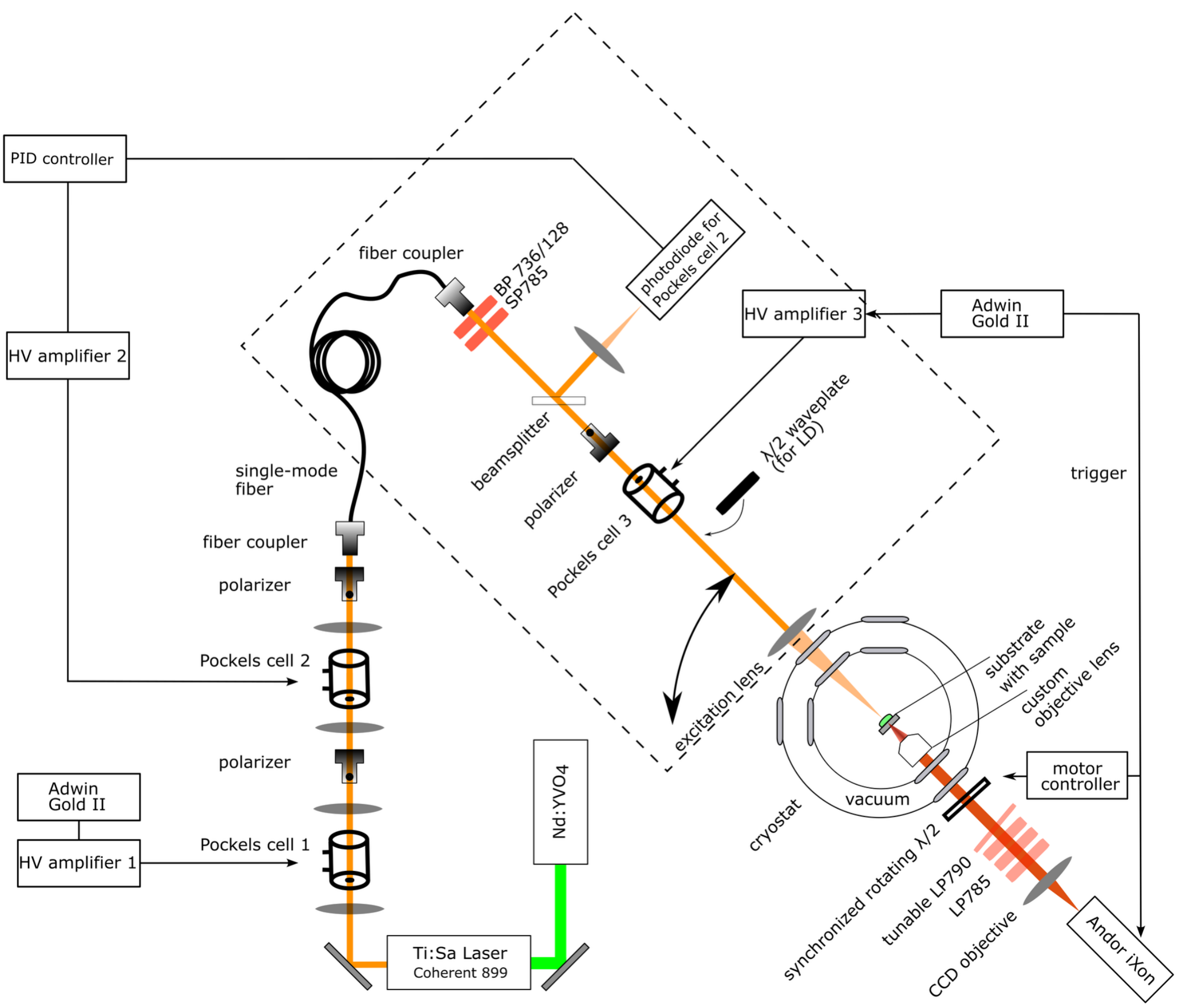

***Fig.2**: Schematic sketch of the setup used for single-object CD spectroscopy. The dashed box indicates the components that are mounted on a rail that can be rotated around an axis perpendicular to the optical table (double-headed bent arrow) centered at the position of the sample. For further details see text.*

The excitation light was provided by a titanium-sapphire (Ti:Sa) laser (899-01, Coherent), pumped by a frequency-doubled continuous-wave Nd:$YVO_4$ laser (Verdi G10, Coherent). The Ti:Sa wavelength was varied from 693 nm to 783 nm with a birefringent filter, and the spectral resolution amounts to 1 $cm^{-1}$. To compensate for wavelength-dependent variations of the laser output, the intensity was measured with a power meter (S120C with PM100D, Thorlabs) in 2 nm increments directly behind the single-mode fiber (see below). The resulting calibration curve was fed into a real-time

I/O system (Adwin Gold II, Jäger), which adjusted the high voltage (HV) of amplifier 1 driving Pockels cell 1 (transversal Pockels cell; Gsänger LM0202P5W), see Fig.2. In combination with a Glan-Laser polarizer (GL10-B, Thorlabs), this acted as a voltage-controlled attenuator, ensuring a constant output power of 10 mW across the entire spectral range.

Temporal fluctuations of the laser intensity were reduced using Pockels cell 2 (transversal Pockels cell; Gsänger LM0202P5W) in combination with a polarizer (GL10-B, Thorlabs). After passing through a polarization-maintaining single-mode fiber serving as a spatial filter, a small fraction of the laser light was picked off by a beamsplitter (BSF10-B, Thorlabs) and monitored with a photodiode. This signal provided feedback for controlling the voltage applied to Pockels cell 2 via a PID controller (10 kHz bandwidth) and HV amplifier 2. The light emerging from the fiber was spectrally cleaned up using two optical filters (BP736/128, SP785; AHF Analysentechnik) and passed through a Glan-Laser polarizer (GL10-B, Thorlabs) to ensure high-purity linear polarization. As a result, the input to Pockels cell 3 (longitudinal Pockels cell, QX 1020, 700 nm to 1000 nm, G&H) consisted of intensity-stabilized, high-purity linearly polarized light with a stability of 0.02 % RMS within a 100 Hz bandwidth. The entire excitation path downstream of the single-mode fiber was mounted on a rail that could be rotated by 45° around the sample position (dashed box in Fig.2).

The substrate carrying the sample was mounted in a home-built bath cryostat serving as a vacuum chamber. The excitation beam entered the cryostat through two anti-reflection-coated (650 nm to 850 nm) glass windows (SF57, Schott) and was weakly focused onto the sample to a spot size of 50 - 100 µm using an air-spaced achromatic doublet ($f\,=$ 100 mm, ACA254-100-B, Thorlabs) positioned in front of the cryostat. Emission from the sample was collected through the rear side of the substrate (UV fused silica, 1 mm thick, broadband single-sided antireflective window, $R\, < 0.25$ %, 400 nm to 1100 nm, W4051FT1, Thorlabs) using a home-built, custom-designed objective lens (see Appendix for details) and propagated through the rear windows of the cryostat toward the detecting system. Transmitted laser light was suppressed using a tunable long-pass filter (TLP01-790-25x36, Semrock) and a stack of three long-pass filters (F76-785, AHF Analysentechnik). Since the transmission of the optical filters can depend on the linear polarization of the emitted light, this is averaged out by placing a rotating zero-order half-waveplate (820 nm, Coherent) operated at 5 Hz in front of the filters. To prevent back-reflections of laser light transmitted through the objective toward the sample, the tunable long-pass filter was tilted by 5° with respect to the optical axis. The other filters were separated spatially by a few centimeters and slightly tilted with respect to each other to avoid interference effects. In total, this ensured an optical density of $OD > 10$ for wavelengths below 785 nm. The signal was finally focused by a tube lens (U-TV1X-2, Olympus) onto an EMCCD camera (iXonUltra DU-897U-CS0-#BV, Andor), which is triggered by the Adwin Gold II system.

For linear polarization-resolved spectroscopy, a half-waveplate (HWP) (AHWP05M-980, Thorlabs) was placed behind Pockels cell 3, which was kept at zero voltage (idle state) to allow experiments with linearly polarized excitation. Spectra were acquired by continuously scanning the laser between 693 nm and 783 nm within 55 s, resulting in an effective spectral resolution of about 0.1 nm (2 $cm^{-1}$). Between successive scans, the linear polarization of the incident light was rotated by 6° using the half-waveplate.

For circular polarization-resolved spectroscopy, the sample was illuminated with alternating LCP and RCP light. Linear polarization was converted into circular polarization by applying appropriate voltages to Pockels cell 3 via HV amplifier 3 (HA3B3-S, hivolt). Switching between LCP and RCP excitation light was performed every 20 ms by the real-time I/O system (Adwin Gold II, Jäger) controlling the HV amplifier. To avoid artifacts arising from the rising and falling edges of the high-voltage pulses, the signal was recorded only during the central 10 ms of each illumination period. The Pockels cell was temperature-stabilized to within 0.1 K using a heating foil wrapped around its metal housing and a temperature controller (LFI-3751, Wavelength Electronics). Precise angular alignment with the optical axis was achieved using a motorized tip-tilt stage (8MTP116, Standa). Two-dimensional lateral positioning perpendicular to the optical axis was achieved using a motorized linear stage (XMS50, Newport) for horizontal movement and a motorized actuator (TRA6CC, Newport) combined with a low-profile vertical translation stage (TSD-603FT, OptoSigma) for vertical adjustment. For circular dichroism spectra, the excitation wavelength was scanned in 0.5 nm steps. At each wavelength, 100 alternating excitation cycles were averaged. Recording one CD spectrum over the full excitation range from 693 nm to 783 nm required 10 to 15 minutes. Typically, ten such spectra were averaged.

## 3. Sources of artifacts in CD spectra of single objects

Ideally, CD spectroscopy is performed with perfectly LCP and RCP light under exactly the same excitation conditions at the spot of the sample. Any deviation from these prerequisites, i.e., imbalanced intensities of LCP and RCP light at the sample position and/or imperfectly circularly polarized light, are sources for artifacts in a CD spectrum. Therefore, understanding and diminishing the effects that contribute artificially to differences in the response of a single object upon illumination with LCP or RCP light is crucial for obtaining correct CD spectra and the interpretation thereof. In this chapter, we will theoretically evaluate the impact of non-ideal excitation conditions on a CD spectrum, followed by a discussion of how the resulting artifacts can be quantified experimentally.

### 3.1 The impact of imbalanced LCP and RCP excitation intensities on CD spectra

To focus on the effects of imbalanced excitation intensities of the LCP and RCP beams on a CD spectrum, we neglect deviations from perfect circular polarization for the moment and assume both beams to be perfectly circularly polarized. For a more quantitative analysis we use a one-dimensional Gaussian laser profile given as $I_{\mathrm{l,r}}(x) = I_{0_{\mathrm{l,r}}} \exp\left(-\frac{1}{2\sigma_{\mathrm{l,r}}^2}\left(x - \mu_{\mathrm{l,r}}\right)^2\right)$ centered at $x = \mu_{\mathrm{l,r}}$, with a variance of $\sigma_{\mathrm{l,r}}^2$ ($\frac{1}{e^2}$-beam radius $\omega_{\mathrm{l,r}}^2 = 4\sigma_{\mathrm{l,r}}^2$), where the subscripts refer to LCP ($\mathrm{l}$) and RCP ($\mathrm{r}$) light. For a difference $\Delta I$ in the excitation intensities $I_{\mathrm{l}} = I_{\mathrm{r}} + \Delta I$ the corresponding absorptions upon LCP and RCP excitation are given as $A_{\mathrm{l}} = c_{\mathrm{l}} \cdot I_{\mathrm{l}}$ and $A_{\mathrm{r}} = c_{\mathrm{r}} \cdot I_{\mathrm{r}}$, where $c_{\mathrm{l}}$ and $c_{\mathrm{r}}$ denote proportionality factors. Usually in an experimental situation $\frac{\Delta I}{I_{\mathrm{l}}}, \frac{\Delta I}{I_{\mathrm{r}}} \ll 1$ and $c_{\mathrm{l}} \approx c_{\mathrm{r}}$ holds. For the sake of simplicity, we omitted the excitation wavelength $\lambda_{\mathrm{ex}}$ in the above expressions and evaluate the dissymmetry factor in eq.(2) as

$$\begin{aligned} g &= 2\frac{A_{\mathrm{l}} - A_{\mathrm{r}}}{A_{\mathrm{l}} + A_{\mathrm{r}}} = 2\frac{I_{\mathrm{l}} \cdot c_{\mathrm{l}} - I_{\mathrm{r}} \cdot c_{\mathrm{r}}}{I_{\mathrm{l}} \cdot c_{\mathrm{l}} + I_{\mathrm{r}} \cdot c_{\mathrm{r}}} \\ &= 2\frac{(I_{\mathrm{r}} + \Delta I) \cdot c_{\mathrm{l}} - I_{\mathrm{r}} \cdot c_{\mathrm{r}}}{(I_{\mathrm{r}} + \Delta I) \cdot c_{\mathrm{l}} + I_{\mathrm{r}} \cdot c_{\mathrm{r}}} \approx \underbrace{2\frac{I_{\mathrm{r}} \cdot c_{\mathrm{l}} - I_{\mathrm{r}} \cdot c_{\mathrm{r}}}{I_{\mathrm{r}} \cdot c_{\mathrm{l}} + I_{\mathrm{r}} \cdot c_{\mathrm{r}}}}_{g_{\mathrm{CD}}} + 2\frac{\Delta I \cdot c_{\mathrm{l}}}{I_{\mathrm{l}} \cdot c_{\mathrm{l}} + I_{\mathrm{r}} \cdot c_{\mathrm{r}}} \\ &\approx g_{\mathrm{CD}} + \underbrace{2\frac{I_{\mathrm{l}} - I_{\mathrm{r}}}{I_{\mathrm{l}} + I_{\mathrm{r}}}}_{g_{\mathrm{I}}} \end{aligned} \tag{6}$$

where $g_{\mathrm{CD}}$ refers to the "real" CD effect, and

$$g_{\mathrm{I}} = 2\frac{I_{\mathrm{l}} - I_{\mathrm{r}}}{I_{\mathrm{l}} + I_{\mathrm{r}}} \tag{7}$$

to the artifact that results from imbalanced excitation intensities at the sample position. As illustrated in Fig.3 this can be caused by (i) a difference in the peak intensities, (ii) a lateral shift of the two beams, and/or (iii) a difference in beam widths.

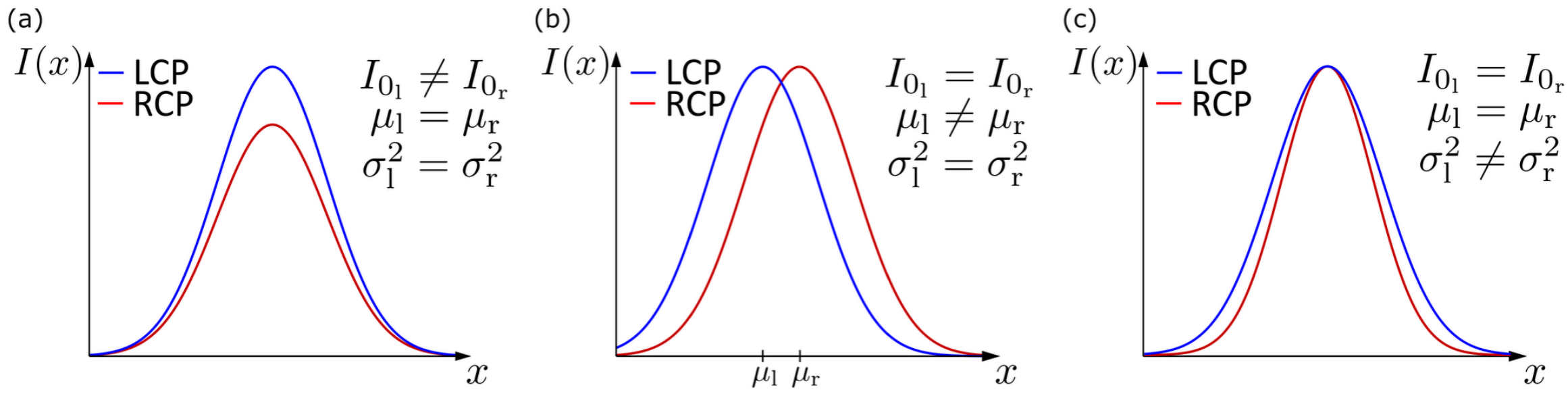

***Fig.3**: Possible sources for imbalanced excitation intensities for LCP ($l$, blue) and RCP ($r$, red) excitation light at the position of a single nano-object. The beam profiles are depicted as Gaussian beam profiles with $I_{l,r}(x) = I_{0_{l,r}}\, exp\left(-\frac{1}{2\sigma_{l,r}^2}\left(x - \mu_{l,r}\right)^2\right)$. (a) Difference in peak intensity. (b) Difference in lateral position. (c) Difference in beam widths.*

It is worth noting, that imbalanced excitation intensities for LCP and RCP light are a specific challenge in single-object CD spectroscopy. In conventional CD spectroscopy of macroscopic samples, the intensity profile is averaged over the spatial extent of the sample, which is typically much larger than the beam cross section. Consequently, a small lateral displacement between the two light beams can be neglected. In contrast, the dimensions of a single object are usually much smaller than the diffraction-limited beam diameter, meaning that the object probes only the local excitation intensity at its position. Under these conditions, even a slight spatial mismatch between the LCP and RCP beams results in unequal excitation intensities and can therefore produce an intensity artifact. To quantify the impact of these variations on the $g$-factor, we write the parameters in the profiles for LCP and RCP excitation light as $I_{0_{\mathrm{l,r}}} = I_0 + \delta I_{0_{\mathrm{l,r}}}$, $\mu_{\mathrm{l,r}} = \mu + \delta\mu_{\mathrm{l,r}}$, and $\sigma_{\mathrm{l,r}}^2 = \sigma^2 + \delta\sigma_{\mathrm{l,r}}^2$, which yields

$$
\begin{aligned}
&I_{\mathrm{l,r}}(\delta I_{0_{\mathrm{l,r}}}, \delta\mu_{\mathrm{l,r}}, \delta\sigma_{\mathrm{l,r}}^2) \\
&= \underbrace{\left[I_0 + \delta I_{0_{\mathrm{l,r}}}\right]}_{I_{0_{\mathrm{l,r}}}} \exp\left(-\frac{1}{2}\frac{1}{\underbrace{\sigma^2 + \delta\sigma_{\mathrm{l,r}}^2}_{\sigma_{\mathrm{l,r}}^2}}\left(x - \underbrace{\left[\mu + \delta\mu_{\mathrm{l,r}}\right]}_{\mu_{\mathrm{l,r}}}\right)^2\right) \\
&= I_{0_{\mathrm{l,r}}} \exp\left(-\frac{1}{2\sigma_{\mathrm{l,r}}^2}\left(x - \mu_{\mathrm{l,r}}\right)^2\right)
\end{aligned}
\tag{8}
$$

For $\frac{\delta I_{0_{\mathrm{l,r}}}}{I_0} \ll 1$, $\frac{\delta\mu_{\mathrm{l,r}}}{\sigma} \ll 1$ and $\frac{\delta\sigma_{\mathrm{l,r}}^2}{\sigma^2} \ll 1$ we can expand eq.(8) up to first order in all three parameters obtaining

$$
\begin{aligned}
&I_{\mathrm{l,r}}(\delta I_{0_{\mathrm{l,r}}}, \delta\mu_{\mathrm{l,r}}, \delta\sigma_{\mathrm{l,r}}^2) \\
&\approx I_{\mathrm{l,r}}(0,0,0) + \left.\frac{\partial I_{\mathrm{l,r}}}{\partial\delta I_{0_{\mathrm{l,r}}}}\right|_{0,0,0} \cdot \delta I_{0_{\mathrm{l,r}}} + \left.\frac{\partial I_{\mathrm{l,r}}}{\partial\delta\mu_{\mathrm{l,r}}}\right|_{0,0,0} \cdot \delta\mu_{l,r} + \left.\frac{\partial I_{\mathrm{l,r}}}{\partial\delta\sigma_{\mathrm{l,r}}^2}\right|_{0,0,0} \cdot \delta\sigma_{\mathrm{l,r}}^2
\end{aligned}
\tag{9}
$$

with

$$
I_{\mathrm{l,r}}(0,0,0) = I_0 \cdot \exp\left(-\frac{1}{2\sigma^2}(x-\mu)^2\right) \tag{10}
$$

Working out the derivatives gives

$$
\begin{aligned}
\left.\frac{\partial I_{\mathrm{l,r}}}{\partial\delta I_{0_{\mathrm{l,r}}}}\right|_{0,0,0} &= \left[\frac{1}{I_0} \cdot (x-\mu)^0\right] \cdot I_{\mathrm{l,r}}(0,0,0) \\
\left.\frac{\partial I_{\mathrm{l,r}}}{\partial\delta\mu_{\mathrm{l,r}}}\right|_{0,0,0} &= \left[\frac{1}{\sigma^2}(x-\mu)^1\right] \cdot I_{\mathrm{l,r}}(0,0,0) \\
\left.\frac{\partial I_{\mathrm{l,r}}}{\partial\delta\sigma_{\mathrm{l,r}}^2}\right|_{0,0,0} &= \left[\frac{1}{2\sigma^4}(x-\mu)^2\right] \cdot I_{\mathrm{l,r}}(0,0,0)
\end{aligned}
\tag{11}
$$

and finally, eq.(9) reads

$$
\begin{aligned}
&I_{\mathrm{l,r}}(\delta I_{0_{\mathrm{l,r}}}, \delta\mu_{\mathrm{l,r}}, \delta\sigma_{\mathrm{l,r}}^2) \\
&\approx I_{\mathrm{l,r}}(0,0,0) \\
&\cdot \left[1 + \frac{1}{I_0} \cdot \underbrace{(I_{0_{\mathrm{l,r}}} - I_0)}_{\delta I_{0_{\mathrm{l,r}}}} \cdot (x-\mu)^0 + \frac{1}{\sigma^2}\underbrace{(\mu_{\mathrm{l,r}} - \mu)}_{\delta\mu_{\mathrm{l,r}}}(x-\mu)^1 + \frac{1}{2\sigma^4}\underbrace{(\sigma_{\mathrm{l,r}}^2 - \sigma^2)}_{\delta\sigma_{\mathrm{l,r}}^2}(x-\mu)^2\right]
\end{aligned}
\tag{12}
$$

Inserting eq.(12) into eq.(7), and approximating

$$
I_{\mathrm{l}} + I_{\mathrm{r}} \approx 2 \cdot I_{\mathrm{l,r}}(0,0,0) \tag{13}
$$

yields for the artifact

$$g_{\mathrm{I}} = 2\frac{I_{\mathrm{l}} - I_{\mathrm{r}}}{I_{\mathrm{l}} + I_{\mathrm{r}}} \approx \frac{I_{\mathrm{l}} - I_{\mathrm{r}}}{I_{\mathrm{l,r}}(0,0,0)}$$

$$\approx \frac{I_{0_{\mathrm{l}}} - I_{0_{\mathrm{r}}}}{I_0} \cdot (x-\mu)^0 + \frac{(\mu_{\mathrm{l}} - \mu_{\mathrm{r}})}{\sigma^2} \cdot (x-\mu)^1 + \frac{(\sigma_{\mathrm{l}}^2 - \sigma_{\mathrm{r}}^2)}{2\sigma^4} \cdot (x-\mu)^2 \tag{14}$$

Using $I_0 \approx \frac{I_{0_{\mathrm{l}}} + I_{0_{\mathrm{r}}}}{2}$, $\Delta\mu_{\mathrm{lr}} = (\mu_{\mathrm{l}} - \mu_{\mathrm{r}})$ and $\Delta\sigma_{\mathrm{lr}}^2 = (\sigma_{\mathrm{l}}^2 - \sigma_{\mathrm{r}}^2)$, this can be written concisely as

$$g_{\mathrm{I}} = g_{\mathrm{I}}^{(0)} + g_{\mathrm{I}}^{(1)} + g_{\mathrm{I}}^{(2)} = 2\frac{I_{0_{\mathrm{l}}} - I_{0_{\mathrm{r}}}}{I_{0_{\mathrm{l}}} + I_{0_{\mathrm{r}}}} \cdot (x-\mu)^0 + \frac{\Delta\mu_{\mathrm{lr}}}{\sigma^2} \cdot (x-\mu)^1 + \frac{\Delta\sigma_{\mathrm{lr}}^2}{2\sigma^4} \cdot (x-\mu)^2 \tag{15a}$$

For two dimensions a similar calculation can be done for the $y$ coordinate. The result also holds for non-perfect rotationally symmetric beam shape $\left(\sigma_x^2 \neq \sigma_y^2\right)$ if the $x$ and $y$ directions are considered to point along the principal axes of the 2D Gaussian profile, but it neglects that the 2D Gaussian beam profiles for LCP and RCP light might be rotated with respect to each other. This approximation is acceptable for quasi-rotationally symmetric beam profiles $\sigma_x^2 \approx \sigma_y^2$. The result of a full calculation in two dimensions including a small rotation between the principal axes of the Gaussian beam profiles for LCP and RCP light is

$$g_{\mathrm{I}} = 2\frac{I_{0_{\mathrm{l}}} - I_{0_{\mathrm{r}}}}{I_{0_{\mathrm{l}}} + I_{0_{\mathrm{r}}}} + \frac{\Delta\mu_{\mathrm{lr},x}}{\sigma_x^2} \cdot (x-\mu_x)^1 + \frac{\Delta\mu_{\mathrm{lr},y}}{\sigma_y^2} \cdot \left(y-\mu_y\right)^1 + \frac{\Delta\sigma_{\mathrm{lr},x}^2}{2\sigma_x^4} \cdot (x-\mu_x)^2 + \frac{\Delta\sigma_{\mathrm{lr},y}^2}{2\sigma_y^4} \cdot \left(y-\mu_y\right)^2 + \frac{\Delta\sigma_{\mathrm{lr},xy}}{\sigma_x^2\sigma_y^2}(x-\mu_x)\left(y-\mu_y\right) \tag{15b}$$

For the details the reader is referred to the Appendix.

## 3.2 The impact of imperfectly circularly polarized light on CD spectra

### 3.2.1 Parametrization of elliptically polarized light

Before we discuss the impact of imperfectly circularly polarized light on a CD spectrum, we briefly recall the mathematical background for describing the polarization status of light[44]. For the electric field $\vec{E}(t,z)$, we consider a monochromatic plane wave with angular frequency $\omega = \frac{2\pi \cdot c}{\lambda}$ that oscillates in the $x$-$y$ plane and propagates along the $z$-direction at the speed $c$.

$$\vec{E}(t,z) = \begin{pmatrix} E_x(t,z) \\ E_y(t,z) \end{pmatrix} = \begin{pmatrix} E_{0x} \cdot \cos\left[\omega\left(t - \frac{z}{c}\right) + \varphi_x\right] \\ E_{0y} \cdot \cos\left[\omega\left(t - \frac{z}{c}\right) + \varphi_y\right] \end{pmatrix} \tag{16}$$

Here, $E_{0x}$ and $E_{0y}$ are the amplitudes, and $\varphi_x$ and $\varphi_y$ are the phases of the $x$ and $y$ components of the electric field, respectively. Perfectly circularly polarized light corresponds to $\frac{E_{0x}}{E_{0y}} = 1$ and $\varphi = \varphi_y - \varphi_x = \mp\frac{\pi}{2}$ (modulo $\pi$), where the "−" sign applies to LCP and the "+" sign to RCP light. Deviations from this will give rise to elliptically polarized light or, in the limiting case $\varphi = \varphi_y - \varphi_x = 0$, to linearly polarized light.

For elliptically polarized light the projection of the electric field vector onto the plane perpendicular to the propagation direction rotates counterclockwise (LCP) or clockwise (RCP) on an ellipse, when looking against the direction of propagation. The shape of the ellipse can be characterized by the ellipticity $\epsilon$, or likewise by two angles $\chi$ and $\psi$, as illustrated in Fig.4.

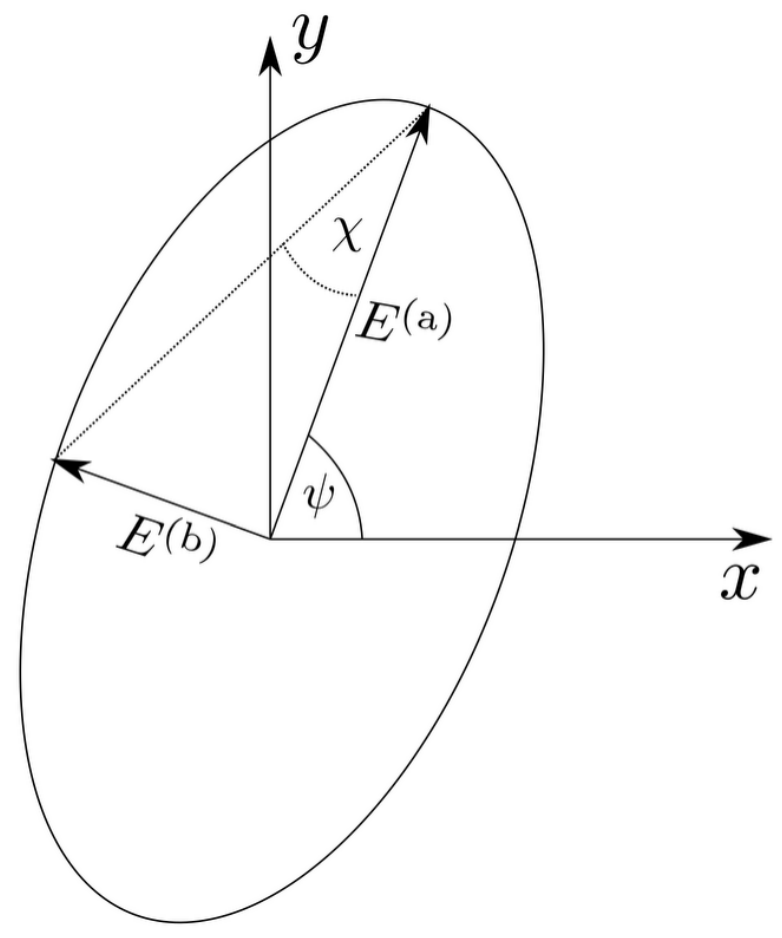

***Fig.4**: Polarization ellipse with semi-major axis $E^{(a)}$ and semi-minor axis $E^{(b)}$ and the definition of the ellipticity angle $\chi$ and azimuth angle $\psi$ relative to the laboratory frame represented by the $x$ and $y$ axes.*

This is given in detail as[45]

$$\epsilon = \frac{\pm E^{(\mathrm{b})}}{E^{(\mathrm{a})}} = \tan\chi$$

$$\tan 2\psi = \frac{2\left(\frac{E_{0y}}{E_{0x}}\right)}{1 - \left(\frac{E_{0y}}{E_{0x}}\right)^2}\cos\varphi \qquad (17)$$

$$\sin 2\chi = \frac{2\left(\frac{E_{0y}}{E_{0x}}\right)}{1 + \left(\frac{E_{0y}}{E_{0x}}\right)^2}\sin\varphi$$

Here, $E^{(\mathrm{a})}$ and $E^{(\mathrm{b})}$ correspond to the amplitudes of the electric field strengths along the semi-major and semi-minor axes of the ellipse, respectively, $E_{0x}$ and $E_{0y}$ correspond to the amplitudes of the $x$ and $y$ components of the electric field projected onto the $x$- and $y$-axes of the laboratory frame, $\varphi$ to the phase difference between the components of the electric field, and $\psi$ to the angle between the semi-major axis of the ellipse and the $x$-axis in the laboratory frame, for details see Fig.4.

In particular $\epsilon = 0$ (equivalently $\chi = 0$) holds for perfect linear polarization, and $\epsilon = \mp 1$ (equivalently $|\chi| = \frac{\pi}{4}$ ) for perfect circular polarization. Conventionally $\epsilon = -1$ $\left(\chi = -\frac{\pi}{4}\right)$ refers to LCP light and $\epsilon = +1$ $\left(\chi = +\frac{\pi}{4}\right)$ to RCP light. For elliptically polarized light, i.e., $0 < |\epsilon| < 1$, the polarization state can be considered as a superposition of circularly and linearly polarized components.
An alternative description of the parameters of the polarization ellipse is provided by the Stokes vector[44], which is for fully polarized light defined as

$$\vec{S} = \begin{pmatrix} S_0 \\ S_1 \\ S_2 \\ S_3 \end{pmatrix} = \begin{pmatrix} E_{0x}^2 + E_{0y}^2 \\ E_{0x}^2 - E_{0y}^2 \\ 2E_{0x}E_{0y}\cos(\varphi) \\ 2E_{0x}E_{0y}\sin(\varphi) \end{pmatrix} = I \cdot \begin{pmatrix} 1 \\ \cos 2\psi \cos 2\chi \\ \sin 2\psi \cos 2\chi \\ \sin 2\chi \end{pmatrix} \tag{18}$$

The second equation follows from the relations given in eq.(17), and the normalization $I = E_{0x}^2 + E_{0y}^2$. The Stokes vector is a powerful tool for describing polarized light, because its components can be directly connected with observables, see eq.(19)[44],

$$\vec{S} = \begin{pmatrix} S_0 \\ S_1 \\ S_2 \\ S_3 \end{pmatrix} = \begin{pmatrix} I = I_{0°} + I_{90°} \\ I_{0°} - I_{90°} \\ I_{45°} - I_{-45°} \\ I_{RCP} - I_{LCP} \end{pmatrix} \tag{19}$$

where $I_{0°}$, $I_{90°}$, $I_{45°}$, $I_{-45°}$ refer to the intensities that can be measured after the light has passed an ideal linear polarizer that is oriented under angles of 0° (horizontal, $x$-axis), 90° (vertical, $y$-axis) and ± 45° in the laboratory frame. The intensities $I_{RCP}$ and $I_{LCP}$ represent the intensities after passing (virtual) right- and left-circular polarizers. In particular for fully polarized light the components of the Stokes vector fulfill the relation

$$S_0^2 = S_1^2 + S_2^2 + S_3^2 \tag{20}$$

and the zeroth component corresponds to the total intensity.

### 3.2.2 Artifacts resulting from imperfect circular polarization

Any deviation from perfectly circularly polarized light will give rise to a finite difference in the absorption spectra for LCP and RCP light. In a real CD experiment the magnitude of the deviations from the preconditions for perfectly circularly polarized light, namely $\frac{E_{0x}}{E_{0y}} = 1$ and $\varphi = \varphi_y - \varphi_x = \mp \frac{\pi}{2}$ (modulo $\pi$) are small yet finite, and result in slightly elliptically polarized light. To avoid confusion, and to make writing more efficient, imperfectly LCP and imperfectly RCP light are hereafter referred to as as (i)-LCP and (i)-RCP light, respectively. The corresponding (exaggerated) polarization ellipses are shown separately in Fig.5. For (i)-LCP and (i)-RCP light, one must consider that their respective principal axes do not necessarily coincide in magnitude or orientation. In other words, the different residual linear polarization components resulting from these imperfections lead to an artificial linear dichroism (LD) signal from the sample, which can exceed the CD effect by orders of magnitude. It is worth noting that this artifact is negligible for CD spectroscopy on an ensemble of randomly oriented objects. Owing to the randomness of the relative orientations between the transition dipole moments and the principal axes of the polarization ellipses this artifact averages out in conventional CD spectroscopy on isotropic macroscopic ensembles. In contrast, for CD spectroscopy on single objects, the relative orientations between the transition dipole moments and the polarization ellipses are fixed and the contribution of this effect to the $g$-factor has to be considered.

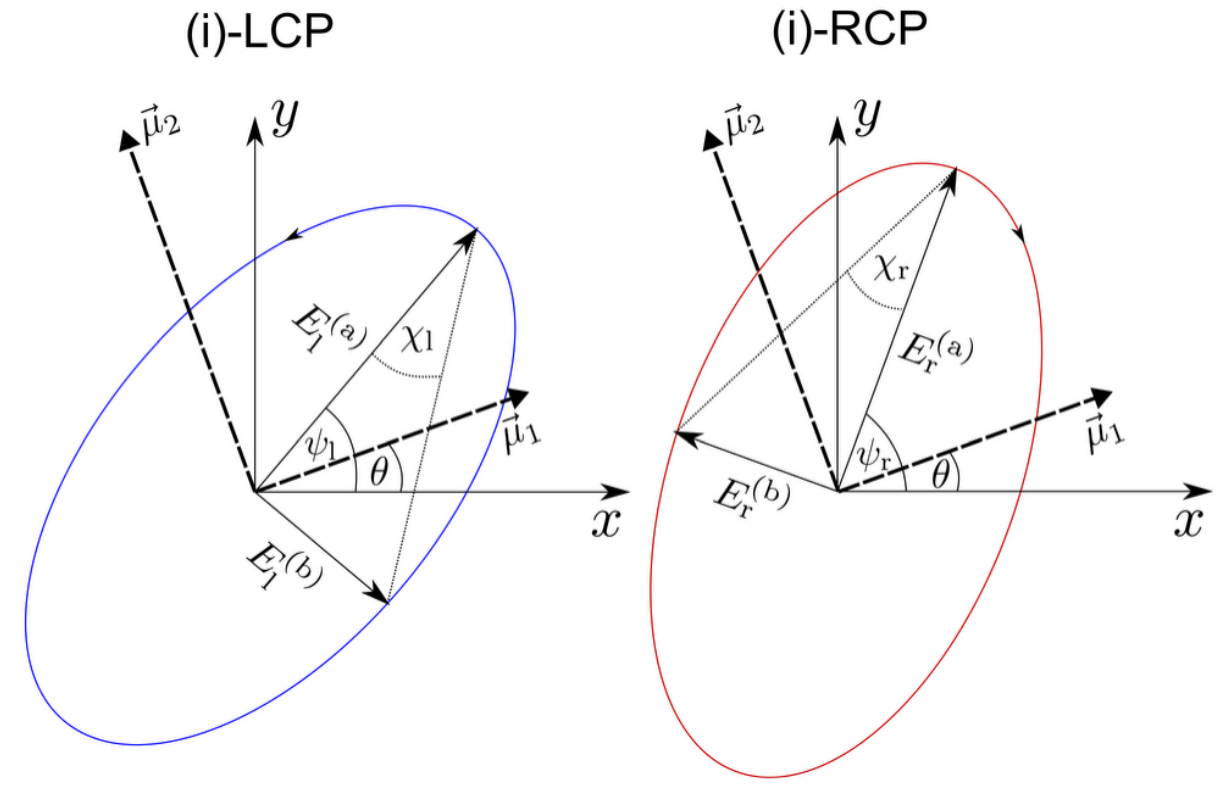

***Fig.5**: Polarization ellipses for (i)-LCP (blue) and (i)-RCP (red) excitation light. For clarity the ellipticity is exaggerated. The sample is represented by two mutually orthogonal transition dipole moments $\vec{\mu}_1$ and $\vec{\mu}_2$ such that $\vec{\mu}_1$ encloses the angle $\theta$ with the $x$-axis.*

To analyze the contribution of spurious linear polarization components to artificial circular dichroism in more detail, we mimic a sample as a cluster of transition dipole moments confined to the $x$-$y$ plane. By projecting each moment onto the $x$- and $y$-axis and summing their respective components, the entire system can be represented by two mutually orthogonal transition dipole moments, $\vec{\mu}_1$ and $\vec{\mu}_2$. In order to get started we assume that these moments are aligned along the $x$- and the $y$-axis of the laboratory frame and can be written as $\vec{\mu}_1 = \vec{\mu}_x = \mu_x \vec{e}_x$ and $\vec{\mu}_2 = \vec{\mu}_y = \mu_y \vec{e}_y$ with magnitudes $\mu_{x,y}$. Later we will allow them to deviate from the laboratory frame by an angle $\theta$ as already indicated in Fig.5. For close-to-perfect circular polarization the measured dissymmetry factor $g$ can be separated into

$$g = g_{\mathrm{CD}} + g_{\mathrm{P}} \tag{21}$$

where $g_{\mathrm{CD}}$ corresponds to the "real" $g$-factor that results from circular dichroism and an artifact $g_{\mathrm{P}}$ that is caused by the deviation from perfect circular polarization for the LCP and RCP light. Such a sample will absorb light as follows:

$$A_{x,y} \propto \left|\mu_{x,y}\vec{e}_{x,y} \cdot \vec{E}\right|^2 \tag{22}$$

Here $A_{x,y}$ refers to the absorption that is associated with the transition dipole moments $\vec{\mu}_{x,y}$. It is important to note that eq.(22) refers to the absorption due to an electric-dipole transition induced by linearly polarized light. It does not describe circular dichroism, because the latter would be related to the electric and the magnetic-dipole operator[5]. A sample described by eq.(22) exhibits linear dichroism

$$LD = \frac{A_x - A_y}{A_x + A_y} = \frac{\mu_x^2 - \mu_y^2}{\mu_x^2 + \mu_y^2} \tag{23}$$

Comparing eq.(22) with the transmission behavior of a linear polarizer[44] given as

$$I'_{x,y} \propto \left|t_{x,y}\vec{e}_{x,y} \cdot \vec{E}\right|^2 \tag{24}$$

where $\vec{E}$ is an incident electric field, $I'_{x,y}$ is the transmitted intensity, and $t_{x,y}$ refers to the amplitude transmission coefficients, $0 \leq t_{x,y} \leq 1$ along the direction $\vec{e}_{x,y}$, reveals the mathematical similarity of eq.(22) and eq.(24), if we normalize the magnitudes of the transition dipole moments to their maximum value $\mu_{x,y}^{(\mathrm{n})} = \frac{\mu_{x,y}}{\max(\mu_x,\mu_y)}$. Then, we can associate the coefficients $t_{x,y}$ with the normalized transition dipole moments $\mu_{x,y}^{(\mathrm{n})}$, and the absorbance of a sample due to a transition dipole moment can be treated similarly as the transmitted intensity of an incident electric field through a linear polarizer with finite transmission characteristics.

This association allows us to take advantage of the Mueller matrix formalism[44] for calculating the absorption properties of a sample represented by two mutually orthogonal transition dipole moments $\vec{\mu}_x$ and $\vec{\mu}_y$ that is illuminated with light of arbitrary polarization characterized by a Stokes vector $\vec{S}$. To avoid confusion with the non-normalized transition dipole moments, the subsequent calculations instead employ the amplitude transmission coefficients $t_x$ and $t_y$, where $0 \leq t_x \leq 1$ and $0 \leq t_y \leq 1$. For a linear polarizer the Mueller matrix, $M_{\mathrm{pol}}$, is given as[44]

$$M_{\mathrm{pol}} = \frac{1}{2}\begin{pmatrix} t_x^2 + t_y^2 & t_x^2 - t_y^2 & 0 & 0 \\ t_x^2 - t_y^2 & t_x^2 + t_y^2 & 0 & 0 \\ 0 & 0 & 2t_x t_y & 0 \\ 0 & 0 & 0 & 2t_x t_y \end{pmatrix} \tag{25}$$

Now we allow the alignment of the transition dipole moments to deviate from the laboratory frame such that the transition dipole moment $\vec{\mu}_1$ encloses an angle $\theta$ with the $x$-axis, see Fig.5. Then, $M_{\mathrm{pol}}$ has to be transformed according to

$$M_{\mathrm{pol,rot}}(\theta) = M_{\mathrm{rot}}(-\theta) \cdot M_{\mathrm{pol}} \cdot M_{\mathrm{rot}}(\theta) \tag{26}$$

where

$$M_{\mathrm{rot}}(\theta) = \begin{pmatrix} 1 & 0 & 0 & 0 \\ 0 & \cos(2\theta) & \sin(2\theta) & 0 \\ 0 & -\sin(2\theta) & \cos(2\theta) & 0 \\ 0 & 0 & 0 & 1 \end{pmatrix} \tag{27}$$

is the corresponding rotation matrix. As will be shown below the first row of the Mueller matrix $M_{\mathrm{pol,rot}}(\theta)$ is sufficient for further discussion. Consequently, for the first row of the Mueller matrix of a linear polarizer whose transmission axis encloses an angle $\theta$ with the $x$-axis of the laboratory frame, eq.(26) yields

$$M_{\mathrm{pol,rot}}(\theta) = \frac{1}{2}\cdot\begin{pmatrix} t_x^2 + t_y^2 & \left(t_x^2 - t_y^2\right)\cdot\cos 2\theta & \left(t_x^2 - t_y^2\right)\cdot\sin 2\theta & 0 \\ \cdots & \cdots & \cdots & \cdots \\ \cdots & \cdots & \cdots & \cdots \\ \cdots & \cdots & \cdots & \cdots \end{pmatrix} \tag{28}$$

Using the equivalence between $\left(\mu_{x,y}^{(\mathrm{n})}\right)^2$ and $t_{x,y}^2$ we can rewrite eq.(23) as

$$LD = \frac{\mu_x^2 - \mu_y^2}{\mu_x^2 + \mu_y^2} = \frac{\left(\mu_x^{(\mathrm{n})}\right)^2 - \left(\mu_y^{(\mathrm{n})}\right)^2}{\left(\mu_x^{(\mathrm{n})}\right)^2 + \left(\mu_y^{(\mathrm{n})}\right)^2} = \frac{t_x^2 - t_y^2}{t_x^2 + t_y^2} \tag{29}$$

and obtain for eq.(28)

$$M_{\mathrm{pol,rot}}(\theta) = \frac{1}{2}\cdot\left(t_x^2 + t_y^2\right)\begin{pmatrix} 1 & LD\cdot\cos 2\theta & LD\cdot\sin 2\theta & 0 \\ \cdots & \cdots & \cdots & \cdots \\ \cdots & \cdots & \cdots & \cdots \\ \cdots & \cdots & \cdots & \cdots \end{pmatrix} \tag{30}$$

Now, we use the Stokes vectors $\vec{S}_{\mathrm{l}}^{(\mathrm{i})}$ and $\vec{S}_{\mathrm{r}}^{(\mathrm{i})}$ to characterize the polarization of (i)-LCP and (i)-RCP incident light. Here, the superscript $(\mathrm{i})$ is used for Stokes vectors that correspond to imperfect circular polarizations. The Stokes vector $\vec{S'}_{\mathrm{l,r}}^{(\mathrm{i})}$ of light transmitted through a linear polarizer follows from

$$\vec{S'}^{(\mathrm{i})}_{\mathrm{l,r}} = M_{\text{pol,rot}}(\theta) \cdot \vec{S}^{(\mathrm{i})}_{\mathrm{l,r}} \tag{31}$$

and the transmitted intensity is given by its zeroth component

$$\left[\vec{S'}^{(\mathrm{i})}_{\mathrm{l,r}}\right]_0 = \left[M_{\text{pol,rot}}(\theta) \cdot \vec{S}^{(\mathrm{i})}_{\mathrm{l,r}}\right]_0 \tag{32}$$

Owing to the mathematical equivalence between the normalized magnitudes $\mu^{(n)}_{x,y}$ and the coefficients $t_{x,y}$, see eq.(22) and (24), this holds also for the corresponding absorptions of the sample, $A^{(\mathrm{i})}_{\mathrm{l,r}}$, due to imperfectly circularly polarized light.

$$\begin{aligned} A^{(\mathrm{i})}_{\mathrm{l}} &\sim \left[\vec{S'}^{(\mathrm{i})}_{\mathrm{l}}\right]_0 = \left[M_{\text{pol,rot}}(\theta) \cdot \vec{S}^{(\mathrm{i})}_{\mathrm{l}}\right]_0 \\ A^{(\mathrm{i})}_{\mathrm{r}} &\sim \left[\vec{S'}^{(\mathrm{i})}_{\mathrm{r}}\right]_0 = \left[M_{\text{pol,rot}}(\theta) \cdot \vec{S}^{(\mathrm{i})}_{\mathrm{r}}\right]_0 \end{aligned} \tag{33}$$

Using eq.(33) we can calculate the spurious contribution of residual linearly polarized light to the circular dichroism as

$$g_{\mathrm{P}} = \frac{A^{(\mathrm{i})}_{\mathrm{l}} - A^{(\mathrm{i})}_{\mathrm{r}}}{\frac{1}{2} \cdot \left(A^{(\mathrm{i})}_{\mathrm{l}} + A^{(\mathrm{i})}_{\mathrm{r}}\right)} \tag{34}$$

Since the whole discussion applies to small deviations from ideal polarization conditions the denominator of eq.(34) can be calculated using the unperturbed Stokes vectors $\vec{S}_{\mathrm{l,r}}$ given as

$$\vec{S}_{\mathrm{l,r}} = \begin{pmatrix} 1 \\ 0 \\ 0 \\ \mp 1 \end{pmatrix} \tag{35}$$

where the negative (positive) sign refers to LCP (RCP) light. Then, we obtain

$$\begin{aligned} A^{(\mathrm{i})}_{\mathrm{l}} + A^{(\mathrm{i})}_{\mathrm{r}} &\propto \left[M_{\text{pol,rot}}(\theta) \cdot \vec{S}^{(\mathrm{i})}_{\mathrm{l}} + M_{\text{pol,rot}}(\theta) \cdot \vec{S}^{(\mathrm{i})}_{\mathrm{r}}\right]_0 \\ &\approx \left[M_{\text{pol,rot}}(\theta) \cdot \vec{S}_{\mathrm{l}} + M_{\text{pol,rot}}(\theta) \cdot \vec{S}_{\mathrm{r}}\right]_0 = t_x^2 + t_y^2 \end{aligned} \tag{36}$$

and together with eq.(33) this yields for eq.(34)

$$g_{\mathrm{P}} \approx \frac{A^{(i)}_{\mathrm{l}} - A^{(i)}_{\mathrm{r}}}{\frac{1}{2}\left(t_x^2 + t_y^2\right)} \approx \frac{1}{\frac{1}{2}\left(t_x^2 + t_y^2\right)} \cdot \left[M_{\text{pol,rot}} \cdot \vec{S}^{(\mathrm{i})}_{\mathrm{l}} - M_{\text{pol,rot}} \cdot \vec{S}^{(\mathrm{i})}_{\mathrm{r}}\right]_0 \tag{37}$$

$M_{\text{pol,rot}}$ is fully defined by the parameters $t_x^2$, $t_y^2$ (or their equivalents $\left(\mu^{(\mathrm{n})}_{x,y}\right)^2$) and the angle $\theta$ between $\vec{\mu}_1$ and the $x$-axis of the laboratory frame. Therefore, the artifact $g_{\mathrm{P}}$ can be calculated for given Stokes vectors $\vec{S}^{(\mathrm{i})}_{\mathrm{l}}$ and $\vec{S}^{(\mathrm{i})}_{\mathrm{r}}$ of the imperfectly circularly polarized excitation light. Consequently, quantifying the artifacts arising from spurious linear polarization requires finding expressions for the Stokes vectors $\vec{S}_{\mathrm{l}}^{(\mathrm{i})}$ and $\vec{S}_{\mathrm{r}}^{(\mathrm{i})}$ to be inserted into eq.(37). If we normalize the intensity of the incident Stokes vector, i.e. $S_0 = 1$, and use eq.(20), we can write

$$S_3 = \pm\sqrt{1 - S_1^2 - S_2^2} \tag{38}$$

and for small deviations from perfectly LCP and RCP light, where $S_3 = \mp 1$ holds, it follows that $|S_1|, |S_2| \ll 1$. Accordingly, the Stokes vectors $\vec{S}_{\mathrm{l,r}}^{(\mathrm{i})}$ are written as

$$\vec{S}_{\mathrm{l,r}}^{(\mathrm{i})} \approx \begin{pmatrix} 1 \\ S_{1_{\mathrm{l,r}}}^{(\mathrm{i})} \\ S_{2_{\mathrm{l,r}}}^{(\mathrm{i})} \\ \mp 1 \end{pmatrix} \tag{39}$$

where $\left|S_{1_{\mathrm{l,r}}}^{(\mathrm{i})}\right|, \left|S_{2_{\mathrm{l,r}}}^{(\mathrm{i})}\right| \ll 1$ refer to the residual linear polarization components of the (i)-LCP and (i)-RCP excitation light. Instead of finding expressions for the full Stokes vector $\vec{S}_{\mathrm{l}}^{(\mathrm{i})}$ and $\vec{S}_{\mathrm{r}}^{(\mathrm{i})}$ it is sufficient to quantify only the 1,2 components $S_{1_{\mathrm{l,r}}}^{(\mathrm{i})}$, $S_{2_{\mathrm{l,r}}}^{(\mathrm{i})}$. For doing so, it is convenient to decompose these contributions into a symmetric part and an antisymmetric part[46] by defining

$$\begin{aligned} S_{1,2}^{(\mathrm{sym})} &= \frac{1}{2}\left(S_{1,2_{\mathrm{l}}}^{(\mathrm{i})} + S_{1,2_{\mathrm{r}}}^{(\mathrm{i})}\right) \\ S_{1,2}^{(\mathrm{asym})} &= \frac{1}{2}\left(S_{1,2_{\mathrm{l}}}^{(\mathrm{i})} - S_{1,2_{\mathrm{r}}}^{(\mathrm{i})}\right) \end{aligned} \tag{40}$$

Then, eq.(39) reads

$$\vec{S}_{\mathrm{l,r}}^{(\mathrm{i})} \approx \begin{pmatrix} 1 \\ S_1^{(\mathrm{sym})} \pm S_1^{(\mathrm{asym})} \\ S_2^{(\mathrm{sym})} \pm S_2^{(\mathrm{asym})} \\ \mp 1 \end{pmatrix} \tag{41}$$

where the positive (negative) sign before $S_{1,2}^{(\mathrm{asym})}$ is valid for $\vec{S}_{\mathrm{l}}^{(\mathrm{i})}$ ($\vec{S}_{\mathrm{r}}^{(\mathrm{i})}$). Inserting $\vec{S}_{\mathrm{l,r}}^{(\mathrm{i})}$ into expression eq.(33) yields absorptions

$$\begin{aligned} A_{\mathrm{l,r}}^{(\mathrm{i})} &\propto \left[M_{\text{pol,rot}}(\theta) \cdot \vec{S}_{\mathrm{l,r}}^{(\mathrm{i})}\right]_0 \\ &\propto \frac{1}{2}\left(t_x^2 + t_y^2\right) \cdot \left(1 + LD \cdot \left(S_1^{(\mathrm{sym})} \pm S_1^{(\mathrm{asym})}\right) \cdot \cos 2\theta + LD \cdot (S_2^{(\mathrm{sym})} \pm S_2^{(\mathrm{asym})}) \cdot \sin 2\theta\right) \end{aligned} \tag{42}$$

and inserting in eq.(37) yields

$$g_{\mathrm{P}} = \left[ \underbrace{2 \cdot LD \cdot S_1^{(\mathrm{asym})} \cdot \cos 2\theta \cdot}_{g_{\mathrm{P}_1}} + \underbrace{2 \cdot LD \cdot S_2^{(\mathrm{asym})} \cdot \sin 2\theta}_{g_{\mathrm{P}_2}} \right] \tag{43}$$

for the artifact. This reveals that only the antisymmetric linear polarization components of the perturbed Stokes vector contribute to the polarization artifact in circular dichroism.

Now we have shifted the problem of finding expressions for the Stokes vectors $\vec{S}_{\mathrm{l}}^{(\mathrm{i})}$ and $\vec{S}_{\mathrm{r}}^{(\mathrm{i})}$ to the problem of finding an expression for $S_{1,2}^{(\mathrm{asym})}$. A general problem arises from the fact that the fraction of unwanted linear polarization components may vary across the Gaussian excitation intensity profile. In other words, $S_{1,2}^{(\mathrm{asym})}$ can additionally depend on the lateral coordinates $x$ and $y$. In order to find quantitative expressions for this variation we neglect intensity variations between the (i)-LCP and (i)-RCP excitation profiles, whose impact on a CD spectrum has been treated in Section 3.1, and focus on differences of the polarization compositions across the excitation profile. Therefore we make the spatial variation of the Stokes vector in eq.(41) more explicit by writing

$$\vec{S}_{\mathrm{l,r}}^{(\mathrm{i})}(x,y) \approx \underbrace{I_0 \cdot \exp\left( -\frac{1}{2\sigma_x^2}(x-\mu_x)^2 - \frac{1}{2\sigma_y^2}\left(y-\mu_y\right)^2 \right)}_{\text{intensity variation } I_0(x,y) \text{ along } x \text{ and } y} \cdot \underbrace{\begin{pmatrix} 1 \\ \pm S_1^{(\mathrm{asym})}(x,y) \\ \pm S_2^{(\mathrm{asym})}(x,y) \\ \mp 1 \end{pmatrix}}_{\text{polarization variation along } x,y} \tag{44}$$

Here we have taken into account that also the intensity prefactor $I_0(x,y)$ may depend on the lateral coordinate $x$ and $y$, and we used only the helicity-odd components, $S_{1,2}^{(\mathrm{asym})}$, of the Stokes vector, because the helicity-even components of the Stokes vector, $S_{1,2}^{(\mathrm{sym})}$, cancel out for calculating the artifact, eq.(43). Inserting the Stokes vector eq.(44) into eq.(33) yields effective absorption profiles

$$\begin{aligned} A_{\mathrm{l,r}}^{(\mathrm{i})}(x,y) &\propto \left[ M_{\mathrm{pol,rot}}(\theta) \cdot \vec{S}_{\mathrm{l,r}}^{(\mathrm{i})}(x,y) \right]_0 \\ &\propto \frac{1}{2}\left(t_x^2 + t_y^2\right) \cdot I_0(x,y) \cdot \left(1 \pm LD \cdot S_1^{(\mathrm{asym})}(x,y) \cdot \cos 2\theta \pm LD \cdot S_2^{(\mathrm{asym})}(x,y) \cdot \sin 2\theta\right) \end{aligned} \tag{45}$$

Now we have an expression for $A_{\mathrm{l,r}}^{(\mathrm{i})}(x,y)$ that can be inserted into eq.(34), which will be done separately for the denominator and the numerator of eq.(34). For the denominator of eq.(34) we use the approximation $S_{1,2}^{(\mathrm{asym})}(x,y) \approx 0$, which reads

$$g_{\mathrm{P}}(x,y) = 2\frac{A_{\mathrm{l}}^{(\mathrm{i})}(x,y) - A_{\mathrm{r}}^{(\mathrm{i})}(x,y)}{\left(A_{\mathrm{l}}^{(\mathrm{i})}(x,y) + A_{\mathrm{r}}^{(\mathrm{i})}(x,y)\right)} \approx 2\frac{A_{\mathrm{l}}^{(\mathrm{i})}(x,y) - A_{\mathrm{r}}^{(\mathrm{i})}(x,y)}{\left(t_x^2 + t_y^2\right) \cdot I_0(x,y)} \tag{46a}$$

Finally, inserting eq.(45) into the numerator of eq.(46a) yields for the polarization artifact

$$g_{\mathrm{P}}(x,y) = \left[\underbrace{2 \cdot LD \cdot S_1^{(\mathrm{asym})}(x,y) \cdot \cos 2\theta \cdot}_{g_{\mathrm{P}_1}} + \underbrace{2 \cdot LD \cdot S_2^{(\mathrm{asym})}(x,y) \cdot \sin 2\theta}_{g_{\mathrm{P}_2}}\right] \quad (46b)$$

**3.3 Experimental characterization of the artifacts**

In a real CD experiment the magnitude of the contributions to $g_{\mathrm{I}}$ (eq.(15)) and $g_{\mathrm{P}}$ (eqs.(46)) due to unavoidable experimental imperfections need to be quantified and the CD spectrum has to be corrected accordingly.

**3.3.1 Intensity artifacts**

Obtaining the contributions of imbalanced intensities for LCP and RCP light requires determining the peak intensities $I_{0_{\mathrm{l}}}$ and $I_{0_{\mathrm{r}}}$, the spatial distance $\Delta\mu_{\mathrm{lr}}$ of the peak positions between both beams, and both the difference in widths, $\Delta\sigma_{\mathrm{lr}}^2$ , of the two beams as well as the width, $\sigma^2$, itself. However, experimentally it is easier to measure the excitation power $P_{\mathrm{l,r}}$ of the two beams rather than the peak intensities $I_{0_{\mathrm{l,r}}}$. For a Gaussian excitation profile these are related via

$$P_{\mathrm{l,r}} = \int_{-\infty}^{+\infty} I_{\mathrm{l,r}}(x)\, dx = \sqrt{2\pi} \cdot I_{0_{\mathrm{l,r}}} \cdot \sigma_{\mathrm{l,r}} \quad (47a)$$

$$I_{0_{\mathrm{l,r}}} = \frac{P_{\mathrm{l,r}}}{\sqrt{2\pi} \cdot \sigma_{\mathrm{l,r}}} \quad (47b)$$

For small intensity differences where $\frac{I_{0_{\mathrm{l}}} - I_{0_{\mathrm{r}}}}{I_{0_{\mathrm{l}}} + I_{0_{\mathrm{r}}}} \ll 1$, it follows that $P_{\mathrm{l,r}} = P + \delta P_{\mathrm{l,r}}$, with $\frac{\delta P_{\mathrm{l,r}}}{P} \ll 1$. By additionally accounting for a slight difference in beam diameter via $\sigma_{\mathrm{l,r}}^2 = \sigma^2 + \delta\sigma_{\mathrm{l,r}}^2$, eq.(47b) reads

$$I_{0_{\mathrm{l,r}}}(\delta P_{\mathrm{l,r}}, \delta\sigma_{\mathrm{l,r}}^2) = \frac{1}{\sqrt{2\pi}} \cdot \frac{P + \delta P_{\mathrm{l,r}}}{\sqrt{\sigma^2 + \delta\sigma_{\mathrm{l,r}}^2}} \quad (48)$$

Expanding this in $\delta\sigma_{\mathrm{l,r}}^2$ and $\delta P_{\mathrm{l,r}}$ up to first order gives

$$
\begin{aligned}
I_{0_{\mathrm{l,r}}} &\approx I_{0_{\mathrm{l,r}}}(0{,}0) + \left.\frac{\partial I_{0_{\mathrm{l,r}}}}{\partial \delta P_{\mathrm{l,r}}}\right|_{0,0} \cdot \delta P_{\mathrm{l,r}} + \left.\frac{\partial I_{0_{\mathrm{l,r}}}}{\partial \delta \sigma_{\mathrm{l,r}}^2}\right|_{0,0} \cdot \delta \sigma_{\mathrm{l,r}}^2 \\
&\approx \frac{1}{\sqrt{2\pi}} \cdot \left[ \frac{P}{\sigma} + \frac{1}{\sigma} \underbrace{(P_{\mathrm{l,r}} - P)}_{\delta P_{\mathrm{l,r}}} - \frac{1}{2} \frac{P}{\sigma^3} \underbrace{(\sigma_{\mathrm{l,r}}^2 - \sigma^2)}_{\delta \sigma_{\mathrm{l,r}}^2} \right]
\end{aligned}
\tag{49}
$$

and hence for the first term of eqs.(15) we find

$$
\begin{aligned}
&2\frac{I_{0_\mathrm{l}} - I_{0_\mathrm{r}}}{I_{0_\mathrm{l}} + I_{0_\mathrm{r}}} \\
&= 2\frac{\left[\frac{P}{\sigma} + \frac{1}{\sigma}(P_\mathrm{l} - P) - \frac{1}{2}\frac{P}{\sigma^3}\left(\sigma_\mathrm{l}^2 - \sigma^2\right)\right] - \left[\frac{P}{\sigma} + \frac{1}{\sigma}(P_\mathrm{r} - P) - \frac{1}{2}\frac{P}{\sigma^3}(\sigma_\mathrm{r}^2 - \sigma^2)\right]}{\left[\frac{P}{\sigma} + \frac{1}{\sigma}(P_\mathrm{l} - P) - \frac{1}{2}\frac{P}{\sigma^3}(\sigma_\mathrm{l}^2 - \sigma^2)\right] + \left[\frac{P}{\sigma} + \frac{1}{\sigma}(P_\mathrm{r} - P) - \frac{1}{2}\frac{P}{\sigma^3}(\sigma_\mathrm{r}^2 - \sigma^2)\right]}
\end{aligned}
\tag{50}
$$

By applying the approximations $\delta\sigma_{\mathrm{l,r}}^2 = \left(\sigma_{\mathrm{l,r}}^2 - \sigma^2\right) \approx 0$ and $\delta P_{\mathrm{l,r}} = \left(P_{\mathrm{l,r}} - P\right) \approx 0$ in the denominator, alongside the typically valid relation $2P \approx P_\mathrm{l} + P_\mathrm{r}$ in the numerator, eq.(50) simplifies to

$$
2\frac{I_{0_\mathrm{l}} - I_{0_\mathrm{r}}}{I_{0_\mathrm{l}} + I_{0_\mathrm{r}}} \approx 2\frac{P_\mathrm{l} - P_\mathrm{r}}{P_\mathrm{l} + P_\mathrm{r}} - \frac{1}{2}\frac{\Delta\sigma_{\mathrm{lr}}^2}{\sigma^2}
\tag{51}
$$

Inserting (51) into (15a) gives

$$
g_\mathrm{I} = \left[2\frac{P_\mathrm{l} - P_\mathrm{r}}{P_\mathrm{l} + P_\mathrm{r}} - \frac{\Delta\sigma_{\mathrm{lr}}^2}{2\sigma^2}\right] + \frac{\Delta\mu_{\mathrm{lr}}}{\sigma^2} \cdot (x - \mu)^1 + \frac{\Delta\sigma_{\mathrm{lr}}^2}{2\sigma^4} \cdot (x - \mu)^2
\tag{52a}
$$

The results for the 2D Gaussian excitation profiles using eq.(15b) is

$$
\begin{aligned}
g_\mathrm{I} = &\left[2\frac{P_\mathrm{l} - P_\mathrm{r}}{P_\mathrm{l} + P_\mathrm{r}} - \frac{\Delta\sigma_{\mathrm{lr},x}^2}{2\sigma_x^2} - \frac{\Delta\sigma_{\mathrm{lr},y}^2}{2\sigma_y^2}\right] + \frac{\Delta\mu_{\mathrm{lr},x}}{\sigma_x^2} \cdot (x - \mu_x)^1 + \frac{\Delta\mu_{\mathrm{lr},y}}{\sigma_y^2} \cdot \left(y - \mu_y\right)^1 + \frac{\Delta\sigma_{\mathrm{lr},x}^2}{2\sigma_x^4} \cdot (x - \mu_x)^2 \\
&+ \frac{\Delta\sigma_{\mathrm{lr},y}^2}{2\sigma_y^4} \cdot \left(y - \mu_y\right)^2 + \frac{\Delta\sigma_{\mathrm{lr},xy}}{\sigma_x^2\sigma_y^2}(x - \mu_x)\left(y - \mu_y\right)
\end{aligned}
\tag{52b}
$$

According to eq.(52b), the artifacts due to an imbalanced excitation intensity in the LCP and RCP beams are experimentally accessible by measuring the differences of the zeroth, first, and second moments of the excitation intensity distributions for LCP and RCP light.

The zeroth moments

$$
P_{\mathrm{l,r}} = \iint_{-\infty}^{+\infty} I_{\mathrm{l,r}}(x, y)\, dxdy
\tag{53}
$$

can be obtained with a simple photodiode.

The first moments correspond to

$$\mu_{\mathrm{l,r};x} = \langle x \rangle_{\mathrm{l,r}} = \frac{\iint_{-\infty}^{\infty} x \cdot I_{\mathrm{l,r}}(x,y)\, dxdy}{P_{\mathrm{l,r}}} \tag{54}$$

and can be measured with a lateral effect sensor or a quadrant photodiode.
Finally the second moments given as

$$\sigma^2_{\mathrm{l,r};x} = \langle x^2 \rangle_{\mathrm{l,r}} - \langle x \rangle^2_{\mathrm{l,r}} = \frac{\iint_{-\infty}^{\infty} \left(x - \mu_{\mathrm{l,r};x}\right)^2 \cdot I_{\mathrm{l,r}}(x,y)\, dxdy}{P_{\mathrm{l,r}}} \tag{55}$$

can be determined with rotatable photodiode array or a CCD camera. The values of $\sigma^2$ and $\sigma^4$ in the prefactors can be obtained from the $\frac{1}{e^2}$ beam waist $\omega^2 = 4\sigma^2$. All equations apply in the principal axis system of the geometrical beam profile, and can equivalently be used for the $y$ direction.

### 3.3.2 Polarization artifacts

The contributions to a CD spectrum from polarization artifacts have been quantified in eq.(46), and require to determine the helicity-odd components $S_{1,2}^{(\mathrm{asym})}(x,y)$ of the Stokes vector. Because of the equivalence between the polarization properties of radiation absorbed by a linear absorber and those transmitted through a linear polarizer, the expression $A_{\mathrm{l,r}}^{(\mathrm{i})} \propto \left[\vec{S'}_{\mathrm{l,r}}^{(\mathrm{i})}\right]_0$ in eq.(33) can be replaced by the expression $I_{\mathrm{l,r}}^{(\mathrm{i})} = \left[\vec{S'}_{\mathrm{l,r}}^{(\mathrm{i})}\right]_0$. Since $I_{\mathrm{l,r}}^{(\mathrm{i})}(x,y)$ denotes the intensity transmitted through a linear polarizer (with LD = 1 by definition) eq.(45) can be rewritten as

$$\begin{aligned} I_{\mathrm{l,r}}^{(\mathrm{i})}(x,y) &= \left[M_{\text{pol,rot}}(\theta) \cdot \vec{S}_{\mathrm{l,r}}^{(\mathrm{i})}(x,y)\right]_0 \\ &= \frac{1}{2}\left(t_x^2 + t_y^2\right) \cdot I_0(x,y) \cdot \left(1 \pm S_1^{(\mathrm{asym})}(x,y) \cdot \cos(2\theta) \pm S_2^{(\mathrm{asym})}(x,y) \cdot \sin(2\theta)\right) \end{aligned} \tag{56}$$

Because $\cos(2\theta)$ and $\sin(2\theta)$ are orthogonal with respect to each other, the coefficients $S_{1,2}^{(\mathrm{asym})}(x,y)$ can be determined independently by two experiments with a linear polarizer that is oriented under $\theta = 0°$, and $\theta = 45°$ with respect to the principal axis of the 2D Gaussian intensity profile, which for $\sigma_x^2 \approx \sigma_y^2$ coincides with the $x$-axis of the laboratory frame. With the abbreviation $T = \frac{1}{2}\left(t_x^2 + t_y^2\right)$ this yields

$$I_{\mathrm{l,r}}^{(\mathrm{i})}(x,y,\theta = 0°) = T \cdot I_0(x,y) \cdot \left[1 \pm S_1^{(\mathrm{asym})}(x,y)\right] \tag{57a}$$

$$I_{\mathrm{l,r}}^{(\mathrm{i})}(x,y,\theta = 45°) = T \cdot I_0(x,y) \cdot \left[1 \pm S_2^{(\mathrm{asym})}(x,y)\right] \tag{57b}$$

For handling the spatial variations, $S_{1,2}^{(\mathrm{asym})}(x,y)$ is expanded to the second order in $x$ and $y$. For convenience, in the following the index 1 refers to $\theta = 0°$ and the index 2 to $\theta = 45°$, and the derivatives in eq.(58) have been abbreviated as $a_{1,2}$ to $f_{1,2}$.

$$
\begin{aligned}
S_{1,2}^{(\mathrm{asym})}(x,y) &= S_{1,2}^{(\mathrm{asym})}(0,0) + \left.\frac{\partial S_{1,2}^{(\mathrm{asym})}}{\partial x}\right|_{0,0} \cdot x + \left.\frac{\partial S_{1,2}^{(\mathrm{asym})}}{\partial y}\right|_{0,0} y \\
&\quad + \frac{1}{2}\left.\frac{\partial^2 S_{1,2}^{(\mathrm{asym})}}{\partial x^2}\right|_{0,0} \cdot x^2 + \frac{1}{2}\left.\frac{\partial^2 S_{1,2}^{(\mathrm{asym})}}{\partial y^2}\right|_{0,0} \cdot y^2 + \left.\frac{\partial^2 S_{1,2}^{(\mathrm{asym})}}{\partial x\,\partial y}\right|_{0,0} \cdot xy \\
&= a_{1,2} + b_{1,2}x + c_{1,2}y + \frac{1}{2}d_{1,2}x^2 + \frac{1}{2}e_{1,2}y^2 + f_{1,2}xy
\end{aligned}
\tag{58}
$$

Since we are only interested in the difference for (i)-LCP and (i)-RCP excitation we choose as origin $\mu_x = \mu_y = 0$, and obtain

$$
I_{\mathrm{l,r}}^{(\mathrm{i})}(x,y,\theta = 0°) \approx T \cdot I_0(x,y) \cdot \left[1 \pm \left(a_1 + b_1 x + c_1 y + \frac{1}{2}d_1 x^2 + \frac{1}{2}e_1 y^2 + f_1 xy\right)\right] \tag{59a}
$$

$$
I_{\mathrm{l,r}}^{(\mathrm{i})}(x,y,\theta = 45°) \approx T \cdot I_0(x,y) \cdot \left[1 \pm \left(a_2 + b_2 x + c_2 y + \frac{1}{2}d_2 x^2 + \frac{1}{2}e_2 y^2 + f_2 xy\right)\right] \tag{59b}
$$

It will turn out convenient again to calculate the moments of the intensity distribution, eq.(53) to (55); for details the reader is referred to the Appendix. This yields

$$
\left(\frac{P_\mathrm{l} - P_\mathrm{r}}{P_\mathrm{l} + P_\mathrm{r}}\right)_{1,2} \approx \left[a_{1,2} + \frac{\sigma_x^2}{2}d_{1,2} + \frac{\sigma_y^2}{2}e_{1,2}\right] \tag{60}
$$

and

$$
\begin{aligned}
\left(\Delta\mu_{\mathrm{lr},x}\right)_{1,2} &= (\langle x\rangle_\mathrm{l} - \langle x\rangle_\mathrm{r})_{1,2} \approx 2\sigma_x^2 \cdot b_{1,2} \\
\left(\Delta\mu_{\mathrm{lr},y}\right)_{1,2} &= (\langle y\rangle_\mathrm{l} - \langle y\rangle_\mathrm{r})_{1,2} \approx 2\sigma_y^2 \cdot c_{1,2} \\
\left(\Delta\sigma_{\mathrm{lr},x}^2\right)_{1,2} &\approx (\langle \mathrm{x}^2\rangle_\mathrm{l} - \langle \mathrm{x}^2\rangle_\mathrm{r})_{1,2} \approx 2\sigma_x^4 \cdot d_{1,2} \\
\left(\Delta\sigma_{\mathrm{lr},y}^2\right)_{1,2} &\approx (\langle y^2\rangle_\mathrm{l} - \langle y^2\rangle_\mathrm{r})_{1,2} \approx 2\sigma_y^4 \cdot e_{1,2} \\
\left(\Delta\sigma_{\mathrm{lr},xy}\right)_{1,2} &\approx (\langle xy\rangle_\mathrm{l} - \langle xy\rangle_\mathrm{r})_{1,2} \approx 2\sigma_x^2\sigma_y^2 \cdot f_{1,2}
\end{aligned}
\tag{61}
$$

and reveals that the moments of the intensity distribution after a polarizer are directly related to the Taylor expansion coefficients of $S_{1,2}^{(\mathrm{asym})}(x,y)$, eq.(58), via

$$
\begin{aligned}
a_{1,2} &\approx \left(\frac{P_\mathrm{l}-P_\mathrm{r}}{P_\mathrm{l}+P_\mathrm{r}}\right)_{1,2} - \frac{\left(\Delta\sigma^2_{\mathrm{lr},x}\right)_{1,2}}{4\sigma_x^2} - \frac{\left(\Delta\sigma^2_{\mathrm{lr},y}\right)_{1,2}}{4\sigma_y^2}\\
b_{1,2} &\approx \frac{1}{2\sigma_x^2}\left(\Delta\mu_{\mathrm{lr},x}\right)_{1,2}\\
c_{1,2} &\approx \frac{1}{2\sigma_y^2}\left(\Delta\mu_{\mathrm{lr},y}\right)_{1,2}\\
d_{1,2} &\approx \frac{1}{2\sigma_x^4}\left(\Delta\sigma^2_{\mathrm{lr},x}\right)_{1,2}\\
e_{1,2} &\approx \frac{1}{2\sigma_y^4}\left(\Delta\sigma^2_{\mathrm{lr},y}\right)_{1,2}\\
f_{1,2} &\approx \frac{1}{2\sigma_x^2\sigma_y^2}\left(\Delta\sigma_{\mathrm{lr},xy}\right)_{1,2}
\end{aligned}
\tag{62}
$$

Inserting the coefficients into eq.(58) yields

$$
\begin{aligned}
S_{1,2}^{(\mathrm{asym})}(x,y) = &\left[\left(\frac{P_\mathrm{l}-P_\mathrm{r}}{P_\mathrm{l}+P_\mathrm{r}}\right)_{1,2} - \frac{\left(\Delta\sigma^2_{\mathrm{lr},x}\right)_{1,2}}{4\sigma_x^2} - \frac{\left(\Delta\sigma^2_{\mathrm{lr},y}\right)_{1,2}}{4\sigma_y^2}\right] + \frac{\left(\Delta\mu_{\mathrm{lr},x}\right)_{1,2}}{2\sigma_x^2}x\\
&+ \frac{\left(\Delta\mu_{\mathrm{lr},y}\right)_{1,2}}{2\sigma_y^2}y + \frac{\left(\Delta\sigma^2_{\mathrm{lr},x}\right)_{1,2}}{4\sigma_x^4}x^2 + \frac{\left(\Delta\sigma^2_{\mathrm{lr},y}\right)_{1,2}}{4\sigma_y^4}y^2 + \frac{\left(\Delta\sigma_{\mathrm{lr},xy}\right)_{1,2}}{2\sigma_x^2\sigma_y^2}xy
\end{aligned}
\tag{63}
$$

and together with eq.(46b) follows

$$
\begin{aligned}
g_{\mathrm{P}_{1,2}} \approx LD \cdot &\left[\left(2\left(\frac{P_\mathrm{l}-P_\mathrm{r}}{P_l+P_r}\right)_{1,2} - \frac{\left(\Delta\sigma^2_{\mathrm{lr},x}\right)_{1,2}}{2\sigma_x^2} - \frac{\left(\Delta\sigma^2_{\mathrm{lr},y}\right)_{1,2}}{2\sigma_y^2}\right) + \frac{\left(\Delta\mu_{\mathrm{lr},x}\right)_{1,2}}{\sigma_x^2}x\right.\\
&\left.+ \frac{\left(\Delta\mu_{\mathrm{lr},y}\right)_{1,2}}{\sigma_y^2}y + \frac{\left(\Delta\sigma^2_{\mathrm{lr},x}\right)_{1,2}}{2\sigma_x^4}x^2 + \frac{\left(\Delta\sigma^2_{\mathrm{lr},y}\right)_{1,2}}{2\sigma_y^4}y^2 + \frac{\left(\Delta\sigma_{\mathrm{lr},xy}\right)_{1,2}}{\sigma_x^2\sigma_y^2}xy\right]\\
&\cdot\begin{cases} g_{\mathrm{P}_1}: \cos 2\theta \\ g_{\mathrm{P}_2}: \sin 2\theta \end{cases}
\end{aligned}
\tag{64}
$$

This equation is the counterpart of eq.(52) that was derived for imbalanced excitation intensity. All contributions can be determined in the same way as before, yet measured behind a linear polarizer oriented at either $\theta = 0°$ or $\theta = 45°$ with respect to the $x$-axis of the laboratory frame.

### 3.3.3 Summary

A complete characterization of possible artifacts arising from imbalanced excitation intensities and/or imperfect circular polarization of LCP and RCP light requires to determine the zeroth, first, and second moment of the intensity distribution of the excitation light for three experimental configurations: (i) without a polarizer, and with a linear polarizer oriented at (ii) 0° and (iii) 45° with respect to the principal $x$-axis of the 2D Gaussian intensity profile. This is depicted in Fig.6, which illustrates the sources for intensity and polarization artifacts and how they can be measured.

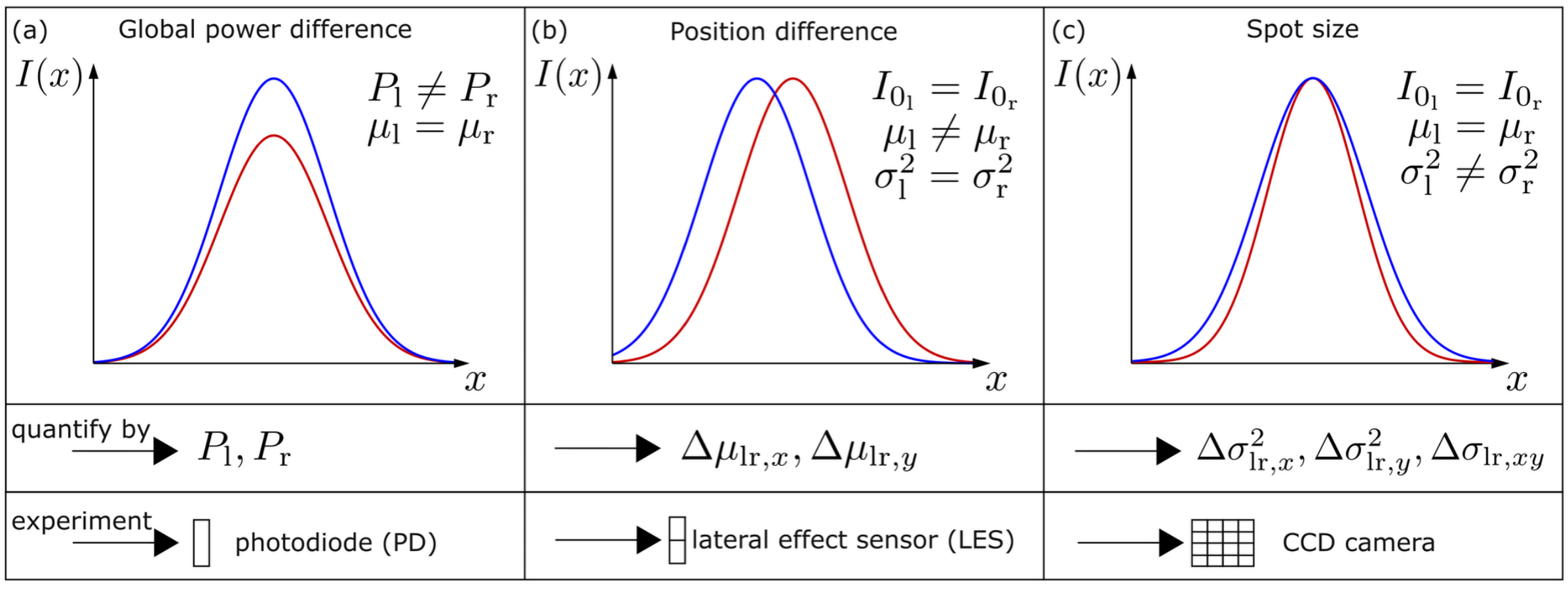

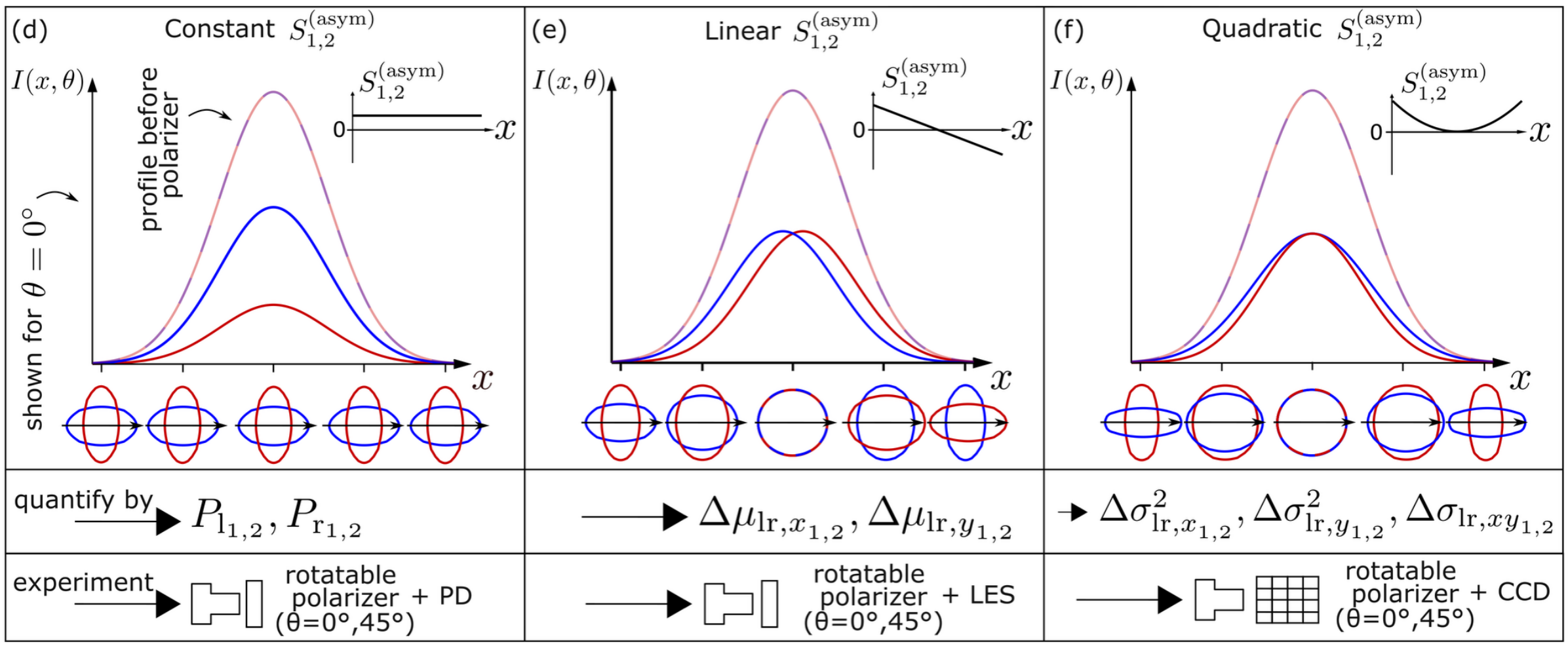

***Fig.6**: Illustration of the sources for intensity (top) and polarization (bottom) artifacts and how their magnitude can be quantified experimentally. For details see text. Inspired by[47].*

This characterization of these artifacts is valid as long as the spatial variation of $S_{1,2}^{(\mathrm{asym})}(x,y)$ can be approximated by a second-order polynomial, and as long as the intensity profile is Gaussian-like and can be described by the second moment method[48]. It is worth noting that in real experiments $\Delta\sigma_{\mathrm{lr}}^2$ is typically very small, and it is sufficient to measure only the artifacts that arise from contributions of the zeroth and first moments.

## 4. Generation of circularly polarized light with a Pockels cell

### 4.1 Mode of operation of a Pockels cell

For generating circularly polarized light, we employ a longitudinal Pockels cell (Pockels cell 3 in Fig.2) based on a KD*P crystal. A Pockels cell is an electro-optic device that allows to alter the index of refraction upon application of a voltage and thereby allowing control of the polarization state of the transmitted light. KD*P is a birefringent material with refractive indices $n_\mathrm{o} = 1.5021$ for the ordinary beam and $n_\mathrm{e} = 1.4638$ for the extraordinary beam (@ $694.3$ nm). Upon application of a voltage $V$ along the short axis of the index ellipsoid ($n_\mathrm{e}$ axis) the crystal symmetry is reduced to orthorhombic, resulting in modified principal refractive indices

$$\begin{aligned} n_1 &= n_\mathrm{o} - \frac{1}{2} n_\mathrm{o}^3 r_{63} \frac{V}{d} \\ n_2 &= n_\mathrm{o} + \frac{1}{2} n_\mathrm{o}^3 r_{63} \frac{V}{d} \\ n_3 &= n_\mathrm{e} \end{aligned} \qquad (65)$$

where $r_{63}$ is the electro-optic coefficient, which is approximately $-25 \cdot 10^{-12}\,\frac{\mathrm{m}}{V}$ for KD*P[49,50], and $d$ is the length of the crystal. The artifacts introduced by a Pockels cell have been systematically studied in[46,47,51–54] and their results have been used as a guide for our work.

For perfect optical alignment such that the propagation direction of the incident light coincides with the $n_\mathrm{e}$ axis of the index ellipsoid, the two other axes of the index ellipsoid lie in the plane perpendicular to the propagation direction and can be aligned such that they are oriented under ±45° with respect to the $x$-axis, see Fig.7. In the following the principal axes associated with the refractive indices $n_1$ and $n_2$ will be referred to as $n_1$ and $n_2$ axes irrespective of the applied voltage.

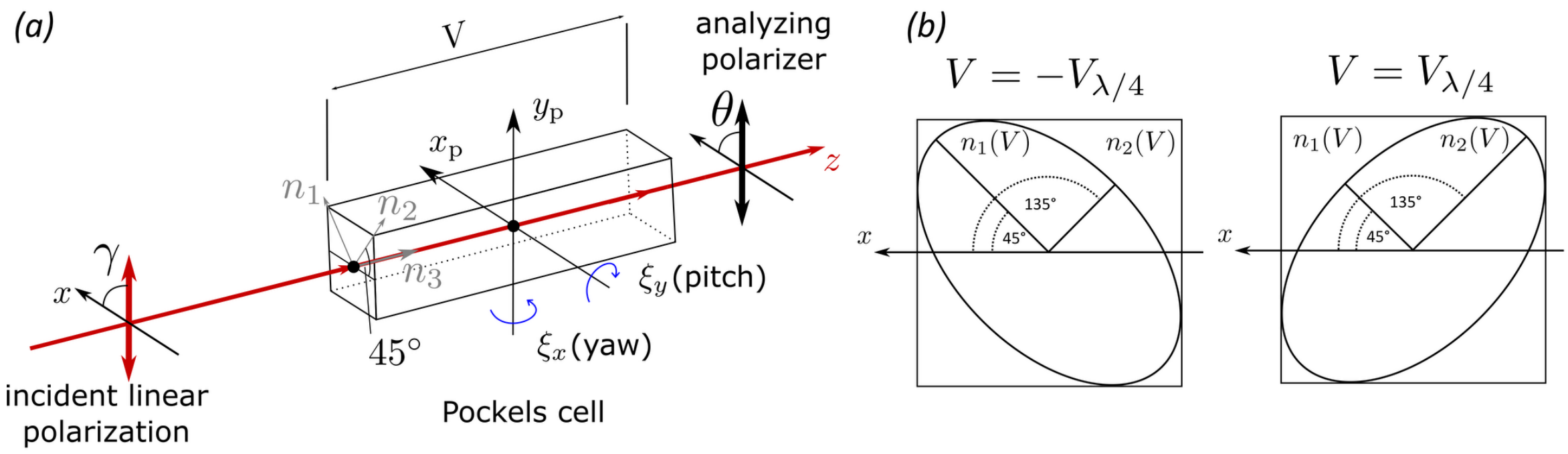

***Fig.7**: (a) Mode of operation of a Pockels cell. Light that is linearly polarized at an angle $\gamma$ with respect to the horizontal axis ($x$-axis) of the laboratory frame propagates along the $z$-direction through a Pockels cell. The cell is positioned, (coordinates: $x_p$ and $y_p$) and oriented (angles: yaw ($\xi_x$) and pitch ($\xi_y$)) such that both its optical axis and the direction of the applied voltage $V$ coincide with the $z$-axis. A linear polarizer oriented at an angle $\theta$ relative to the $x$-axis serves to quantify residual linearly polarized components (b) Front view of the Pockels cell and principal axes of the index ellipsoid perpendicular to the propagation direction of the incident light as a function of the applied voltage. The illustration shows the (exaggerated) situation when applying $V = \mp V_{\lambda/4}$, which corresponds to the voltages that induce a phase shift of $\varphi = \mp\frac{\pi}{2}$ between the beams polarized along the $n_1$ and $n_2$ axes.*

Hence, light that is linearly polarized along the $y$-direction entering the cell will be split into two mutually orthogonal polarized beams that accumulate a phase difference φ given as

$$\varphi = \frac{2\pi d}{\lambda}(n_2 - n_1) = \frac{2\pi}{\lambda} n_o^3 r_{63} V \tag{66}$$

while propagating through the medium. However, since the optical medium is a crystal, spatial and angular gradients in the refractive index of the crystal, induced by the applied electric fields required to generate LCP and RCP light, can lead to a small angular separation between the beams[47,51]. The magnitude of this effect depends on the specific cell used, but it can be minimized by translational adjustment toward the electrical center of the KD*P crystal[47]. It can be quantified by measuring the separation of the centroids of the two beam profiles $\Delta\mu_{\mathrm{lr}}$ (see Section 3.3) at a distance $l$ behind the Pockels cell. This yields the angular beam deviation (BD) defined as

$$BD_{x,y} = \arctan\left(\frac{\Delta\mu_{\mathrm{lr},x,y}}{l}\right) \approx \frac{\Delta\mu_{\mathrm{lr},x,y}}{l} \tag{67}$$

for the two lateral coordinates $x$ and $y$. For KD*P-based cells, the BD typically ranges from several tens to a few hundreds of nanoradians (nrad), which can lead to non-negligible intensity differences between LCP and RCP light at the sample position.

According to eq.(52b) $\Delta\mu_{\mathrm{lr}} \neq 0$ is directly related to the first-order intensity artifact of the $g_{\mathrm{I}}$ artifact

$$g_{\mathrm{I}} = \ldots + \frac{\Delta\mu_{\mathrm{lr},x}}{\sigma_x^2} \cdot (x - \mu_x)^1 + \frac{\Delta\mu_{\mathrm{lr},y}}{\sigma_y^2} \cdot (y - \mu_y)^1 + \ldots \tag{68}$$

### 4.2 A Pockels cell as a voltage-controlled optical retarder

According to eq.(66) the cell can act as a voltage-controlled optical retarder. Application of voltages $\mp V_{\lambda/4}$ that induce phase shifts of $\varphi = \mp\frac{\pi}{2} \,\hat{=}\, \mp\frac{\lambda}{4}$ convert incident linearly polarized light into LCP and RCP light. For a single longitudinal KD*P Pockels cell operated with light at around 700 nm typical values for $V_{\lambda/4}$ are in the range 2000 to 3000 V. CD spectroscopy requires inversion of the polarity of the applied voltages to switch between LCP and RCP light. In a real experiment, however, reversing the polarity requires a finite time $dt$, and both the voltage waveform as well as the amplitude may deviate from the ideal shape and the values of $\mp V_{\lambda/4}$, Fig.8. We first consider the influence of these deviations on the polarization state of the light exiting the Pockels cell, assuming perfect parallel alignment of the optical axis of the cell and the propagation direction of the incident light ($z$-axis). Subsequently, we analyze the impact of a small, unavoidable misalignment of the cell on the polarization state of the light exiting the cell.

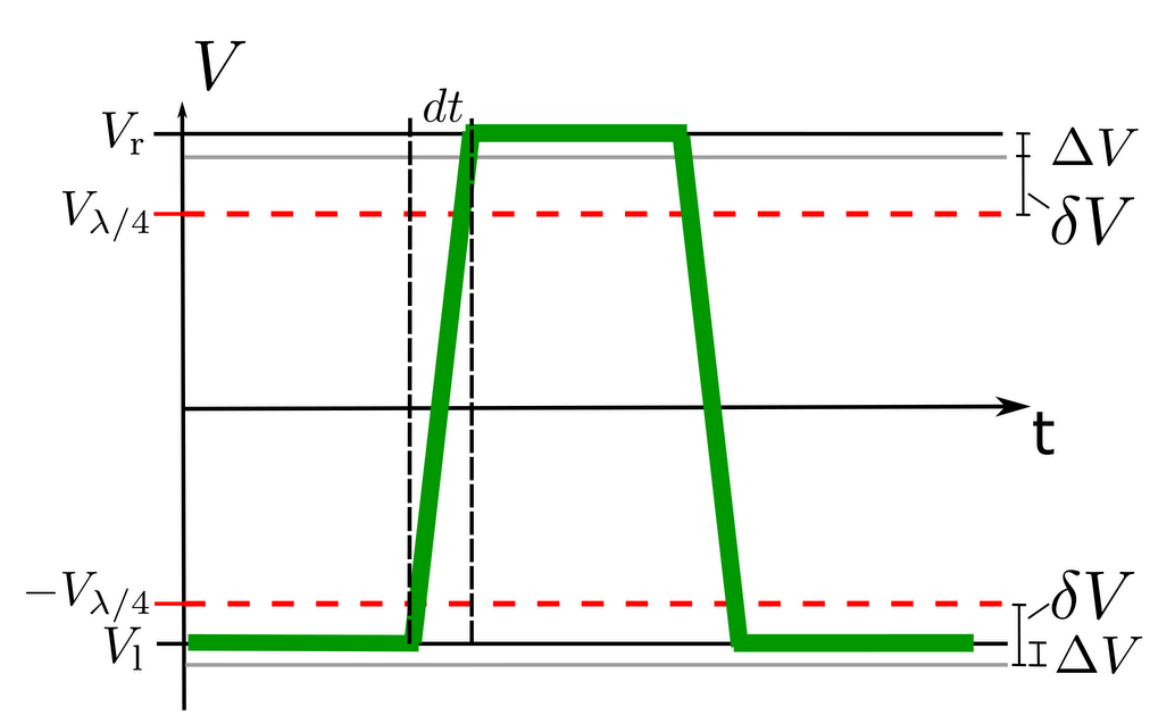

***Fig.8**: The dashed red lines represent the ideal voltages $\mp V_{\lambda/4}$ that would result in perfect phase shifts of $\varphi = \mp\frac{\pi}{2}$ between the two beams propagating through the cell. The experimentally applied voltages for obtaining LCP and RCP light, $V_l$ and $V_r$, deviate from the ideal voltages and are shown by the green line. The deviation from the ideal voltages are split into a symmetric part, $\delta V$, and an antisymmetric part, $\Delta V$, for details see text. The time it takes to invert the polarity of the voltages is $dt$.*

If we denote the voltages applied for obtaining LCP and RCP light $V_{\mathrm{l}}$ and $V_{\mathrm{r}}$, respectively, and their (small) deviations from the ideal voltages $V_{\mathrm{l}}'$ and $V_{\mathrm{r}}'$, we have

$$\begin{aligned} V_{\mathrm{l}} &= -V_{\lambda/4} + V_{\mathrm{l}}' \\ V_{\mathrm{r}} &= +V_{\lambda/4} + V_{\mathrm{r}}' \end{aligned} \tag{69}$$

It will turn out to be beneficial to define

$$\delta V = \frac{1}{2}(V_\mathrm{l}' - V_\mathrm{r}')$$
$$\Delta V = \frac{1}{2}(V_\mathrm{l}' + V_\mathrm{r}') \tag{70a}$$

which allows to transform eq.(69) into

$$V_\mathrm{l} = -\left(V_{\lambda/4} - \delta V\right) + \Delta V$$
$$V_\mathrm{r} = +\left(V_{\lambda/4} - \delta V\right) + \Delta V \tag{70b}$$

and similarly from eq.(66) for the corresponding phases, $\varphi_\mathrm{l}$ and $\varphi_\mathrm{r}$,

$$\varphi_\mathrm{l} \overset{(66)}{=} \frac{2\pi}{\lambda} n_\mathrm{o}^3 r_{63} V_\mathrm{l} \overset{(70)}{=} \frac{2\pi}{\lambda} n_\mathrm{o}^3 r_{63}\left[-\left(V_{\lambda/4} - \delta V\right) + \Delta V\right] = -\left(\frac{\pi}{2} - \delta\varphi\right) + \Delta\varphi$$
$$\varphi_\mathrm{r} \overset{(66)}{=} \frac{2\pi}{\lambda} n_\mathrm{o}^3 r_{63} V_\mathrm{r} \overset{(70)}{=} \frac{2\pi}{\lambda} n_\mathrm{o}^3 r_{63}\left[\left(V_{\lambda/4} - \delta V\right) + \Delta V\right] = \left(\frac{\pi}{2} - \delta\varphi\right) + \Delta\varphi \tag{71}$$

where the deviations of $\varphi_\mathrm{l}$ and $\varphi_\mathrm{r}$ from $\mp\frac{\pi}{2}$ can also be associated with a symmetric part $\delta\varphi$ and an antisymmetric part $\Delta\varphi$ of the phase shift. Eq. (71) reveals that symmetric deviations change the absolute phases of both beams by the same amount, causing their phases to shift in opposite directions. In contrast, antisymmetric deviations decrease the absolute phase of one beam while increasing the other, effectively shifting both phases in the same direction. According to eq.(19) the Stokes vector for light that is perfectly linearly polarized light along the $y$-axis of the laboratory frame reads

$$S_\mathrm{lin} = \begin{pmatrix} S_0 \\ S_1 \\ S_2 \\ S_3 \end{pmatrix} = \begin{pmatrix} I_{0^\circ} + I_{90^\circ} \\ I_{0^\circ} - I_{90^\circ} \\ I_{45^\circ} - I_{-45^\circ} \\ I_{RCP} - I_{LCP} \end{pmatrix} = \begin{pmatrix} 1 \\ -1 \\ 0 \\ 0 \end{pmatrix} \tag{72}$$

Since the Pockels cell acts like a retarder the Stokes vectors of the LCP and RCP light behind the cell are given by

$$\vec{S}_\mathrm{l}^{(\mathrm{i})} = M_\mathrm{ret,l} \cdot S_\mathrm{lin} \tag{73a}$$

$$\vec{S}_\mathrm{r}^{(\mathrm{i})} = M_\mathrm{ret,r} \cdot S_\mathrm{lin} \tag{73b}$$

where $M_\mathrm{ret,l,r}$ refer to the Mueller matrices of a retarder generating LCP or RCP light. The generic Mueller matrix of a retarder is given as

$$M_\mathrm{ret}(\phi,\varphi) = \begin{pmatrix} 1 & 0 & 0 & 0 \\ 0 & \cos^2(2\phi) + \sin^2(2\phi)\cos(\varphi) & (1-\cos(\varphi))\sin(2\phi)\cos(2\phi) & -\sin(\varphi)\sin(2\phi) \\ 0 & (1-\cos(\varphi))\sin(2\phi)\cos(2\phi) & \sin^2(2\phi) + \cos(\varphi)\cos^2(2\phi) & \sin(\varphi)\cos(2\phi) \\ 0 & \sin(\varphi)\sin(2\phi) & -\sin(\varphi)\cos(2\phi) & \cos(\varphi) \end{pmatrix} \tag{74}$$

where $\phi$ denotes the angle of the $n_2$ axis of the retarder relative to the horizontal axis of the laboratory frame ($x$-axis) and $\varphi$ corresponds to the resulting phase shift. Using $\phi = 135°$ , see Fig.7b, we find the Mueller matrices for the induced phases $\varphi_{\mathrm{l}}$ and $\varphi_{\mathrm{r}}$:

$$M_{\mathrm{ret,l}} = M_{\mathrm{ret}}(135°, \varphi_{\mathrm{l}}) = \begin{pmatrix} 1 & 0 & 0 & 0 \\ 0 & \cos(\varphi_{\mathrm{l}}) & 0 & \sin(\varphi_{\mathrm{l}}) \\ 0 & 0 & 1 & 0 \\ 0 & -\sin(\varphi_{\mathrm{l}}) & 0 & \cos(\varphi_{\mathrm{l}}) \end{pmatrix}$$
$$M_{\mathrm{ret,r}} = M_{\mathrm{ret}}(135°, \varphi_{\mathrm{r}}) = \begin{pmatrix} 1 & 0 & 0 & 0 \\ 0 & \cos(\varphi_{\mathrm{r}}) & 0 & \sin(\varphi_{\mathrm{r}}) \\ 0 & 0 & 1 & 0 \\ 0 & -\sin(\varphi_{\mathrm{r}}) & 0 & \cos(\varphi_{\mathrm{r}}) \end{pmatrix} \tag{75}$$

Evaluating the trigonometric functions for $\delta\varphi \ll \frac{\pi}{2}$ and $\Delta\varphi \ll \frac{\pi}{2}$ up to first order yields

$$\cos(\varphi_{\mathrm{l}}) = \cos\left(-\frac{\pi}{2} + \delta\varphi + \Delta\varphi\right) \approx \delta\varphi + \Delta\varphi$$
$$\cos(\varphi_{\mathrm{r}}) = \cos\left(+\frac{\pi}{2} - \delta\varphi + \Delta\varphi\right) \approx \delta\varphi - \Delta\varphi$$
$$\sin(\varphi_{\mathrm{l}}) = \sin\left(-\frac{\pi}{2} + \delta\varphi + \Delta\varphi\right) \approx -1$$
$$\sin(\varphi_{\mathrm{r}}) = \sin\left(+\frac{\pi}{2} - \delta\varphi + \Delta\varphi\right) \approx 1 \tag{76}$$

and finally

$$M_{\mathrm{ret,l}} = M_{\mathrm{ret}}(135°, \varphi_{\mathrm{l}}) \approx \begin{pmatrix} 1 & 0 & 0 & 0 \\ 0 & \delta\varphi + \Delta\varphi & 0 & -1 \\ 0 & 0 & 1 & 0 \\ 0 & 1 & 0 & \delta\varphi + \Delta\varphi \end{pmatrix}$$
$$M_{\mathrm{ret,r}} = M_{\mathrm{ret}}(135°, \varphi_{\mathrm{r}}) \approx \begin{pmatrix} 1 & 0 & 0 & 0 \\ 0 & \delta\varphi - \Delta\varphi & 0 & 1 \\ 0 & 0 & 1 & 0 \\ 0 & -1 & 0 & \delta\varphi - \Delta\varphi \end{pmatrix} \tag{77}$$

for the Mueller matrices. Inserting these into eq.(73a) and eq.(73b) the Stokes vectors for LCP and RCP light produced by the Pockels cell are

$$\vec{S}_{\mathrm{l}}^{(\mathrm{i})} = M_{\mathrm{ret,l}} \cdot S_{\mathrm{lin}} = \begin{pmatrix} 1 \\ -\delta\varphi - \Delta\varphi \\ 0 \\ -1 \end{pmatrix}$$
$$\vec{S}_{\mathrm{r}}^{(\mathrm{i})} = M_{\mathrm{ret,r}} \cdot S_{\mathrm{lin}} = \begin{pmatrix} 1 \\ -\delta\varphi + \Delta\varphi \\ 0 \\ 1 \end{pmatrix} \tag{78}$$

Comparing these Stokes vectors with that for perfectly circularly polarized light

$$\vec{S}_{\mathrm{l,r}} = \begin{pmatrix} 1 \\ 0 \\ 0 \\ \mp 1 \end{pmatrix} \tag{79}$$

reveals a deviation in the $S_1$ component indicating the presence of a residual linearly polarized component. This originates from the small deviations of $\varphi_{\mathrm{l,r}}$ from $\mp\frac{\pi}{2}$ and results in elliptically polarized light. Comparison of eq.(41) and eq.(78)

$$\vec{S}_{\mathrm{l,r}}^{(\mathrm{i})} \overset{(41)}{\approx} \begin{pmatrix} 1 \\ S_1^{(\mathrm{sym})} \pm S_1^{(\mathrm{asym})} \\ S_2^{(\mathrm{sym})} \pm S_2^{(\mathrm{asym})} \\ \mp 1 \end{pmatrix} \overset{(78)}{=} \begin{pmatrix} 1 \\ -(\delta\varphi \pm \Delta\varphi) \\ 0 \\ \mp 1 \end{pmatrix} \tag{80}$$

uncovers the correspondences $-\delta\varphi = S_1^{(\mathrm{sym})}$ and $-\Delta\varphi = S_1^{(\mathrm{asym})}$.

Up to now we discussed to what extend imperfections in the applied voltages can lead to deviations of the induced phases $\varphi_{\mathrm{l,r}}$ from $\mp\frac{\pi}{2}$, while assuming perfect parallel alignment between the propagation direction of the incident light and the optical axis of the Pockels cell. However, any deviation from perfect alignment modifies the phases $\varphi_{\mathrm{l}}$ and $\varphi_{\mathrm{r}}$ as well and introduces alignment-dependent terms to the Stokes vector. These additional contributions to the Stokes vector reflect (among other effects) the involvement of the $n_3 = n_{\mathrm{e}}$ component of the refractive index in the birefringence of the cell due to the mismatch of the propagation direction of the light and the optical axis of the cell. A detailed treatment of these effects can be found in[46,47,51,52], and here we briefly summarize the outcome. In addition to an extra contribution to the $S_1$ component, this also leads to a deviation in the $S_2$ component of the resulting Stokes vector from the ideal vector for perfectly circularly polarized light. As before, separating the additional terms into a symmetric ($\delta S_{1,2}$) and an antisymmetric ($\Delta S_{1,2}$) part these can be added to the symmetric and antisymmetric components of the Stokes vector given in eq.(80) which yields

$$\begin{aligned} S_1^{(\mathrm{sym})} &= -\delta\varphi + \delta S_1 \\ S_1^{(\mathrm{asym})} &= -\Delta\varphi + \Delta S_1 \\ S_2^{(\mathrm{sym})} &= \delta S_2 \\ S_2^{(\mathrm{asym})} &= \Delta S_2 \end{aligned} \tag{81}$$

for the components, and

$$\vec{S}_{\mathrm{l}}^{(\mathrm{i})} = \begin{pmatrix} 1 \\ S_1^{(\mathrm{sym})} + S_1^{(\mathrm{asym})} \\ S_2^{(\mathrm{sym})} + S_2^{(\mathrm{asym})} \\ -1 \end{pmatrix} = \begin{pmatrix} 1 \\ -\delta\varphi - \Delta\varphi + \delta \mathrm{S}_1 + \Delta \mathrm{S}_1 \\ \delta \mathrm{S}_2 + \Delta \mathrm{S}_2 \\ -1 \end{pmatrix}$$

$$\vec{S}_{\mathrm{r}}^{(\mathrm{i})} = \begin{pmatrix} 1 \\ S_1^{(\mathrm{sym})} - S_1^{(\mathrm{asym})} \\ S_2^{(\mathrm{sym})} - S_2^{(\mathrm{asym})} \\ 1 \end{pmatrix} = \begin{pmatrix} 1 \\ -\delta\varphi + \Delta\varphi + \delta \mathrm{S}_1 - \Delta \mathrm{S}_1 \\ \delta \mathrm{S}_2 - \Delta \mathrm{S}_2 \\ 1 \end{pmatrix} \tag{82}$$

for the Stokes vectors. The additional terms $\delta S_{1,2}$ and $\Delta S_{1,2}$ are mainly determined by the adjustment of the Pockels cell. As has been shown in Section 3.2.2 only the antisymmetric contributions $S_{1,2}^{(\mathrm{asym})}$ to the Stokes vector are relevant for determining the artifact.

In order to find an expression for this contribution, we will first consider a perfectly collimated beam traveling along the $z$-direction incident on the Pockels cell. The optical axis of the cell is inclined at angles $\xi_x$ and $\xi_y$ relative to the $z$-axis, thereby enclosing angles $\eta_x$ and $\eta_y$ relative to the propagation direction of the light, Fig.9.

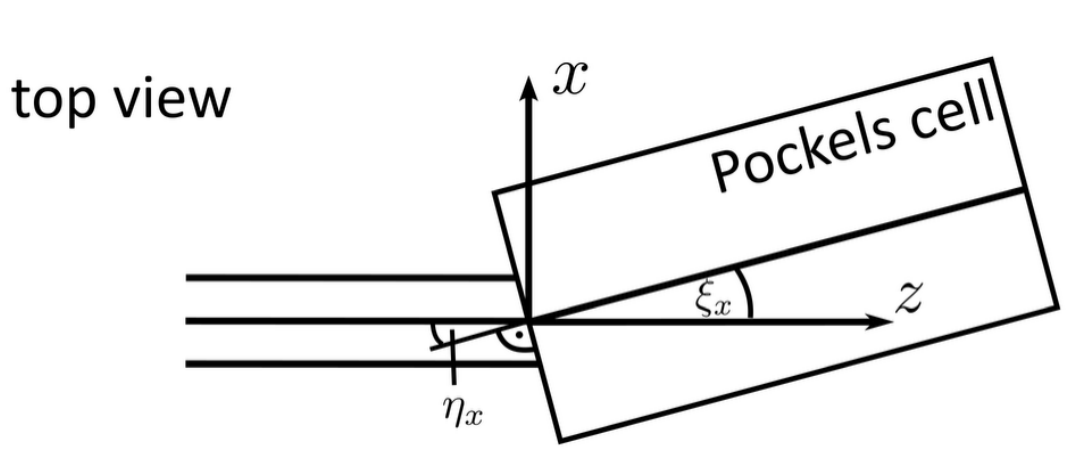

***Fig.9**: Schematic sketch of a collimated light beam propagating along the $z$-direction incident on a Pockels cell. The optical axis of the Pockels cell encloses the angles $\xi_x$ and $\xi_y$ (not shown) with the $z$-axis. The angles between the incident beam and the optical axis of the cell are denoted as $\eta_x$ and $\eta_y$ (not shown). For the situation depicted here $\eta_x = \xi_x$ and $\eta_y = \xi_y$ applies. Note that the misalignment of the Pockels cell is highly exaggerated. In reality the angles $\xi_x$ and $\xi_y$ are on the order of 0.02°.*

According to[47,51] the contributions to $\Delta S_{1,2}$ can be approximately decomposed in a term $\Delta \mathrm{S}_{1,2}^{(\mathrm{ang})}(\eta_x, \eta_y)$ arising from an angular misalignment $\eta_{x,y} \neq 0$ between the optical axis of the cell and the propagation direction of the incident light, and a term $\Delta \mathrm{S}_{1,2}^{(\mathrm{pos})}(x, y)$ that depends on the translational alignment of the Pockels cell relative to the incident beam. Hence,

$$\begin{aligned} \Delta \mathrm{S}_1(\eta_x, \eta_y, x, y) &\approx \Delta \mathrm{S}_1^{(\mathrm{ang})}(\eta_x, \eta_y) + \Delta \mathrm{S}_1^{(\mathrm{pos})}(x, y) \\ \Delta \mathrm{S}_2(\eta_x, \eta_y, x, y) &\approx \Delta \mathrm{S}_2^{(\mathrm{ang})}(\eta_x, \eta_y) + \Delta \mathrm{S}_2^{(\mathrm{pos})}(x, y) \end{aligned} \tag{83}$$

The term $\Delta S_{1,2}^{(\mathrm{pos})}(x,y)$ corresponds to a phase induced by the (background) birefringence of the Pockels cell and cannot be predicted a priori, as it depends on the specific device in use. The contribution due to angular misalignment $\Delta S_{1}^{(\mathrm{ang})}$ results from the involvement of the $n_3 = n_\mathrm{e}$ component of the refractive index in the birefringence and can be quantified to first order as[47,51]

$$\begin{aligned} \Delta S_{1}^{(\mathrm{ang})}(\eta_x,\eta_y) &= \kappa_1 \cdot 2\eta_x\eta_y \\ \Delta S_{2}^{(\mathrm{ang})}(\eta_x,\eta_y) &= \kappa_2 \cdot \left[{\eta_x}^2 - {\eta_y}^2\right] \end{aligned} \tag{84}$$

with constants $\kappa_1$ and $\kappa_2$ that depend on $n_\mathrm{e}$. For a collimated beam propagating along the $z$-direction, see Fig.9, $\eta_x = \xi_x$ and $\eta_y = \xi_y$ holds, and thus

$$\begin{aligned} \Delta S_{1}^{(\mathrm{ang})}(\eta_x,\eta_y) &= \kappa_1 \cdot 2\xi_x\xi_y \\ \Delta S_{2}^{(\mathrm{ang})}(\eta_x,\eta_y) &= \kappa_2 \cdot \left[{\xi_x}^2 - {\xi_y}^2\right] \end{aligned} \tag{85}$$

which for a perfectly aligned Pockels cell would result in $\xi_x = \xi_y = 0$ yielding $\Delta S_{1}^{(\mathrm{ang})} = \Delta S_{2}^{(\mathrm{ang})} = 0$.

Next, we take the divergence of the incident beam into account, Fig.10. We explicitly consider the variation of the angle between the propagation direction and the cell's optical axis as a function of the lateral coordinates $x$ and $y$, i.e. $\eta_x = \eta_x(x)$ and $\eta_y = \eta_y(y)$.

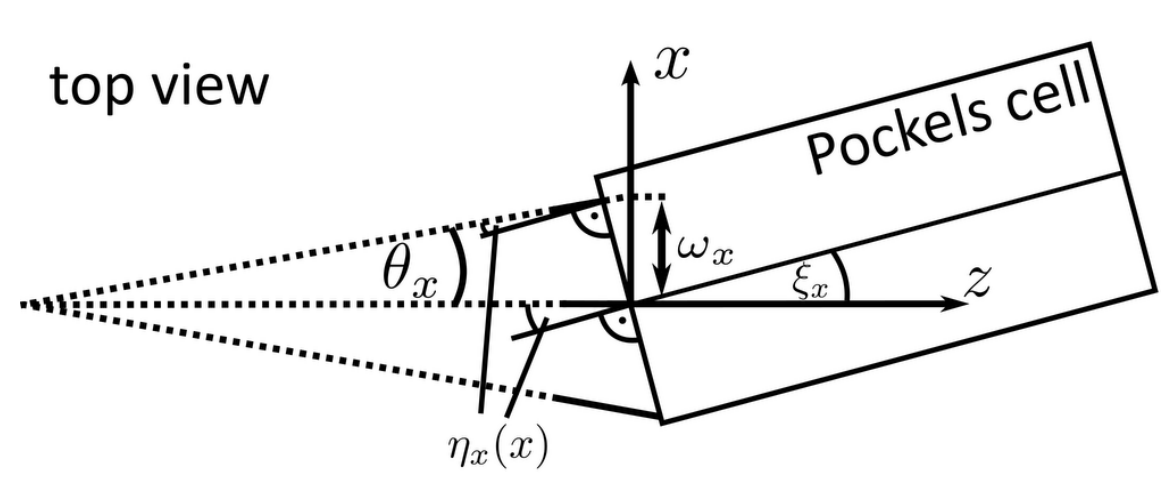

***Fig.10****: Same situation as in Fig.9 for a divergent beam with finite opening angles $\theta_x$ and $\theta_y$ (not shown). Now the angles between the propagation direction and the cell's optical axis $\eta_x$ and $\eta_y$ (not shown) will depend on the lateral coordinates $x$ and $y$ (not shown). Note that the misalignment of the Pockels cell is highly exaggerated. In reality the angles $\xi_x$ and $\xi_y$ are on the order of 0.02°.*

For small opening angles $\theta_x$ and $\theta_y$ of the divergent light, with the principal axes of the beam profile aligned with the $x$- and $y$- axes, these dependencies can be expressed as[51]

$$\begin{aligned} \eta_x &\approx \left(\xi_x + \frac{\theta_x}{\omega_x}\cdot x\right) \\ \eta_y &\approx \left(\xi_y + \frac{\theta_y}{\omega_y}\cdot y\right) \end{aligned} \tag{86}$$

Inserting this subsequently into eqs.(84), (83), and (81) yields for the antisymmetric contributions to the $S_1$ and $S_2$ components of the Stokes vector

$$S_1^{(\mathrm{asym})} = \underbrace{-\Delta\varphi}_{a)} + \kappa_1 \cdot 2\left(\underbrace{\xi_x}_{b)} + \underbrace{\frac{\theta_x}{\omega_x}\cdot x}_{c)}\right)\cdot\left(\underbrace{\xi_y}_{b)} + \underbrace{\frac{\theta_y}{\omega_y}\cdot y}_{c)}\right) + \underbrace{\Delta S_1^{(\mathrm{pos})}(x,y)}_{d)}$$

$$S_2^{(\mathrm{asym})} = \kappa_2 \cdot \left[\left(\underbrace{\xi_x}_{b)} + \underbrace{\frac{\theta_x}{\omega_x}\cdot x}_{c)}\right)^2 - \left(\underbrace{\xi_y}_{b)} + \underbrace{\frac{\theta_y}{\omega_y}\cdot y}_{c)}\right)^2\right] + \underbrace{\Delta S_2^{(\mathrm{pos})}(x,y)}_{d)} \tag{87}$$

To summarize: Deviations from perfectly circularly polarized light manifest themselves as antisymmetric contributions to the $S_1$ and $S_2$ components of the Stokes vector. These contributions have distinct physical origins and arise from a) deviations from the ideal voltages $\mp V_{\lambda/4}$ applied to the Pockels cell ($-\Delta\varphi$), b) misalignment between the optical axis of the Pockels cell and the propagation direction of the incident light ($\xi_x$, $\xi_y$), c) beam non-collimation ($\theta_x$, $\theta_y$), and d) background birefringence of the KD*P crystal ($\Delta S_{1,2}^{(\mathrm{pos})}$). The results of eq.(87) can be inserted directly into eq.(46b) to provide the artifacts that will be measured in a CD experiment giving

$$\begin{aligned}
g_{\mathrm{P}_1} &= 2\cdot LD\cdot S_1^{(\mathrm{asym})}(x,y)\cdot\cos 2\theta \\
&= 2LD\left(-\Delta\varphi + 2\kappa_1\cdot\left[\xi_x\xi_y + \xi_y\frac{\theta_x}{\omega_x}x + \xi_x\frac{\theta_y}{\omega_y}y + \frac{\theta_x}{\omega_x}\frac{\theta_y}{\omega_y}xy\right] + \Delta S_1^{(\mathrm{pos})}(x,y)\right) \\
&\qquad \cdot\cos 2\theta \\
g_{\mathrm{P}_2} &= 2\cdot LD\cdot S_2^{(asym)}(x,y)\cdot\sin 2\theta \\
&= 2LD\cdot\left(\kappa_2\left[\left({\xi_x}^2 - {\xi_y}^2\right) + 2\xi_x\frac{\theta_x}{\omega_x}x - 2\xi_y\frac{\theta_y}{\omega_y}y + \frac{{\theta_x}^2}{{\omega_x}^2}x^2 - \frac{{\theta_y}^2}{{\omega_y}^2}y^2\right]\right. \\
&\qquad \left. + \Delta S_2^{(\mathrm{pos})}(x,y)\right)\cdot\ \sin 2\theta
\end{aligned} \tag{88}$$

In order to classify the individual terms according to their order in the lateral coordinates it is useful to expand the terms $\Delta S_1^{(\mathrm{pos})}$ and $\Delta S_2^{(\mathrm{pos})}$ up to second order (defining the point of incidence as the origin $x = y = 0$), which yields

$$\Delta \mathrm{S}_{1,2}^{(\mathrm{pos})}(x,y) \approx \Delta \mathrm{S}_{1,2}^{(\mathrm{pos})}(0,0) + \left.\frac{\partial \Delta \mathrm{S}_{1,2}^{(\mathrm{pos})}}{\partial x}\right|_{0,0} \cdot x + \left.\frac{\partial \Delta \mathrm{S}_{1,2}^{(\mathrm{pos})}}{\partial y}\right|_{0,0} \cdot y + \frac{1}{2}\left.\frac{\partial^2 \Delta \mathrm{S}_{1,2}^{(\mathrm{pos})}}{\partial x^2}\right|_{0,0} \cdot x^2 + \frac{1}{2}\left.\frac{\partial^2 \Delta \mathrm{S}_{1,2}^{(\mathrm{pos})}}{\partial y^2}\right|_{0,0} \cdot y^2 + \left.\frac{\partial^2 \Delta \mathrm{S}_{1,2}^{(\mathrm{pos})}}{\partial x\,\partial y}\right|_{0,0} \cdot xy \tag{89}$$

Using eqs.(88) and (89) this expansion allows the various contributions to the artifact to be grouped into zeroth-, first-, and second-order terms in the lateral coordinates $x$ and $y$, as summarized in Table 1.

***Table 1**: Artifacts in the CD effect at zeroth, first, and second order of the lateral coordinates $x$ and $y$.*

| | $\frac{1}{2LD\cos 2\theta}\cdot g_{\mathrm{P}_1}$ | $\frac{1}{2LD\sin 2\theta}\cdot g_{\mathrm{P}_2}$ |
|---|---|---|
| constant term | $-\Delta\varphi + 2\kappa_1\xi_x\xi_y + \Delta S_1^{(\mathrm{pos})}(0,0)$ | $\kappa_2\left({\xi_x}^2 - {\xi_y}^2\right) + \Delta S_2^{(\mathrm{pos})}(0,0)$ |
| $x$ | $\left(2\kappa_1\xi_y\frac{\theta_x}{\omega_x} + \left.\frac{\partial\Delta S_1^{(\mathrm{pos})}}{\partial x}\right\|_{0,0}\right)x$ | $\left(2\kappa_2\xi_x\frac{\theta_x}{\omega_x} + \left.\frac{\partial\Delta S_2^{(\mathrm{pos})}}{\partial x}\right\|_{0,0}\right)x$ |
| $y$ | $\left(2\kappa_1\xi_x\frac{\theta_y}{\omega_y} + \left.\frac{\partial\Delta S_1^{(\mathrm{pos})}}{\partial y}\right\|_{0,0}\right)y$ | $\left(-2\kappa_2\xi_y\frac{\theta_y}{\omega_y} + \left.\frac{\partial\Delta S_2^{(\mathrm{pos})}}{\partial y}\right\|_{0,0}\right)y$ |
| $x^2$ | $\left.\frac{1}{2}\frac{\partial^2\Delta S_1^{(\mathrm{pos})}}{\partial x^2}\right\|_{0,0}\cdot x^2$ | $\left(\kappa_2\frac{{\theta_x}^2}{{\omega_x}^2} + \left.\frac{1}{2}\frac{\partial^2\Delta S_2^{(\mathrm{pos})}}{\partial x^2}\right\|_{0,0}\right)x^2$ |
| $y^2$ | $\left.\frac{1}{2}\frac{\partial^2\Delta S_1^{(\mathrm{pos})}}{\partial y^2}\right\|_{0,0}\cdot y^2$ | $\left(-\kappa_2\frac{{\theta_y}^2}{{\omega_y}^2} + \left.\frac{1}{2}\frac{\partial^2\Delta S_2^{(\mathrm{pos})}}{\partial y^2}\right\|_{0,0}\right)y^2$ |
| $xy$ | $\left(2\kappa_1\frac{\theta_x}{\omega_x}\frac{\theta_y}{\omega_y} + \left.\frac{\partial^2\Delta S_1^{(\mathrm{pos})}}{\partial x\,\partial y}\right\|_{0,0}\right)xy$ | $\left.\frac{\partial^2\Delta S_2^{(\mathrm{pos})}}{\partial x\,\partial y}\right\|_{0,0}\cdot xy$ |

Ideally, complete suppression of artifacts would require all terms listed in Table 1 to vanish simultaneously. Unfortunately, in practice this cannot be fully achieved because this leads to conflicting demands. For example the influence of $\left.\frac{\partial\Delta S_1^{(\mathrm{pos})}}{\partial x}\right|_{0,0}$ in $\left(2\kappa_1\xi_y\frac{\theta_x}{\omega_x} + \left.\frac{\partial\Delta S_1^{(\mathrm{pos})}}{\partial x}\right|_{0,0}\right)$ can be compensated by adjusting $\xi_y$ such that $2\kappa_1\xi_y\frac{\theta_x}{\omega_x} = -\left.\frac{\partial\Delta S_1^{(\mathrm{pos})}}{\partial x}\right|_{0,0}$. The finite $\xi_y$ however, prevents the zeroth-order artifacts from becoming zero.

Note, that these calculations are done in ray optics approximation and neglect diffraction effects. If the profile of the light beam deviates from spherical symmetry ($\omega_x^2 \neq \omega_y^2$), and if the principal axes of the light beam are aligned along the $x$- and $y$-axes of the laboratory frame, this introduces an additional offset into $\Delta S_2$, which is proportional to $\frac{\kappa_2}{k^2}\left(\frac{1}{\omega_x^2} - \frac{1}{\omega_y^2}\right)$ [47,51]. For an arbitrary orientation of the principal axis of the light beam similar terms arise in $\Delta S_1$. This effect is only relevant for significant deviations of the beam profile from spherical symmetry and stronger focusing, i.e.

small beam radius. Furthermore, shortcomings of the hardware such as non-perfect face cuts of the KD*P crystal, or effects related to non-uniformity of the applied electric field (voltage) have not been taken into account. A comprehensive discussion of the influence of these parameters can be found in[47,51].

### 4.3 Transformation of the corrections from coordinates in the detection plane to coordinates in the sample plane

In practice, it is often not feasible to characterize the artifacts directly at the sample position (lateral coordinates $x_\mathrm{s}, y_\mathrm{s}$). Instead, the control measurements are performed in an external plane located between the Pockels cell and the sample, at the position where the focusing lens is subsequently inserted for the experiment. Consequently, the measured results must be transformed to the sample plane, taking into account the effect of the lens that focuses the excitation light (here $f = 100$ mm). According to the results of Section 3.3 the artificial contributions to the $g$-factor can be separated into a part $g_\mathrm{I}$, eq.(52b), that considers intensity differences between LCP and RCP light and a part $g_{\mathrm{P}_{1,2}}$, eq.(64) that considers polarization artifacts expressed as antisymmetric components $S_{1,2}^{(\mathrm{asym})}$ in the Stokes vector. Using the coordinates $x_\mathrm{s}, y_\mathrm{s}$ in the sample plane eqs.(52b) and (64) are given as

$$g_\mathrm{I} = \left(2\frac{P_\mathrm{l,s} - P_\mathrm{r,s}}{P_\mathrm{l,s} + P_\mathrm{r,s}} - \frac{\Delta\sigma_{\mathrm{lr},x,\mathrm{s}}^2}{2\sigma_{x,\mathrm{s}}^2} - \frac{\Delta\sigma_{\mathrm{lr},y,\mathrm{s}}^2}{2\sigma_{y,\mathrm{s}}^2}\right) + \frac{\Delta\mu_{\mathrm{lr},x,\mathrm{s}}}{\sigma_{x,\mathrm{s}}^2} x_\mathrm{s} + \frac{\Delta\mu_{\mathrm{lr},y,\mathrm{s}}}{\sigma_{y,\mathrm{s}}^2} y_s + \frac{\Delta\sigma_{\mathrm{lr},x,s}^2}{2\sigma_{x,\mathrm{s}}^4} {x_\mathrm{s}}^2 + \frac{\Delta\sigma_{\mathrm{lr},y,s}^2}{2\sigma_{y,\mathrm{s}}^4} {y_\mathrm{s}}^2 + \frac{\Delta\sigma_{\mathrm{lr},xy,s}}{\sigma_{x,\mathrm{s}}^2\sigma_{y,\mathrm{s}}^2} x_\mathrm{s} y_\mathrm{s} \tag{90}$$

$$g_{\mathrm{P}_{1,2}} \approx LD \cdot \left[\left(2\left(\frac{P_\mathrm{l,s} - P_\mathrm{r,s}}{P_\mathrm{l,s} + P_\mathrm{r,s}}\right)_{1,2} - \frac{\left(\Delta\sigma_{\mathrm{lr},x,\mathrm{s}}^2\right)_{1,2}}{2\sigma_{x,\mathrm{s}}^2} - \frac{\left(\Delta\sigma_{\mathrm{lr},y,\mathrm{s}}^2\right)_{1,2}}{2\sigma_{y,\mathrm{s}}^2}\right) + \frac{\left(\Delta\mu_{\mathrm{lr},x,\mathrm{s}}\right)_{1,2}}{\sigma_{x,\mathrm{s}}^2} x_\mathrm{s} + \frac{\left(\Delta\mu_{\mathrm{lr},y,\mathrm{s}}\right)_{1,2}}{\sigma_{y,\mathrm{s}}^2} y_\mathrm{s} + \frac{\left(\Delta\sigma_{\mathrm{lr},x,\mathrm{s}}^2\right)_{1,2}}{2\sigma_{x,\mathrm{s}}^4} {x_\mathrm{s}}^2 + \frac{\left(\Delta\sigma_{\mathrm{lr},y,\mathrm{s}}^2\right)_{1,2}}{2\sigma_{y,\mathrm{s}}^4} {y_\mathrm{s}}^2 + \frac{\left(\Delta\sigma_{\mathrm{lr},xy,\mathrm{s}}\right)_{1,2}}{\sigma_{x,\mathrm{s}}^2\sigma_{y,\mathrm{s}}^2} x_\mathrm{s} y_\mathrm{s}\right] \cdot \begin{cases} g_{\mathrm{P}_1}: \cos 2\theta \\ g_{\mathrm{P}_2}: \sin 2\theta \end{cases} \tag{91}$$

The quantities $\Delta\mu_\mathrm{lr}$ and $\Delta\sigma_\mathrm{lr}^2$ are measured in front of the focal lens (with and without a polarizer), as described above. We now require the transformations of $\Delta\mu_\mathrm{lr}$ and $\Delta\sigma_\mathrm{lr}^2$ from the measurement plane to the sample plane, i.e., $\Delta\mu_\mathrm{lr} \rightarrow \Delta\mu_\mathrm{lr,s}$ and $\Delta\sigma_\mathrm{lr}^2 \rightarrow \Delta\sigma_\mathrm{lr,s}^2$. This must be carried out separately for the intensity artifacts, eq.(90), and the polarization artifacts, eq.(91). It is important to note that the transformations for the two cases are not equivalent, as illustrated in Fig.11.

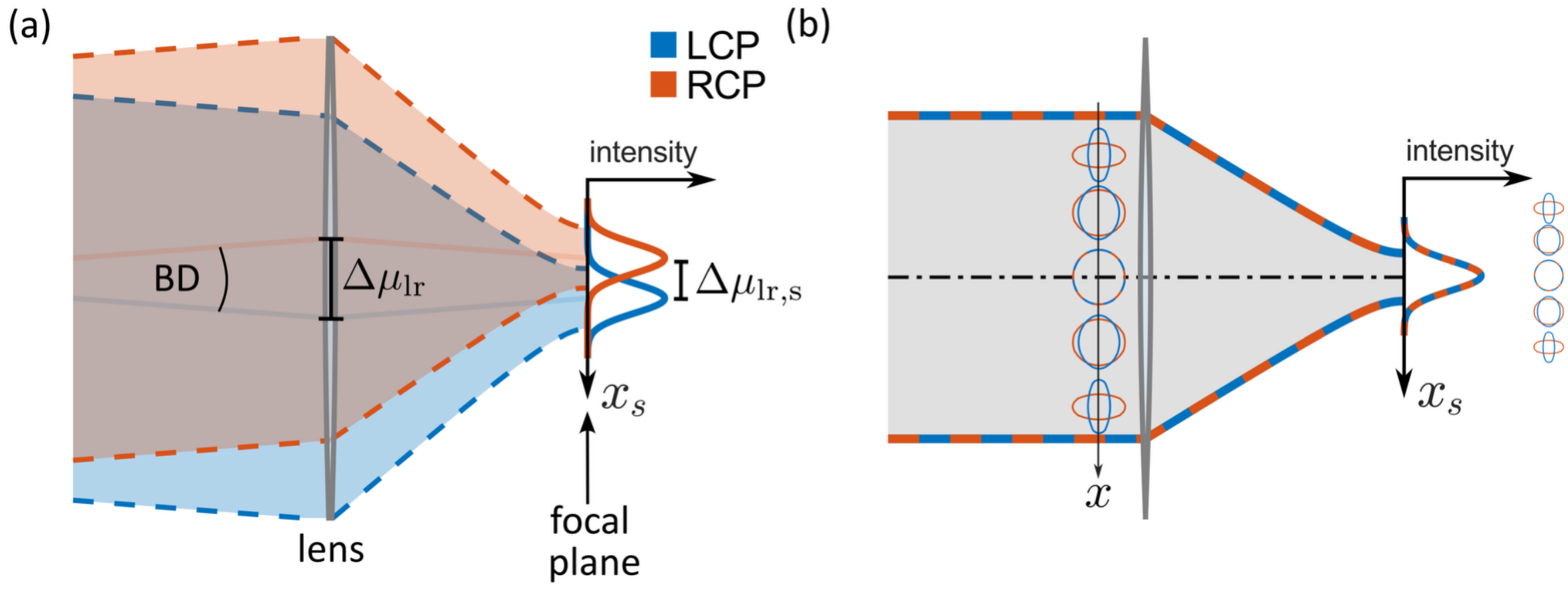

***Fig.11**: Schematic illustration of the mechanisms underlying the transformation of (a) intensity artifacts and (b) polarization artifacts from the measurement plane at the position of the lens to the focal plane at the position of the sample. (a) Propagation of LCP (blue) and RCP (red) Gaussian beams through the focusing lens. At the lens position an angular beam deviation ($BD$) between both beams results in a displacement of the two beams denoted as $\Delta\mu_{lr}$. This, in turn, is transformed by the lens into the lateral beam displacement $\Delta\mu_{lr,s}$ in the focal plane and corresponds to a spatial mismatch of the excitation beams along the lateral focal coordinate $x_s$ as illustrated by the one-dimensional intensity profiles. (b) Propagation of perfectly overlapping LCP and RCP beams (indicated by the dashed red-blue envelope) toward the focusing lens. The spatial variation of the local polarization state along the lateral coordinate $x$ in front of the lens is illustrated by ellipses. In the focal plane the local polarization state varies as a function of $x_s$ across the intensity profile, while the intensity profiles overlap perfectly.*

### 4.3.1 Transformation of the intensity artifacts

The zeroth-order correction to the $g$-factor, given by $\frac{P_{\mathrm{l,s}}-P_{\mathrm{r,s}}}{P_{\mathrm{l,s}}+P_{\mathrm{r,s}}}$, is not affected by focusing and remains invariant. As follows from Fig.11 the first-order corrections of the intensity artifacts stem from the fact that the LCP and RCP Gaussian beams propagate in slightly different directions, resulting in a mutual beam deviation (BD) as described by eq.(67). Using a lens with focal length $f$ the spatial separation of the centroids of the beams in the sample plane is given as

$$\Delta\mu_{\mathrm{lr,s}} = f \cdot \frac{\Delta\mu_{\mathrm{lr}}}{l} = f \cdot BD \qquad (92)$$

Using ray optics, which is justified for small beam divergence and weak focusing, the ratio of the beam waists in the sample plane $\omega_{x,s}$ and in the measurement plane $\omega_x$ can be expressed as an effective magnification $m = \frac{\omega_{x,s}}{\omega_x}$. Hence the second moments scale as

$$\sigma_{x,\mathrm{s}}^2 = m^2 \cdot \sigma_x^2$$
$$\Delta\sigma_{\mathrm{lr},x,\mathrm{s}}^2 = m^2 \cdot \Delta\sigma_{\mathrm{lr},x}^2 \tag{93}$$

Substituting (92) and (93) into eq.(90) yields

$$g_\mathrm{I} = \left(2\frac{P_\mathrm{l} - P_\mathrm{r}}{P_\mathrm{l} + P_\mathrm{r}} - \frac{\Delta\sigma_{\mathrm{lr},x}^2}{2\sigma_x^2} - \frac{\Delta\sigma_{\mathrm{lr},y}^2}{2\sigma_y^2}\right) + \frac{1}{m^2}\frac{f \cdot BD_x}{\sigma_x^2}x_\mathrm{s} + \frac{1}{m^2}\frac{f \cdot BD_y}{\sigma_y^2}y_\mathrm{s} + \frac{1}{m^2}\frac{\Delta\sigma_{\mathrm{lr},x}^2}{2\sigma_x^4}{x_\mathrm{s}}^2 + \frac{1}{m^2}\frac{\Delta\sigma_{\mathrm{lr},y}^2}{2\sigma_y^4}{y_\mathrm{s}}^2 + \frac{1}{m^2}\frac{\Delta\sigma_{\mathrm{lr},xy}}{\sigma_x^2\sigma_y^2}x_\mathrm{s}y_\mathrm{s} \tag{94}$$

With these transformations the contributions to the intensity artifact in the sample plane can be obtained by measuring the differences in the zeroth and second moments of the beam profiles for LCP and RCP light in front of the excitation lens.

### 4.3.2 Transformation of the polarization artifacts

The polarization artifact is caused by the spatial variation of the polarization state across the beam profile (characterized by $S_{1,2}^{(\mathrm{asym})}(x,y)$) and the different absorption strength of the sample with respect to (i)-LCP and (i)-RCP light. A rigorous treatment would require a full Fourier transform of the electric field distribution from the measurement plane into the sample plane. However, this can be approximated with ray optics. Consequently, the zeroth moment remains unchanged, and the differences in the first and second moments scale as

$$\left(\Delta\mu_{\mathrm{lr},x,\mathrm{s}}\right)_{1,2} = m \cdot \left(\Delta\mu_{\mathrm{lr},x}\right)_{1,2}$$
$$\left(\Delta\sigma_{\mathrm{lr},x,\mathrm{s}}^2\right)_{1,2} = m^2 \cdot \left(\Delta\sigma_{\mathrm{lr},x}^2\right)_{1,2}, \tag{95}$$

respectively. Substitution into eq.(91) yields

$$g_{P_{1,2}} \approx LD \cdot \left[\left(2\left(\frac{P_\mathrm{l} - P_\mathrm{r}}{P_\mathrm{l} + P_\mathrm{r}}\right)_{1,2} - \frac{\left(\Delta\sigma_{\mathrm{lr},x}^2\right)_{1,2}}{2\sigma_x^2} - \frac{\left(\Delta\sigma_{\mathrm{lr},y}^2\right)_{1,2}}{2\sigma_y^2}\right) + \frac{1}{m}\frac{\left(\Delta\mu_{\mathrm{lr},x}\right)_{1,2}}{\sigma_x^2}x_\mathrm{s} + \frac{1}{m}\frac{\left(\Delta\mu_{\mathrm{lr},y}\right)_{1,2}}{\sigma_y^2}y_\mathrm{s} + \frac{1}{m^2}\frac{\left(\Delta\sigma_{\mathrm{lr},x}^2\right)_{1,2}}{2\sigma_x^4}{x_\mathrm{s}}^2 + \frac{1}{m^2}\frac{\left(\Delta\sigma_{\mathrm{lr},y}^2\right)_{1,2}}{2\sigma_y^4}{y_\mathrm{s}}^2 + \frac{1}{m^2}\frac{\left(\Delta\sigma_{\mathrm{lr},xy}\right)_{1,2}}{\sigma_x^2\sigma_y^2}x_\mathrm{s}y_\mathrm{s}\right] \cdot \begin{cases} g_{\mathrm{P}_1}: \cos 2\theta \\ g_{\mathrm{P}_2}: \sin 2\theta \end{cases} \tag{96}$$

With these transformations the contributions to the polarization artifact in the sample plane can be obtained by measuring the differences in the zeroth, first and second moments of the beam profiles for (i)-LCP and (i)-RCP light behind an analyzer but in front of the excitation lens.

## 5. Alignment of the Pockels cell

In Chapter 3 a general analysis of artifacts that can occur in CD spectra was presented. This analysis showed the need to distinguish between intensity artifacts $g_{\mathrm{I}}$ and polarization artifacts $g_{\mathrm{P}_{1,2}}$. It was further shown that all artifacts could be decomposed into contributions arising from the differences of the zeroth, first, and second moments of the intensity distributions for LCP and RCP light (in the case of $g_{\mathrm{P}_{1,2}}$, where an analyzer is placed behind the Pockels cell), see eqs.(52b), (64). In Chapter 4, these contributions were related to alignment-dependent parameters, eqs.(67), (68), (88), for a Pockels cell used to generate circularly polarized light.

In the present chapter, the individual steps required to align the KD*P Pockels cell within the setup are described, with the aim of minimizing artifacts that may affect the CD spectrum. In our setup artifacts due to differences in the second moments of the intensity distribution are negligible and are therefore not considered in the alignment optimization. For ease of reference, the key equations mentioned above are reproduced here for the intensity artifacts

$$g_{\mathrm{I}} = \underbrace{\left[2\frac{P_{\mathrm{l}} - P_{\mathrm{r}}}{P_{\mathrm{l}} + P_{\mathrm{r}}}\right]}_{g_I^{(0)}} + \underbrace{\frac{\Delta\mu_{\mathrm{lr},x}}{\sigma_x^2} \cdot (x - \mu_x)^1 + \frac{\Delta\mu_{\mathrm{lr},y}}{\sigma_y^2} \cdot \left(y - \mu_y\right)^1}_{g_I^{(1)}} \qquad \text{from (52b)}$$

$$g_{\mathrm{I}} = \frac{l \cdot BD_x(x_p, y_p)}{\sigma_x^2} \cdot (x - \mu_x)^1 + \frac{l \cdot BD_y(x_p, y_p)}{\sigma_y^2} \cdot \left(y - \mu_y\right)^1 \qquad \text{from (67) (68)}$$

and for the polarization artifacts

$$g_{\mathrm{P}_{1,2}} \approx LD \cdot \left[\underbrace{\left(2\left(\frac{P_{\mathrm{l}} - P_{\mathrm{r}}}{P_{\mathrm{l}} + P_{\mathrm{r}}}\right)_{1,2}\right)}_{g_{\mathrm{P}_{1,2}}^{(0)}} + \underbrace{\frac{\left(\Delta\mu_{\mathrm{lr},x}\right)_{1,2}}{\sigma_x^2} x + \frac{\left(\Delta\mu_{\mathrm{lr},y}\right)_{1,2}}{\sigma_y^2} y}_{g_{\mathrm{P}_{1,2}}^{(1)}}\right] \cdot \begin{cases} \cos 2\theta \;\; (g_{\mathrm{P}_1}) \\ \sin 2\theta \;\; (g_{\mathrm{P}_2}) \end{cases} \qquad \text{from (64)}$$

$$g_{\mathrm{P}_1} = 2LD\left(-\Delta\varphi + 2\kappa_1 \cdot \left[\xi_x\xi_y + \xi_y\frac{\theta_x}{\omega_x}x + \xi_x\frac{\theta_y}{\omega_y}y\right] + \Delta S_1^{(\mathrm{pos})}(x, y)\right) \cdot \cos 2\theta$$

$$g_{\mathrm{P}_2} = 2LD \cdot \left(\kappa_2\left[\left(\xi_x{}^2 - \xi_y{}^2\right) + 2\xi_x\frac{\theta_x}{\omega_x}x - 2\xi_y\frac{\theta_y}{\omega_y}y\right] + \Delta S_2^{(\mathrm{pos})}(x, y)\right) \cdot \sin 2\theta \qquad \text{from (88)}$$

Within these approximations, the artifacts $g_{\mathrm{I}}$ and $g_{\mathrm{P}_{1,2}}$ can be grouped into $g^{(0)}$ and $g^{(1)}$ based on the differences in their zeroth and first moments of their underlying intensity distributions after a polarizer. The adjustment of the Pockels cell to reduce artifacts affecting the CD spectrum is typically performed in six steps. After a coarse lateral alignment of the cell (step 1), the procedure continues with a qualitative assessment of how the angles $\xi_x$ and $\xi_y$ influence the polarization artifacts $g^{(0)}_{\mathrm{P}_{1,2}}$, before performing a preliminary alignment of the optical axis (step 2). Subsequently, the remaining steps comprise characterizing the intensity artifact $g^{(1)}_{\mathrm{I}}$ (step 3), qualitatively characterizing the polarization artifacts $g^{(0)}_{\mathrm{P}_{1,2}}$ (step 4), characterizing the polarization artifacts $g^{(1)}_{\mathrm{P}_{1,2}}$ (step 5), and finally minimizing $g_{\mathrm{I}}$ and $g_{\mathrm{P}_{1,2}}$ based on the results of steps 3 to 5, while maximizing the ellipticity $\epsilon_{\mathrm{l,r}}$ (step 6). Finally, steps 3 to 6 are iterated until convergence. In order to address either $g_{\mathrm{P}_1}$ or $g_{\mathrm{P}_2}$in steps 2, 4, and 5 the measurements must be performed with an analyzer behind the Pockels cell set to $\theta = 0°$ or $\theta = 45°$, respectively.

### 5.1 Step 1: Rough lateral alignment of the cell

The angle $\gamma$ of the linear polarization of the incident beam in front of the Pockels cell is roughly adjusted (within 1°) to be vertical with respect to the optical table. The cell is positioned laterally such that the light beam passes through its center. The yaw ($\xi_x$) and pitch ($\xi_y$) angles (see Fig.7) are adjusted to ensure that the back-reflected beam overlaps with the incident beam. Subsequently, the driving voltages are applied for inducing phase shifts of $|\varphi_{\mathrm{l,r}}| \approx \frac{\pi}{2}$, which typically yields an ellipticity of about $|\epsilon_{\mathrm{l,r}}| \approx 0.97$ - 0.99 for both LCP and RCP light.

### 5.2 Step 2: Qualitative assessment of the influence of jaw and pitch on polarization artifacts $g^{(0)}_{\mathrm{P}_{1,2}}$

Since a mismatch in the yaw and pitch angles strongly affects the resulting polarization state due to the unwanted contribution of the extraordinary refraction index $n_{\mathrm{e}}$ it is natural to begin the fine alignment procedure by analyzing the artifacts $g^{(0)}_{\mathrm{P}_{1,2}}$, see eq.(64). Therefore the quantity $2\left(\frac{P_{\mathrm{l}}-P_{\mathrm{r}}}{P_{\mathrm{l}}+P_{\mathrm{r}}}\right)$ is determined as a function of the angles $\xi_x$ and $\xi_y$. These were varied between $-150$ mdeg and $150$ mdeg in increments of about 15 mdeg. The results are shown in Fig.12 for the analyzer behind the Pockels cell set to $\theta = 0°$ and $\theta = 45°$, see Fig.7. The profiles of both dependencies are in excellent agreement with the predicted relations $g_{\mathrm{P}_1}^{(0)} \propto \xi_x \xi_y$ and $g_{\mathrm{P}_2}^{(0)} \propto (\xi_x{}^2 - \xi_y{}^2)$, see eq.(88). The aim of this qualitative experiment is to identify the saddle points in $\xi_x$ and $\xi_y$, rather than determining the absolute magnitude of $2\left(\frac{P_\mathrm{l}-P_\mathrm{r}}{P_\mathrm{l}+P_\mathrm{r}}\right)$. For both analyzer settings the positions of the saddle points coincide within 1 mdeg, corresponding to the angular resolution of the hardware used for scanning. From here on, the common saddle point, obtained by averaging the results from the two polarization settings, defines the reference positions $\xi_x = 0$ and $\xi_y = 0$ relative to which all subsequent angular adjustments of the cell are performed.

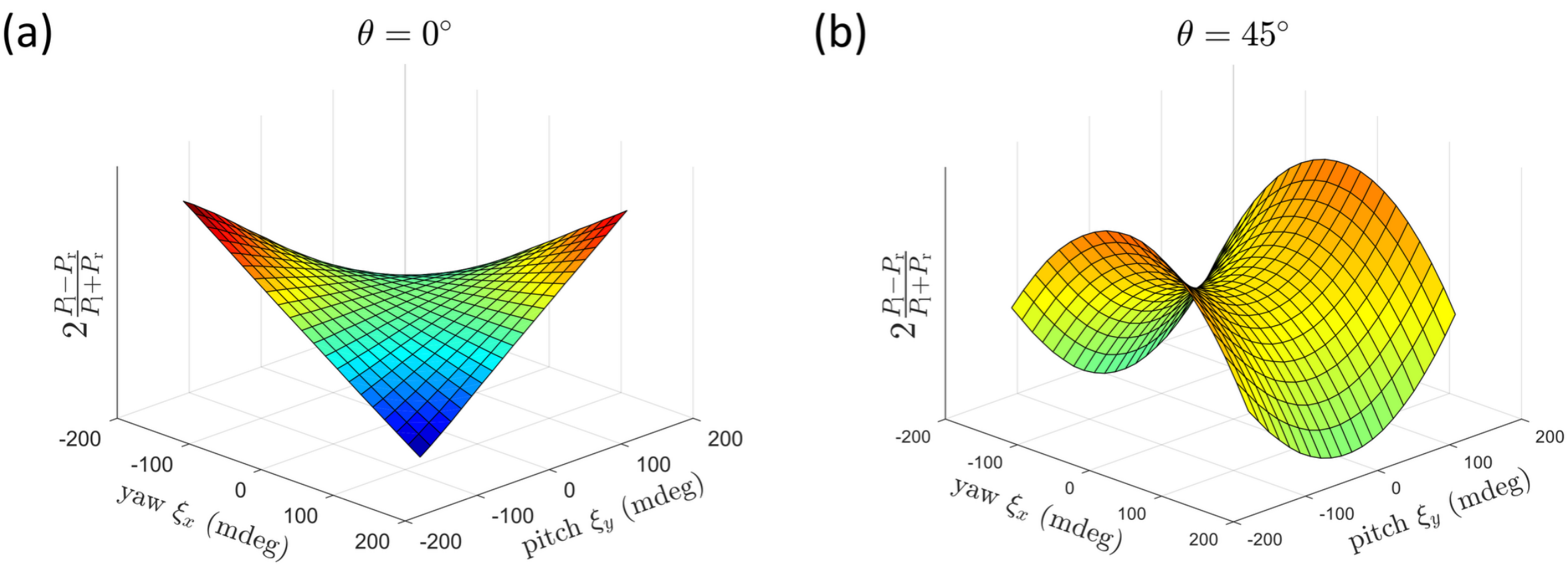

***Fig.12**: Maps of the measured power difference $2\left(\frac{P_l-P_r}{P_l+P_r}\right)$ as a function of yaw ($\xi_x$) and pitch ($\xi_y$) adjustments of the Pockels cell for the analyzer set to (a) $\theta = 0°$ and (b) $\theta = 45°$. The contour colors serve as a guide for the eye.*

### 5.3 Step 3: Characterizing intensity artifacts $g_\mathrm{I}^{(1)}$

Since intensity artifacts $g_\mathrm{I}^{(0)}$ are negligible for KD*P Pockels cells, we focus on the intensity artifacts arising from a difference in the first moments of the intensity distribution, i.e. $\Delta\mu_{\mathrm{lr},x} \neq 0$ and $\Delta\mu_{\mathrm{lr},y} \neq 0$, see eq.(52b). This mismatch originates from the angular beam deviation (BD) of LCP and RCP light, which is in two dimensions given as

$$BD = \sqrt{BD_x^2 + BD_y^2} = \frac{\sqrt{\Delta\mu_{\mathrm{lr},x}^2 + \Delta\mu_{\mathrm{lr},y}^2}}{l} \tag{97}$$

The BD is measured by scanning the lateral position $(x_p, y_p)$ of the Pockels cell over a range of 4 mm for both directions, in steps of 0.2 mm, while keeping $\xi_x = \xi_y = 0$ fixed. The result is shown in Fig.13. The minimum of the BD is indicated by a white square, whose position is defined as $(x_\mathrm{p}, y_\mathrm{p})$ = (0,0). From here on, the reference configuration relative to which all subsequent translational and angular adjustments of the cell are performed is denoted as $(x_\mathrm{p}, y_\mathrm{p}, \xi_x, \xi_y)$ = (0,0,0,0)

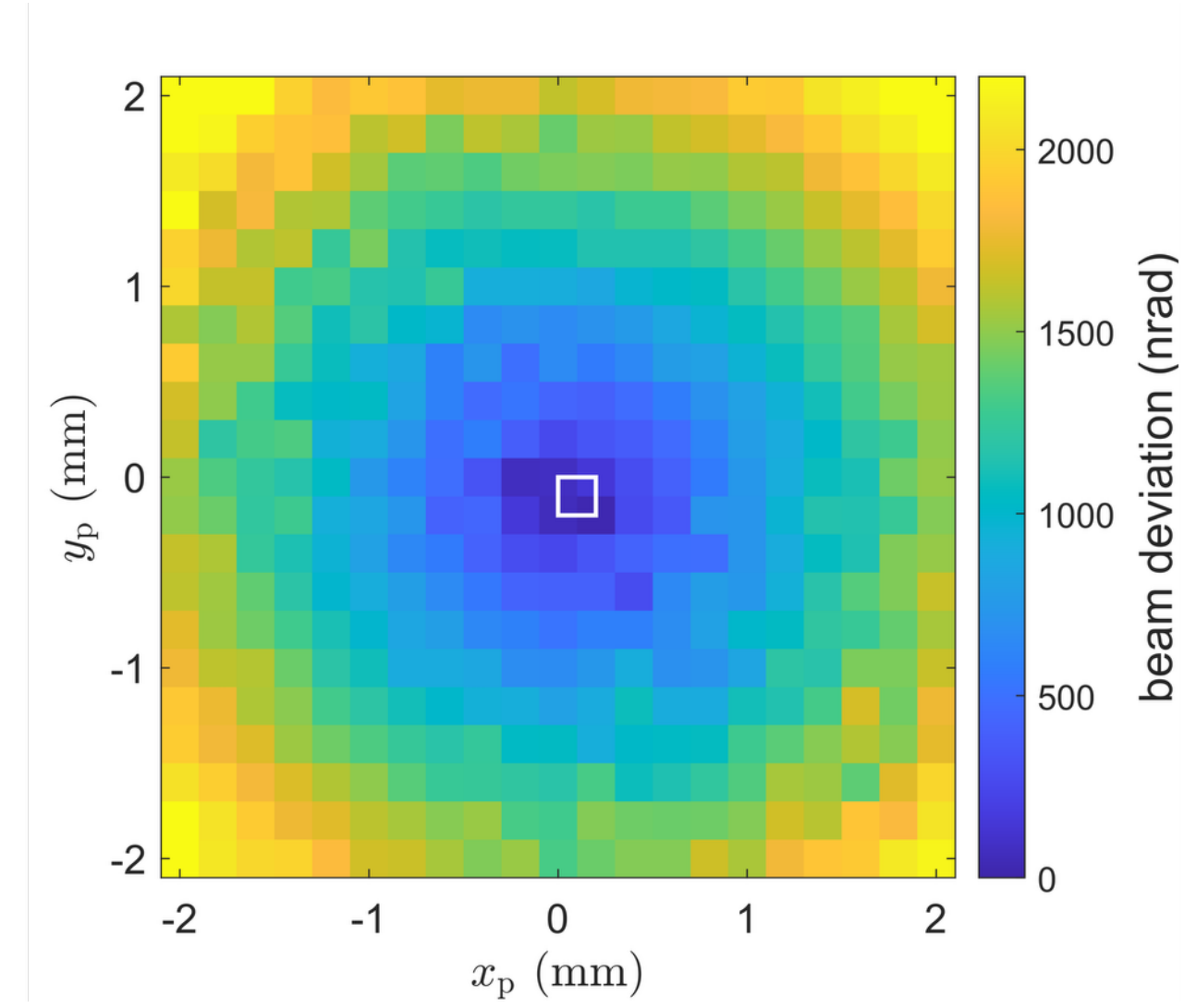

***Fig.13**: Angular beam deviation between the LCP and RCP beams as a function of the lateral position $(x_p,\ y_p)$ of the Pockels cell for $(\xi_x, \xi_y) = (0,0)$. The white box indicates the minimum of BD, and defines the reference position ($x_p$, $y_p$) = (0,0).*

According to eq.(94) the contribution of the difference in the first moments of the intensity distribution in the sample plane $(x, y)$ = $(x_\mathrm{s},\ y_\mathrm{s})$ is given as

$$g_\mathrm{I}^{(1)} \approx \frac{1}{m^2}\frac{f \cdot BD_x}{\sigma_x^2}x_\mathrm{s} + \frac{1}{m^2}\frac{f \cdot BD_y}{\sigma_y^2}y_\mathrm{s} \tag{98}$$

Using $m^2 = \frac{\omega_\mathrm{s}^2}{\omega^2}$, which is equivalent to $m^2 = \frac{\sigma_\mathrm{s}^2}{\sigma^2}$ (due to the relation $\omega^2 = 4\sigma^2$ between the beam waist and the variance of a Gaussian beam), and approximating $\sigma^2 \approx \sigma_x^2 \approx \sigma_y^2$, as well as $BD_x \approx BD_y$ gives $BD = \sqrt{2}\,BD_{x,y}$. In the actual experiment the region of interest in the sample plane covers an area of 60 µm x 60 µm such that the maximum values $x_\mathrm{s}$ and $y_\mathrm{s}$ are $s_\mathrm{max} = \pm 30$ µm. Under these assumptions eq.(98) can be simplified to

$$g_\mathrm{I}^{(1)} \approx 2\frac{f \cdot BD/\sqrt{2}}{\sigma_\mathrm{s}^2}s_\mathrm{max} = \sqrt{2}\frac{f \cdot BD}{\sigma_\mathrm{s}^2}s_\mathrm{max} \overset{\omega_\mathrm{s}^2=4\sigma_\mathrm{s}^2}{=} 4\sqrt{2}\frac{f \cdot BD}{\omega_\mathrm{s}^2}s_\mathrm{max} \tag{99}$$

Allowing an artifact level of $g_\mathrm{I} \approx\ 0.001$ with $\omega \approx 400$ µm, $\omega_\mathrm{s} \approx 50$ µm and $f\ = 100$ mm corresponds to a tolerable angular beam deviation of about $BD \approx 200\ \mathrm{nrad}$. This, in turn, see Fig.13, requires a positioning precision of a few tenths of a millimeter for the lateral coordinates $(x_\mathrm{p},\ y_\mathrm{p})$ of the Pockels cell.

### 5.4 Step 4: Quantitative characterization of polarization artifacts $g_{\mathrm{P}_{1,2}}^{(0)}$

After completing the alignments described in steps 1 - 3, we reexamine the polarization artifacts $g_{\mathrm{P}_{1,2}}^{(0)}$ for a more quantitative analysis. These artifacts arise from differences between the zeroth moments of the intensity distributions of LCP and RCP light after the analyzer. They can be quantified behind an analyzer by measuring $2\left(\frac{P_\mathrm{l}-P_\mathrm{r}}{P_\mathrm{l}+P_\mathrm{r}}\right)$. For doing so the powers $P_\mathrm{l}$ and $P_\mathrm{r}$ for LCP and RCP light are recorded as a function of the angles $\xi_x$ and $\xi_y$ while keeping $x_\mathrm{p} = y_\mathrm{p} = 0$, Fig.14, and as a function of the lateral position $x_\mathrm{p}$ and $y_\mathrm{p}$ while keeping $\xi_x = \xi_y = 0$, Fig.15. In both cases, measurements are performed for the analyzer set to $\theta = 0°$ and $\theta = 45°$.

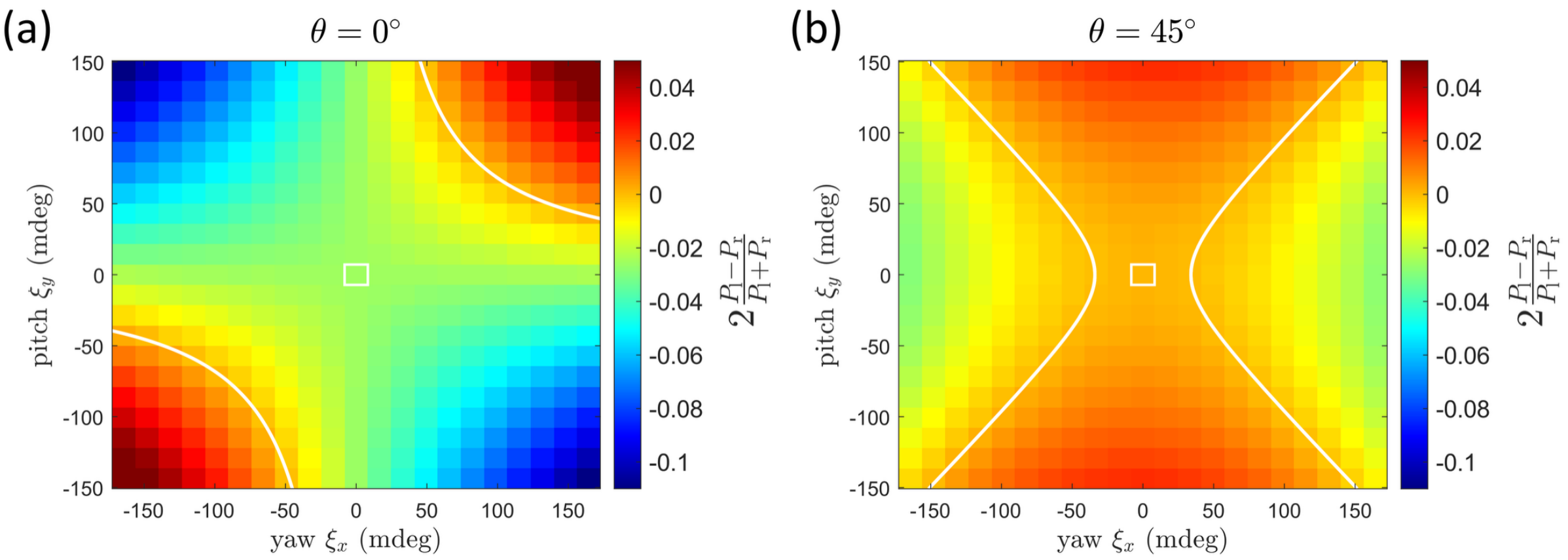

***Fig.14**: Normalized difference of the measured powers $P_l$ and $P_r$ for LCP and RCP light, $2\left(\frac{P_l-P_r}{P_l+P_r}\right)$ measured behind an analyzer, as a function of the yaw ($\xi_x$) and pitch ($\xi_y$) adjustments for keeping the lateral position at $(x_p, y_p)$ = (0,0). The analyzer behind the Pockels cell was set to (a) $\theta = 0°$ and (b) $\theta = 45°$. The white boxes correspond to $(\xi_x, \xi_y)$=(0,0). The white lines are hyperbolas that indicate where $2\left(\frac{P_l-P_r}{P_l+P_r}\right) \approx 0$ holds.*

The quantitative relationships describing $g^{(0)}_{\mathrm{P}_{1,2}}$ as a function of the angles $\xi_x, \xi_y$ are given in eq.(88) and read

$$\begin{aligned} g^{(0)}_{P_1} &= 2\left[\left(\Delta S_1^{(\mathrm{pos})}(x_\mathrm{p}=0, y_\mathrm{p}=0) - \Delta\varphi\right) + 2\kappa_1\xi_x\xi_y\right] \\ g^{(0)}_{P_2} &= 2\left[\Delta S_2^{(\mathrm{pos})}(x_\mathrm{p}=0, y_\mathrm{p}=0) + \kappa_2\left({\xi_x}^2 - {\xi_y}^2\right)\right] \end{aligned} \tag{100}$$

Fitting these expressions yields

$\Delta S_1^{(\mathrm{pos})}(x_p = 0, y_p = 0) - \Delta\varphi \approx -0.0125,$ $\kappa_1 \approx 0.92 \cdot 10^{-6}$ mdeg$^{-2}$,

$\Delta S_2^{(\mathrm{pos})}(x_p = 0, y_p = 0) \approx 0.0007,$ $\kappa_2 \approx -0.53 \cdot 10^{-6}$ mdeg$^{-2}$

The fit reveals non-zero offsets at the saddle point $(\xi_x, \xi_y)$ = (0,0). The offset in $\Delta S_1^{(\mathrm{pos})}$ can be compensated by adjusting $\Delta\varphi$ via $\Delta V$, while the offset in $\Delta S_2^{(\mathrm{pos})}$ can only be compensated by adjusting the angles $\xi_x$ and $\xi_y$.

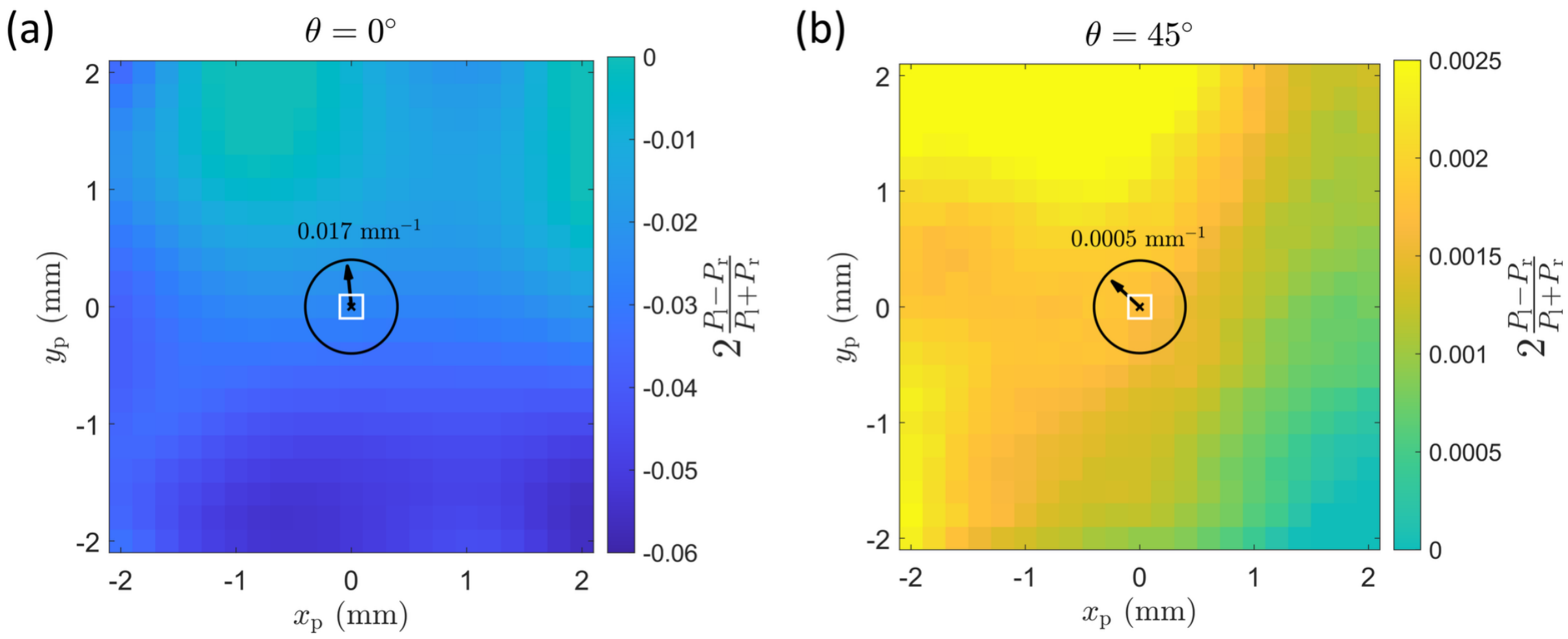

***Fig.15**: Normalized difference of the measured powers $P_l$ and $P_r$ for LCP and RCP light, $2\left(\frac{P_l-P_r}{P_l+P_r}\right)$ measured behind an analyzer, as a function of lateral position $(x_p, y_p)$ of the Pockels cell for keeping the adjustment angles $(\xi_x, \xi_y)$=(0,0). The analyzer behind the Pockels cell was set to (a) $\theta = 0°$ and (b) $\theta = 45°$. The white boxes correspond to $(x_p, y_p)$ = (0,0), and the black circle indicates the $\frac{1}{e^2}$-beam diameter. The arrows point into the direction of the largest gradient whose magnitude is given by the numbers.*

### 5.5 Step 5: Characterizing polarization artifacts $g^{(1)}_{\mathrm{P}_{1,2}}$

The polarization artifacts $g^{(1)}_{\mathrm{P}_{1,2}}$ arise from differences between the first moments of the intensity distributions of LCP and RCP light after the analyzer. They can be characterized behind an analyzer by the corresponding beam centroid displacements, $\Delta\mu_{\mathrm{lr},x}$ and $\Delta\mu_{\mathrm{lr},y}$, through

$$g^{(1)}_{\mathrm{P}_{1,2}} \approx LD \cdot \left[\frac{(\Delta\mu_{\mathrm{lr},x})_{1,2}}{\sigma_x^2}x + \frac{(\Delta\mu_{\mathrm{lr},y})_{1,2}}{\sigma_y^2}y\right] \cdot \begin{cases} \cos 2\theta \ \ (g_{\mathrm{P}_1}) \\ \sin 2\theta \ \ (g_{\mathrm{P}_2}) \end{cases} \qquad \text{from (64)}$$

According to eq.(88) these are related to the adjustment-dependent parameters $\xi_x$ and $\xi_y$ via

$$g^{(1)}_{\mathrm{P}_1} \approx \left(2\kappa_1\xi_y\frac{\theta_x}{\omega_x} + \frac{\partial\Delta S_1^{(\mathrm{pos})}}{\partial x}\bigg|_{0,0}\right)x + \left(2\kappa_1\xi_x\frac{\theta_y}{\omega_y} + \frac{\partial\Delta S_1^{(\mathrm{pos})}}{\partial y}\bigg|_{0,0}\right)y$$
$$g^{(1)}_{\mathrm{P}_2} \approx \left(2\kappa_2\xi_x\frac{\theta_x}{\omega_x} + \frac{\partial\Delta S_2^{(\mathrm{pos})}}{\partial x}\bigg|_{0,0}\right)x + \left(-2\kappa_2\xi_y\frac{\theta_y}{\omega_y} + \frac{\partial\Delta S_2^{(\mathrm{pos})}}{\partial y}\bigg|_{0,0}\right)y \qquad \text{from (88)}$$

Thus, the total beam displacement $\Delta\mu_{\mathrm{lr}} = \sqrt{\Delta\mu_{\mathrm{lr},x}^2 + \Delta\mu_{\mathrm{lr},y}^2}$, is recorded as a function of $\xi_x$ and $\xi_y$ while keeping $x_\mathrm{p} = y_\mathrm{p} = 0$. For both analyzer settings ($\theta = 0°$ and $\theta = 45°$) the results are shown in Fig.16.

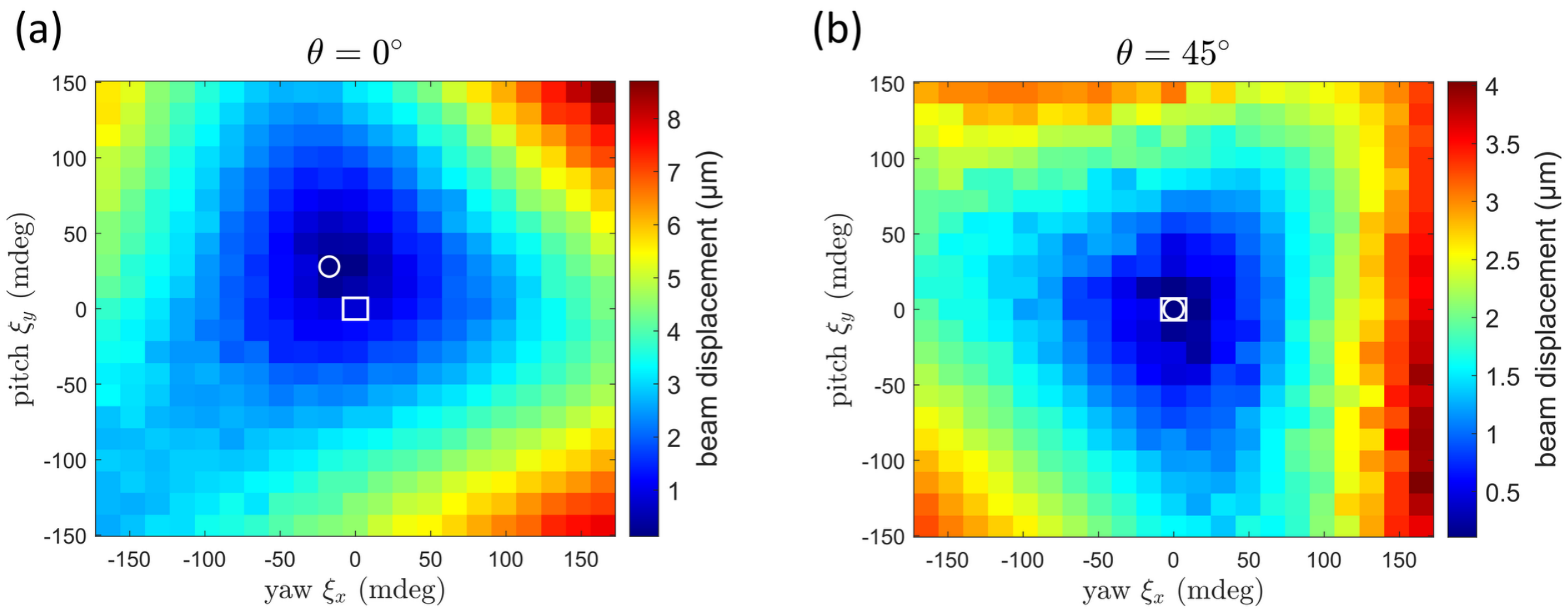

***Fig.16**: Beam displacement $\Delta\mu_{lr}$ as a function of the yaw ($\xi_x$) and pitch ($\xi_y$) adjustments for keeping the lateral position at ($x_p, y_p$) = (0,0). The analyzer behind the Pockels cell was set to (a) $\theta = 0°$ and (b) $\theta = 45°$. The white circles indicate the angles of minimum beam displacement and the white boxes correspond to ($\xi_x, \xi_y$)=(0,0).*

As evident from Fig.16a, for $\theta = 0°$ the angles at which the minimum beam displacement of 0.1 μm (indicated by the white circle in Fig.16a) is observed do not coincide with the saddle point ($\xi_x, \xi_y$)=(0,0). At the saddle point, this displacement amounts to 0.8 μm. In contrast, for $\theta = 45°$ the minimum beam displacement of 0.1 μm occurs at the saddle point, as shown in Fig.16b. Since in a real experiment the orientation of the sample (a single object) is generally unknown it is best to adopt a compromise between both contributions and to ensure $g_{\mathrm{P}_1}^{(1)} \approx g_{\mathrm{P}_2}^{(1)}$. This requires adjustments of the angles $\xi_x$ and $\xi_y$, which may, however, be in conflict with the optimization of $g_{\mathrm{P}_2}^{(0)}$ (step 4). Ideally, one would vary $x_{\mathrm{p}}$ and $y_{\mathrm{p}}$ so that $\frac{\partial \Delta S_{1,2}^{(\mathrm{pos})}}{\partial x}$ and $\frac{\partial \Delta S_{1,2}^{(\mathrm{pos})}}{\partial y}$ are minimized in eq.(88). In our case, however, this cannot be achieved due to the demands on the beam deviation, as shown in Fig.13 and eq.(99).

### 5.6 Step 6: Minimizing the artifacts and maximizing ellipticity

The interdependence of the individual contributions to the artifacts in the $g$-factor prevents the simultaneous minimization of all contributions through adjustment of the alignment parameters, as this would impose mutually conflicting requirements. For example, in step 5, Fig.16, for minimizing $g_{\mathrm{P}_{1,2}}^{(1)}$, it would be required to adjust the angles $\xi_x$ and $\xi_y$ to hit the white circle. However, this condition is mutually exclusive for $\theta = 0°$ and $\theta = 45°$. Moreover, it would conflict with step 4, Fig.14b, where minimizing $g_{\mathrm{P}_2}^{(0)}$ requires adjusting $\xi_x$ and $\xi_y$ to coincide with the white lines.

To resolve these conflicting requirements, an iterative compromise must be found. Building on the initial coordinates ($x_\mathrm{p}$, $y_\mathrm{p}$) obtained in step 3 after the pre-alignment (steps 1 and 2), the actual optimization procedure is carried out in the following sequence (step 6):

A. Angle adjustment: First, $g_{\mathrm{P}_2}^{(0)}$ and $g_{\mathrm{P}_{1,2}}^{(1)}$ are minimized through suitable adjustment of the angles $\xi_x$ and $\xi_y$.

B. Voltage adjustment: Next, $g_{\mathrm{P}_1}^{(0)}$ is minimized by appropriately adjusting $\Delta V$. This establishes a reliable set of starting values for ($x_\mathrm{p}$, $y_\mathrm{p}$, $\xi_x$, $\xi_y$, $\Delta V$).

C. Ellipticity maximization: Finally, using this initial configuration, the angle $\gamma$ of the linear polarization of the incident beam and the symmetric voltage component applied to the Pockels cell, $\delta V$, are adjusted to maximize the ellipticity, $\epsilon$, of the LCP and RCP light. The ellipticity is monitored with a polarimeter.

Adjustment steps 3 to 6 are then repeated iteratively two to three times. The steps required for addressing and minimizing the various artifacts are summarized in Table 2.

***Table 2**: Overview of the adjustment steps that address the various artifacts PD refers to a photodiode, LES to a lateral effect sensor.*

| | artifact | measured quantity | parameters varied | polarizer setting | detector |
|---|---|---|---|---|---|
| step 2 | $g_{\mathrm{P}_{1,2}}^{(0)}$ (qualitative) | $2\frac{P_\mathrm{l}-P_\mathrm{r}}{P_\mathrm{l}+P_\mathrm{r}}$ | $\xi_x, \xi_y$ | $\theta = 0°, 45°$ | PD |
| step 3 | $g_\mathrm{I}^{(1)}$ | $\Delta\mu_{\mathrm{lr},x,y}$ | $x_\mathrm{p}$, $y_\mathrm{p}$ | none | LES |
| step 4 | $g_{\mathrm{P}_{1,2}}^{(0)}$ | $2\frac{P_\mathrm{l}-P_\mathrm{r}}{P_\mathrm{l}+P_\mathrm{r}}$ | $\xi_x, \xi_y$ and $x_\mathrm{p}$, $y_\mathrm{p}$ | $\theta = 0°, 45°$ | PD |
| step 5 | $g_{\mathrm{P}_{1,2}}^{(1)}$ | $\Delta\mu_{\mathrm{lr},x,y}$ | $\xi_x, \xi_y$ | $\theta = 0°, 45°$ | LES |
| step 6 | | $\epsilon_{\mathrm{l,r}}$ | $\gamma, \delta V$ | none | Polarimeter |

### 5.7 Achievements

The results that can be achieved for our KD*P Pockels cell by following the adjustment procedure detailed above are summarized in Table 3. The table provides the data in the measurement plane directly behind the Pockels cell, and the data that result after transformation to the sample plane using eqs.(94) and (96) with $\omega \approx 400$ µm and $\omega_s \approx 50$ µm.

***Table 3****: Magnitude of the remaining artifacts with our Pockels cell after minimization. The entries (0) and (1) refer to the contributions from differences of the zeroth and first moments of the intensity distributions for LCP and RCP light to the artifacts.*

| | Measurement plane | | Sample plane | | Sample plane (at $s_{\max} = \pm 30$ µm) | |
|---|---|---|---|---|---|---|
| moment order | (0) | (1) | (0) | (1) | (0) | (1) |
| $g_{\mathrm{I}}$ | <$10^{-5}$ | $\mathrm{BD} \approx 200$ nrad | <$10^{-5}$ | $\Delta\mu_{\mathrm{lr,s}} \approx 0.02$ µm | <$10^{-5}$ | $\pm 0.001$ |
| $g_{\mathrm{P}_1}$ | <0.001 | $\Delta\mu_{\mathrm{lr}} \approx 0.4$ µm | <0.001 | $\Delta\mu_{\mathrm{lr,s}} \approx 0.05$ µm | <0.001 | $\pm 0.003$ |
| $g_{\mathrm{P}_2}$ | <0.001 | $\Delta\mu_{\mathrm{lr}} \approx 0.4$ µm | <0.001 | $\Delta\mu_{\mathrm{lr,s}} \approx 0.05$ µm | <0.001 | $\pm 0.003$ |
| | | | | | | |
| $\epsilon_{\mathrm{l,r}}$ | $> 0.996$ | | | | | |

Hence, for this Pockels cell the maximum artifact in the sample plane amounts to $\approx |g_I| + \max(|g_{P_{1,2}}|) \approx 0.004$. However, it should be noted that even lower artifact levels can be achieved by using carefully selected Pockels cells with high optical homogeneity and, preferably, larger clear apertures.

## 6. Impact of other optical elements on the polarization status

### 6.1 Stress-induced birefringence from an optical element

The polarization state of the light that emerges from the Pockels cell is affected by any birefringent optical element in the beam path. For excitation, an air-spaced achromatic doublet lens (ACA254-100-B, Thorlabs) was used, as such designs typically exhibit significantly lower stress-induced birefringence compared to cemented doublets. In addition, the retaining ring holding the doublet was slightly loosened to reduce mechanically induced stress, thereby diminishing the stress-induced birefringence to less than 0.0005 rad. Consequently, the birefringence of the excitation lens can be neglected in our setup.

However, the sample is mounted in a cryostat that serves as a vacuum chamber, and the cryostat windows exhibit a small but finite stress-induced birefringence, which cannot be neglected. From an optical point of view, such a window can be regarded as a retarder that introduces a phase shift $\varphi$ between the ordinary and extraordinary components propagating through it. Evaluating the Mueller matrix of a retarder, eq.(74), whose fast axis is oriented at an angle $\phi$ with respect to the horizontal axis of the laboratory frame ($x$-axis) yields for $\varphi \ll \frac{\pi}{2}$ in first order

$$M_{\text{window}} = \begin{pmatrix} 1 & 0 & 0 & 0 \\ 0 & 1 & 0 & -\varphi\sin(2\phi) \\ 0 & 0 & 1 & \varphi cos(2\phi) \\ 0 & \varphi\sin(2\phi) & -\varphi\cos(2\phi) & 1 \end{pmatrix} \tag{101}$$

If the incident (i)-LCP and (i)-RCP light is described by Stokes vectors $\vec{S}_{\mathrm{l}}^{(\mathrm{i})}$ and $\vec{S}_{\mathrm{r}}^{(\mathrm{i})}$, respectively, these are transformed to $M_{\text{window}} \cdot \vec{S}_{\mathrm{l}}^{(\mathrm{i})}$ and $M_{\text{window}} \cdot \vec{S}_{\mathrm{r}}^{(\mathrm{i})}$ where $\vec{S}_{\mathrm{l,r}}^{(\mathrm{i})}$ corresponds to the Stokes vector given in eq.(41). According to the formalism detailed in Section 3.2 the transformed Stokes vector is inserted into eq.(37), which yields for the artifact

$$\begin{aligned} g_P &\approx \frac{1}{\frac{t_x^2 + t_y^2}{2}} \cdot \left[M_{\text{pol,rot}} \cdot M_{\text{window}} \cdot \vec{S}_{\mathrm{l}}^{(\mathrm{i})} - M_{\text{pol,rot}} \cdot M_{\text{window}} \cdot \vec{S}_{\mathrm{r}}^{(\mathrm{i})}\right]_0 \\ &\approx 2LD \cdot \left[S_1^{(\mathrm{asym})} \cdot \cos(2\theta) + S_2^{(\mathrm{asym})} \cdot \sin(2\theta) \underbrace{+\varphi \cdot \sin(2\phi - 2\theta)}_{g_{\mathrm{P}}^{(\mathrm{w})}}\right] \\ &\approx 2LD \cdot \left[\left(S_1^{(\mathrm{asym})} + \varphi\sin(2\phi)\right) \cdot \cos(2\theta) + \left(S_2^{(\mathrm{asym})} - \varphi\cos(2\phi)\right) \cdot \sin(2\theta)\right] \end{aligned} \tag{102}$$

Typically, the window can be rotated around the optical axis ($z$-axis), allowing one to freely choose the relative angle $\phi$ between the fast axis of the birefringent window and the horizontal axis of the laboratory frame ($x$-axis). From eq.(102) follows that setting $\phi = 90°$ or $\phi = 45°$ determines whether the additional birefringence of the window contributes to the $S_2^{(\mathrm{asym})}$ component or the $S_1^{(\mathrm{asym})}$ component of the Stokes vector, respectively. Choosing $\phi = 45°$ is preferable, since in this configuration the additional artifact can be compensated by adjusting the $S_1^{(\mathrm{asym})}$ component via the voltage applied to the Pockels cell, see eq.(80).

## 6.2 Compensation for the birefringence from the cryostat windows

In our setup the light emerging from the Pockels cell passes through two cryostat windows, each of which can be modeled as a retarder as described above. The window material was specifically chosen for its low stress-optical coefficient of $0.02 \cdot 10^{-6}\,\mathrm{mm}^2 N^{-1}$ (SF57,[55]) [56–59]. For comparison, fused silica exhibits a significantly larger coefficient of approximately $3.5 \cdot 10^{-6}\,\mathrm{mm}^2 N^{-1}$ [60]. This deliberate material choice ensures that the static birefringence of the windows, amounting to $\varphi \approx 0.005$ to 0.01 rad at 700 nm (corresponding to a retardance between $\frac{\lambda}{1250}$ and $\frac{\lambda}{600}$), changes, as measured experimentally, by no more than 0.0005 rad upon reducing the pressure inside the cryostat from ambient conditions to $10^{-5}$ mbar. Importantly, this pressure change does not alter the orientation of the fast axis.

In the following, we number the cryostat windows as 1 and 2, and denote the induced phase shifts and orientations of their fast axis (with respect to the horizontal axis of the laboratory frame) as $\varphi^{(\mathrm{w}1,2)}$ and $\phi^{(\mathrm{w}1,2)}$, respectively, inducing the artifacts $g_{\mathrm{P}}^{(\mathrm{w}1,2)}$. The windows can be rotated around the optical axis such that the fast axes of the two windows are oriented perpendicular with respect to each other, i.e. $\phi^{(\mathrm{w}1)} = 45°$ and $\phi^{(\mathrm{w}2)} = 135°$ giving

$$\begin{aligned} g_{\mathrm{P}}^{(\mathrm{w}1)}(\varphi^{(\mathrm{w}1)}, 45°) &= +2LD \cdot \varphi^{(\mathrm{w}1)} \cdot \cos(2\theta) \\ g_{\mathrm{P}}^{(\mathrm{w}2)}(\varphi^{(\mathrm{w}2)}, 135°) &= -2LD \cdot \varphi^{(\mathrm{w}2)} \cdot \cos(2\theta) \end{aligned} \qquad (103)$$

and for the total artifact after passing both windows

$$g_{\mathrm{P}} \approx 2LD \cdot [\left(S_1^{(\mathrm{asym})} + \left(\varphi^{(\mathrm{w}1)} - \varphi^{(\mathrm{w}2)}\right)\right) \cdot \cos(2\theta) + S_2^{(\mathrm{asym})} \cdot \sin(2\theta)] \qquad (104)$$

Out of the 40 windows that were available for the setup the two that fulfilled $\varphi^{(\mathrm{w}1)} \approx \varphi^{(\mathrm{w}2)}$ have been chosen. Hence, only a small residual contribution to $g_{\mathrm{P}}$ is added to $S_1^{(\mathrm{asym})}$, which can be minimized by appropriately adjusting the antisymmetric voltage bias $\Delta V$ applied to the Pockels cell, see. eq.(80). That this compensation works is demonstrated in Fig.17, which shows the artifact as a function of the angle $\theta$. For each window separately this amounts to $\left|g_{\mathrm{P}}^{(\mathrm{w1})}\right| \approx \left|g_{\mathrm{P}}^{(\mathrm{w2})}\right| \approx 0.02$, whereas for both windows together these compensate each other and $\left(g_{\mathrm{P}}^{(\mathrm{w1})} + g_{\mathrm{P}}^{(\mathrm{w2})}\right) < 0.002$.

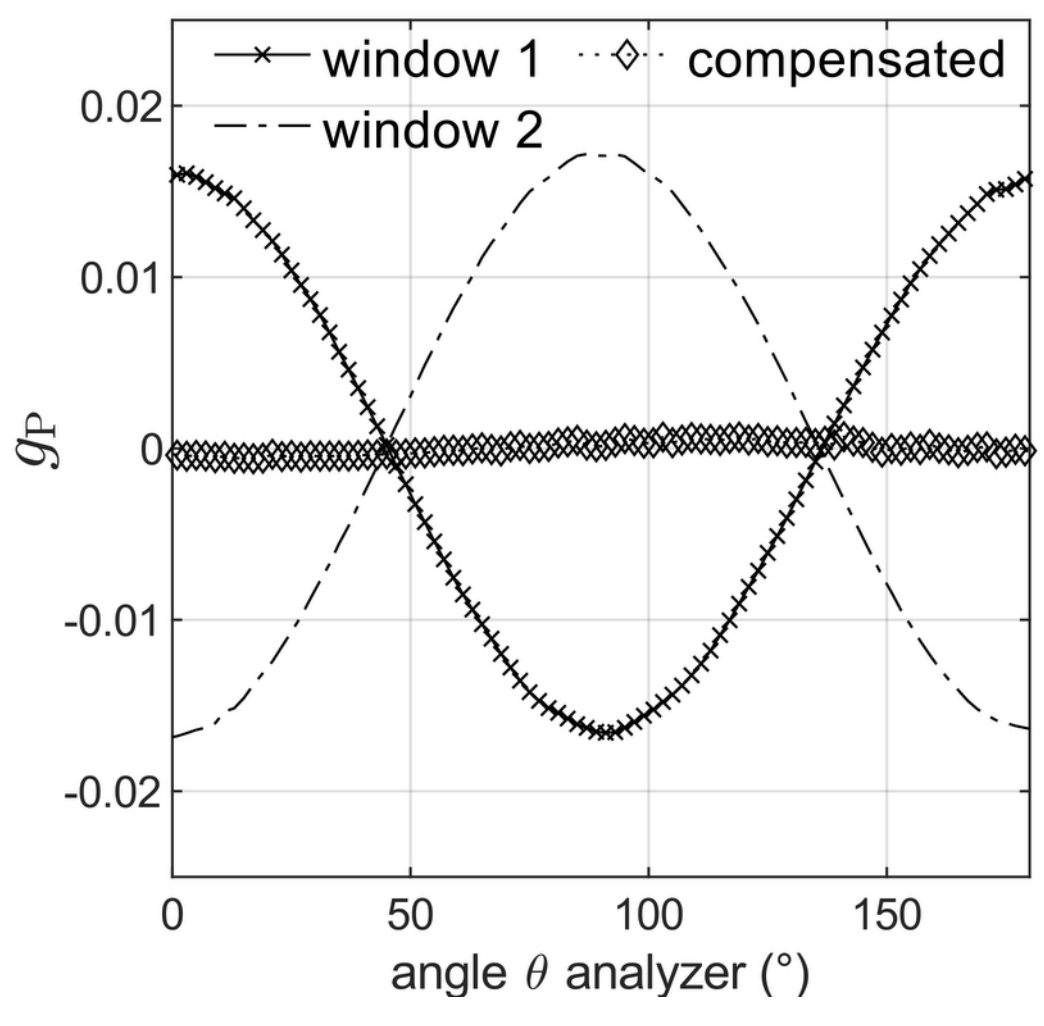

***Fig.17***: *Artifacts $g_P^{(w1)}$ (full line) and $g_P^{(w2)}$ (dash-dotted line) as a function of the angle $\theta$. The dotted line shows the resulting total artifact $g_P$ after compensating the individual contributions by appropriately aligning the windows.*

### 6.3 Wavelength dependence

Since CD spectroscopy requires measuring the dissymmetry factor $g$ as a function of wavelength it is crucial that the conditions for reducing the artifacts to a tolerable magnitude are fulfilled across the whole spectral range of interest. Fig.18 illustrates to what extent this is achieved for various parameters over the spectral range from 693 nm to 783 nm.

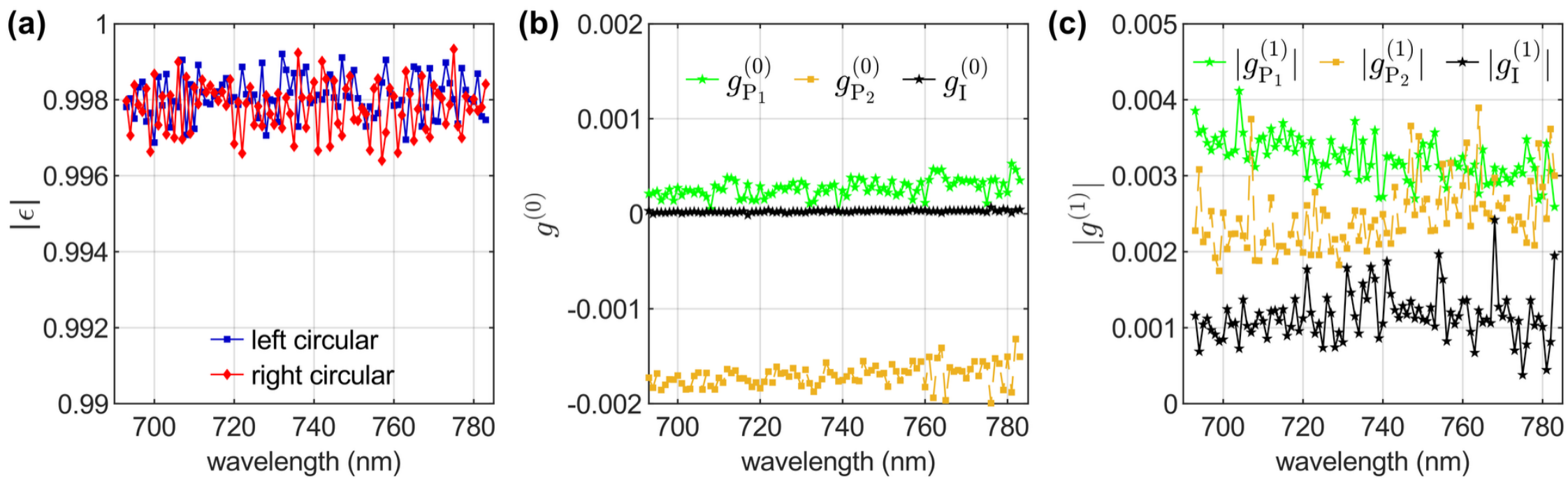

***Fig.18**: (a) Wavelength dependence of the ellipticity for (i)-LCP (blue) and (i)-RCP (red) light. (b) Wavelength dependence of the artifacts from differences in the zeroth moments of the underlying intensity distributions, $g_I^{(0)}$ (black), $g_{P_1}^{(0)}$ (green), and $g_{P_2}^{(0)}$ (yellow). (c) Wavelength dependence of the magnitudes of the artifacts from differences in the first moments of the underlying intensity distributions, $\left|g_I^{(1)}\right|$ (black), $\left|g_{P_1}^{(1)}\right|$ (green), and $\left|g_{P_2}^{(1)}\right|$ (yellow). The values for (b) and (c) are converted to the sample plane, see eqs.(94) and (96) and Table 3, and are evaluated at $x_s = y_s = \pm 30$ µm.*

This reveals $\epsilon_{\mathrm{l,r}} > 0.996$ for the ellipticity of the (i)-LCP and (r)-RCP light over the entire wavelength range, Fig.18a. For the same wavelength range the contributions of intensity and polarization artifacts $g_{\mathrm{I}}^{(0)}$ and $g_{\mathrm{P}_{1,2}}^{(0)}$ resulting from differences in the zeroth moments of the underlying intensity distributions, and converted to the sample plane, are diminished to $g_{\mathrm{I}}^{(0)} < 10^{-5}$ and $\left|g_{\mathrm{P}_{1,2}}^{(0)}\right| < 0.002$, Fig.18b. And finally, the contributions of the intensity and polarization artifacts $g_{\mathrm{I}}^{(1)}$ and $g_{\mathrm{P}_{1,2}}^{(1)}$ resulting from differences in the first moments of the underlying intensity distributions, and converted to the sample plane, are smaller than 0.004 for all wavelengths of interest. All together the artifacts induced by the optical elements in the beam path contribute at most 0.006 to the dissymmetry factor across the whole wavelength range from 693 nm to 783 nm.

### 6.4 What has not been considered

There are additional factors that could introduce artifacts such as:

- Strong focusing of the excitation light, which can produce complex polarization patterns at the spot of the sample[61].
- Non-normal angles of incidence relative to the substrate, potentially causing interference effects at the surface of the substrate and polarization artifacts[61].

- Inappropriate switching times of the Pockels cell: Switching the voltage applied to the KD*P crystal too rapidly (typically $< 100$ µs) excites piezoelectric resonances, leading to acoustic ringing that strongly modulates the birefringence for several milliseconds. Conversely, the high internal resistance of the KD*P crystal (typically about $100\ \mathrm{G\Omega}$) combined with the capacitance of the whole circuitry (about 10 to 100 pF) forms an unintended macroscopic RC circuit. This results in a slow parasitic RC time constant in the sub-second to second range. Consequently, the polarization state undergoes a slow drift due to leakage currents even after the active switching phase is completed. To avoid artifacts, the switching frequency between LCP and RCP light must be carefully balanced to neither overlap with the acoustic ringing nor fall into the regime of this slow dielectric relaxation[51].

## 7. Verification Experiments

In the procedures described so far, a linear polarizer oriented at $\theta = 0°$ and $\theta = 45°$ with respect to the $x$-axis in the laboratory frame (see Fig.7) was used to mimic a sample exhibiting linear dichroism (LD = 1, by definition). Experimental verification of the predicted artifacts, however, requires an actual sample that shows linear dichroism while exhibiting negligible circular dichroism. Such a sample can be realized by preparing a thin layer of fluorescent dye molecules in which a subset of molecules is irreversibly bleached using a linearly polarized laser with high intensity. This introduces the desired anisotropy through photoselection during bleaching, without introducing chirality. For this purpose, Atto 725 (Leica Microsystems) was dissolved in a 1:3 (vol/vol) EtOH:$H_2O$ containing 0.3 % PVA at a concentration of 40 µM. A small drop of this solution ($20$ µL) was spin-coated (adsorption for 30 s, spin-coated at 2000 rpm for 90 s) on a UV fused silica substrate. Linear dichroism was then induced by bleaching the sample with light that was linearly polarized at $\theta = 0°$ (for obtaining the artifact $g_{\mathrm{P}_1}$) or $\theta = 45°$ (for obtaining the artifact $g_{\mathrm{P}_2}$) using an intensity of 50 - 100 W/cm$^2$ for 30 minutes. The contribution to the intensity artifact $g_{\mathrm{I}}$ was determined from an experiment prior to bleaching, where the transition dipole moments of the dye molecules are randomly distributed. The results of this approach are shown in Fig.19

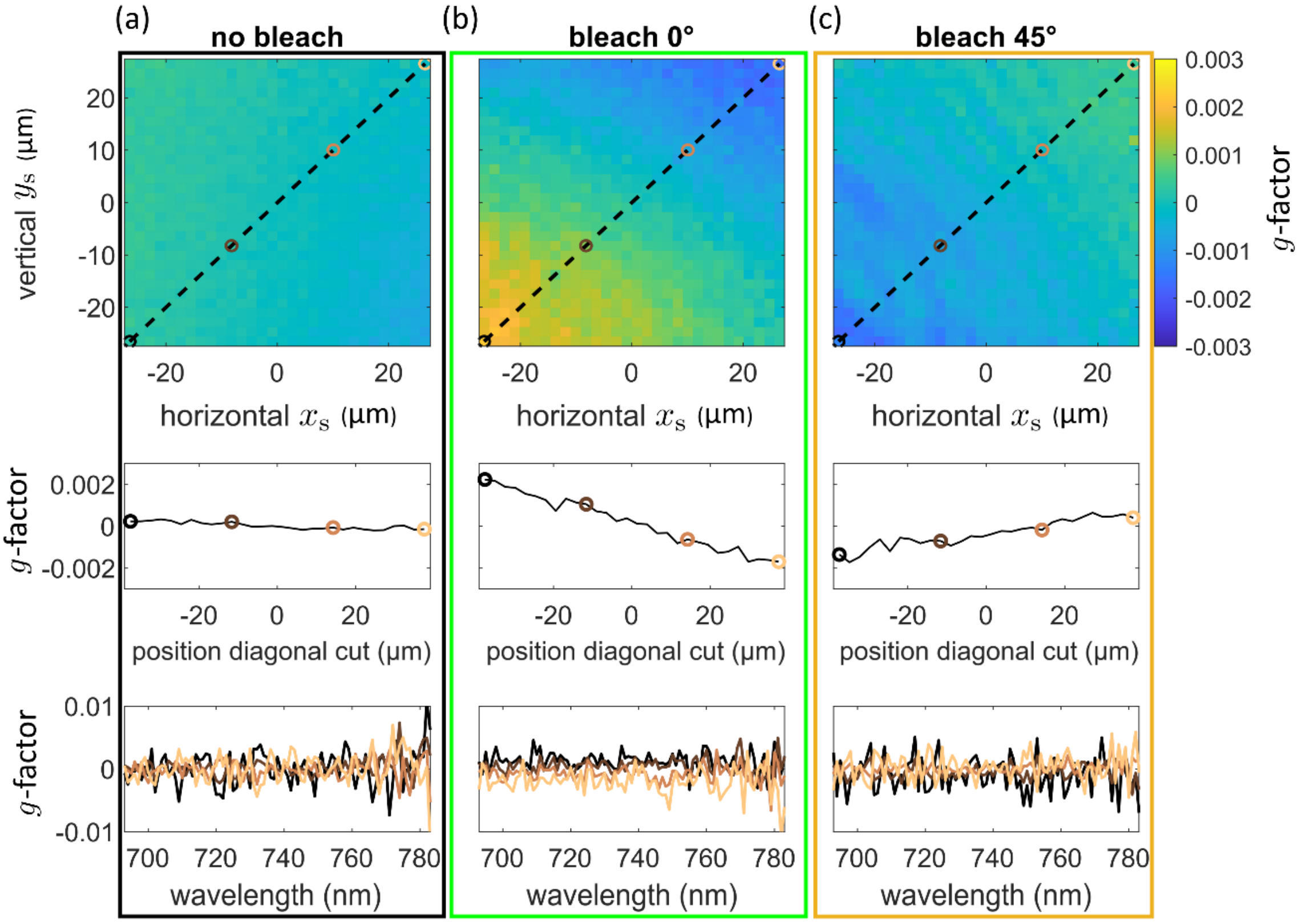

***Fig.19**: Verification of the artifacts $g_I$ (black), $g_{P_1}$ (green), and $g_{P_2}$ (yellow) from left to right for a concentrated layer of Atto 725 dye molecules embedded in PVA. (a) Top: Spatial variation of the $g$-factor (color-coded) across the dye layer. The data acquisition time was 30 minutes. Center: $g$-factors at four distinct sample positions along the diagonal as indicated by the circles. Bottom: $g$-factors for the four sample positions as a function of the excitation wavelength. The data acquisition time was 30 seconds per data point. (b) Same as (a), but prior to measuring the $g$-factor the sample was illuminated for 30 min with light at 740 nm (50 - 100 W/cm$^2$) that was linearly polarized under an angle $\theta = 0°$. (c) Same as (a), but prior to measuring the $g$-factor the sample was illuminated for 30 min with light at 740 nm (50 - 100 W/cm$^2$) that was linearly polarized under an angle $\theta = 45°$. All measurements were done at room temperature inside the cryostat in vacuum. The frames of panels (a), (b) and (c) are in the same colors as the corresponding artifacts in Fig.18b,c.*

The pronounced signal variations observed in Fig.19 at wavelengths above 770 nm are attributed to noise, as the absorption of Atto 725 becomes very small in this spectral range. The quantitative results are summarized in Table 4, which lists the ranges of variations for the different artificial contributions to the $g$-factor. These are compared with the worst-case artifacts predicted using a polarizer that mimics an oriented sample, showing reasonable agreement. It should be noted that the polarizer has by definition LD = 1, whereas the dye layer exhibits LD ≈ 0.5. Therefore, when comparing the results for $g_{\mathrm{P}_1}$ and $g_{\mathrm{P}_2}$ from Figs.18 and 19, the values from Fig.18 must be divided by a factor of 2.

***Table 4****: Ranges of variations of the artifacts obtained from the results shown in Fig.19, and the estimated maximum artifacts for using a polarizer as "sample".*

| | Measured (in Fig.19) | Expected (from Fig.18) |
|---|---|---|
| $g_{\mathrm{I}}$ | $-0.0002$ to $+0.0002$ | $-0.001$ to $+0.001$ |
| $g_{\mathrm{P}_1}$ | $-0.0022$ to $+0.0018$ | $-0.0015$ to $+0.0015$ |
| $g_{\mathrm{P}_2}$ | $-0.0014$ to $+0.0006$ | $-0.002$ to $+0.000$ |

## 8. Demonstration experiment

Finally, we present example CD spectra of single objects to demonstrate the functionality of the constructed setup. For convenience, we briefly recall the definitions of the (absolute) circular dichroism $CD(\lambda)$, eq.(1) and the dissymmetry factor $g(\lambda)$, eq.(2) as introduced in Chapter 1:

$$CD(\lambda) \; = A_{\mathrm{l}}(\lambda) - A_{\mathrm{r}}(\lambda) \qquad \text{from (1)}$$

$$g(\lambda) = \frac{A_{\mathrm{l}}(\lambda) - A_{\mathrm{r}}(\lambda)}{\frac{1}{2}\left(A_{\mathrm{l}}(\lambda) + A_{\mathrm{r}}(\lambda)\right)} \qquad \text{from (2)}$$

In single-object CD spectroscopy, fluorescence-excitation spectra are measured instead of absorption spectra.

$$A(\lambda) \; \propto \; F(\lambda) \qquad \text{from (4)}$$

In this context it is important to realize that λ refers to the excitation wavelength. Provided that the fluorescence quantum yield is independent of the excitation wavelength and that saturation effects are avoided, the spectral profiles of the absorption and excitation spectra are identical and proportional to each other. Consequently,

$$CD(\lambda) \; = \left(A_{\mathrm{l}}(\lambda) - A_{\mathrm{r}}(\lambda)\right) \propto \left(F_{\mathrm{l}}(\lambda) - F_{\mathrm{r}}(\lambda)\right) \qquad (105)$$

$$g(\lambda) = \frac{F_{\mathrm{l}}(\lambda) - F_{\mathrm{r}}(\lambda)}{\frac{1}{2}\left(F_{\mathrm{l}}(\lambda) + F_{\mathrm{r}}(\lambda)\right)} \qquad \text{from (5)}$$

the measured signal can be expressed as proportional to the absorption, where the proportionality factor depends both on photophysical properties of the object, such as the fluorescence quantum yield, and on instrumental parameters, including light collection efficiency and detector sensitivity. Nevertheless, the spectral shapes of CD spectra obtained via either method remain identical.

From eqs.(105) and (5) it follows that $CD(\lambda)$, and $g(\lambda)$ are closely related, differing only by a normalization factor. By choosing $\frac{1}{\max\left(\frac{1}{2}\left(F_{\mathrm{l}}(\lambda)+F_{\mathrm{r}}(\lambda)\right)\right)}$ as the normalization for the fluorescence-excitation CD spectrum, one obtains

$$CD_{\mathrm{norm}}(\lambda) = \frac{\left(F_{\mathrm{l}}(\lambda) - F_{\mathrm{r}}(\lambda)\right)}{\max\left(\frac{1}{2}\left(F_{\mathrm{l}}(\lambda) + F_{\mathrm{r}}(\lambda)\right)\right)} \qquad (106)$$

This can be directly related to the dissymmetry factor

$$
\begin{aligned}
CD_{\mathrm{norm}}(\lambda) &= \frac{\left(F_{\mathrm{l}}(\lambda) - F_{\mathrm{r}}(\lambda)\right)}{\max\left(\frac{1}{2}\left(F_{\mathrm{l}}(\lambda) + F_{\mathrm{r}}(\lambda)\right)\right)} \\
&= \frac{\left(F_{\mathrm{l}}(\lambda) - F_{\mathrm{r}}(\lambda)\right)}{\frac{1}{2}\left(F_{\mathrm{l}}(\lambda) + F_{\mathrm{r}}(\lambda)\right)} \cdot \frac{\frac{1}{2}\left(F_{\mathrm{l}}(\lambda) + F_{\mathrm{r}}(\lambda)\right)}{\max\left(\frac{1}{2}\left(F_{\mathrm{l}}(\lambda) + F_{\mathrm{r}}(\lambda)\right)\right)} \\
&= g(\lambda) \cdot \underbrace{\frac{\frac{1}{2}\left(F_{\mathrm{l}}(\lambda) + F_{\mathrm{r}}(\lambda)\right)}{\max\left(\frac{1}{2}\left(F_{\mathrm{l}}(\lambda) + F_{\mathrm{r}}(\lambda)\right)\right)}}_{\leq 1}
\end{aligned}
\tag{107}
$$

for which we provided a quantitative analysis of potential artifacts.

For demonstrating the functionality of our setup we chose chlorosomes from *Chlorobaculum tepidum*. Chlorosomes are specialized light-harvesting antenna organelles found in photosynthetic green sulfur bacteria. A characteristic feature of chlorosomes is the self-assembly of hundreds of thousands of bacteriochlorophyll (BChl) molecules into large supramolecular aggregates without the aid of a protein scaffold. The close packing of these molecular building blocks gives rise to delocalized electronically excited states, commonly referred to as Frenkel excitons or molecular excitons. However, the precise internal architecture of these secondary structural elements is not fully understood to date. Because the circular dichroism (CD) signal depends on the relative orientations of coupled transition-dipole moments within extended molecular assemblies, chlorosomes provide an attractive model system for single-object CD spectroscopy. The CD spectra from single chlorosomes were recorded inside a cryostat under vacuum conditions. Rather than providing a detailed interpretation of the observed spectral features, these spectra in Fig.20 are presented solely as proof-of-principle measurements, demonstrating the feasibility of performing CD spectroscopy on individual, weakly fluorescing objects.

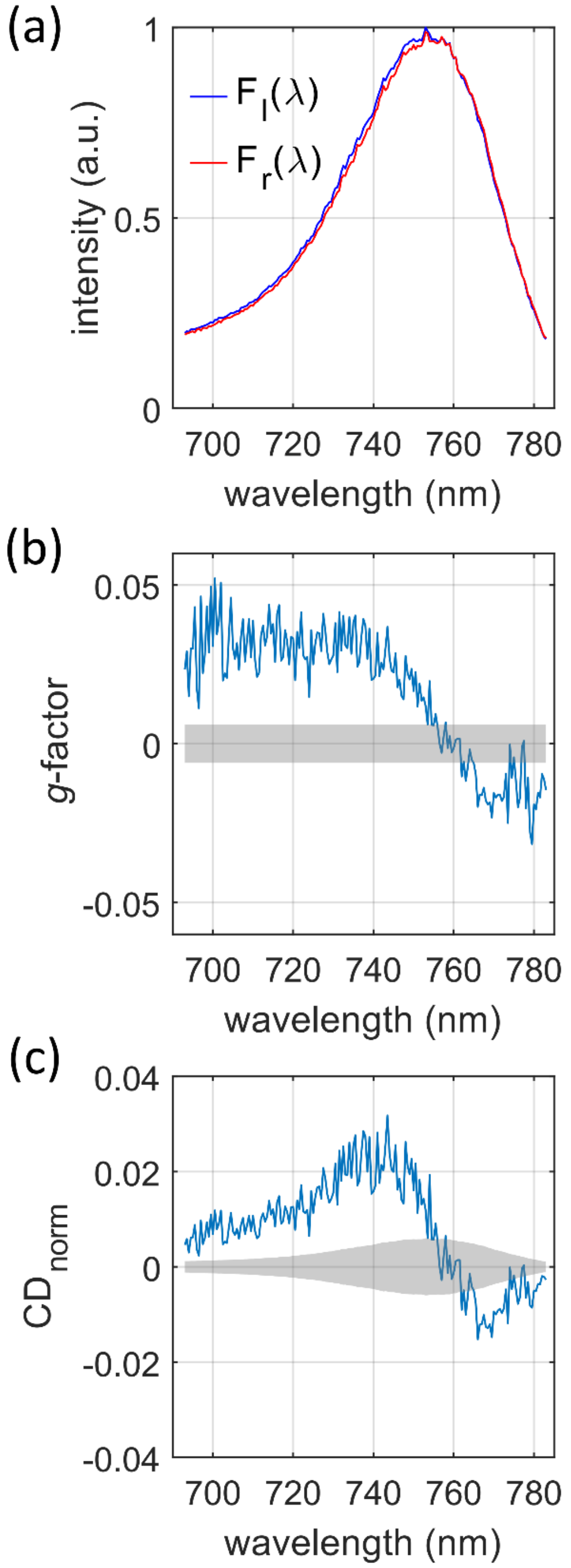

***Fig.20***: *(a) Example of fluorescence-excitation spectra from a single chlorosome from Chlorobaculum tepidum for LCP (blue) and RCP (red) light. For this particular chlorosome the linear dichroism amounted to* $LD \approx 0.5$ *(b) Dissymmetry factor* $g$ *calculated from the difference of the fluorescence-excitation spectra shown in (a). (c) CD spectrum calculated from the fluorescence-excitation spectra according to eq.(106). The bands in light gray in (b) and (c) indicate the range of possible artifacts for the* $g$*-factor and for the CD spectrum, respectively.*

## 9. Alternative experimental approaches for generating LCP and RCP light

The setup ultimately used for CD measurements on single objects, as described in the preceding chapters, has been developed over the course of many years. During this process, several alternative experimental approaches were investigated. In this chapter, we discuss these concepts and explain why they were eventually abandoned.

### 9.1 Waveplates

A straightforward method for generating circularly polarized light is the use of a waveplate as a retarder[44]. A waveplate typically consists of one or more layers of anisotropic birefringent material that introduces a relative phase shift $\varphi$ between light components polarized along its two orthogonal principal axes, commonly referred to as the fast and the slow axis. Circularly polarized light is obtained when the fast axis of the retarder is oriented under $\mp 45°$ with respect to the linear polarization direction of the incident beam. An orientation of $-45°$ generates LCP light, whereas $+45°$ produces RCP light. The Mueller matrix $M_{\text{ret}}(\phi,\varphi)$ of a retarder is given by eq.(74), where $\phi$ denotes the angle of the fast axis relative to the laboratory $x$-axis, and $\varphi$ is the phase retardation. For horizontally polarized incident light and a retarder whose fast axis is oriented at $\phi = \mp 45°$, the Stokes vector of the transmitted beam, $\vec{S}_{\text{l,r}}^{(\text{i})}$, is given by:

$$\begin{aligned} \vec{S}_{\text{l}}^{(\text{i})} &= M_{\text{ret}}(-45°,\varphi)\begin{pmatrix}1\\1\\0\\0\end{pmatrix} = \begin{pmatrix}1\\ \cos(\varphi)\\ 0\\ -\sin(\varphi)\end{pmatrix} \\ \vec{S}_{\text{r}}^{(\text{i})} &= M_{\text{ret}}(+45°,\varphi)\begin{pmatrix}1\\1\\0\\0\end{pmatrix} = \begin{pmatrix}1\\ \cos(\varphi)\\ 0\\ +\sin(\varphi)\end{pmatrix} \end{aligned} \tag{108}$$

Comparison of eq.(41) and eq.(108)

$$\vec{S}_{\text{l,r}}^{(\text{i})} \overset{(41)}{\approx} \begin{pmatrix}1\\ S_1^{(\text{sym})} \pm S_1^{(\text{asym})}\\ S_2^{(\text{sym})} \pm S_2^{(\text{asym})}\\ \mp 1\end{pmatrix} \overset{(108)}{=} \begin{pmatrix}1\\ \cos(\varphi)\\ 0\\ \mp\sin(\varphi)\end{pmatrix} \tag{109}$$

yields the correspondences $S_1^{(\text{sym})} = \cos(\varphi)$ and $S_{1,2}^{(\text{asym})} = 0$. According to eq.(46b), the resulting polarization artifact is then

$$g_{\mathrm{P}} = 0 \tag{110}$$

This result may appear counterintuitive, given that a non-ideal retarder generates slightly elliptically instead of perfectly circularly polarized light. However, it can be understood by noting that rotating the retarder from $-45°$ to $+45°$ changes only the handedness of the polarization, leaving both the shape of the polarization ellipse and the orientation of its principal axes unchanged (see Fig. 21). This result holds exactly for all values of the phase retardation $\varphi$, and does not rely on any approximation.

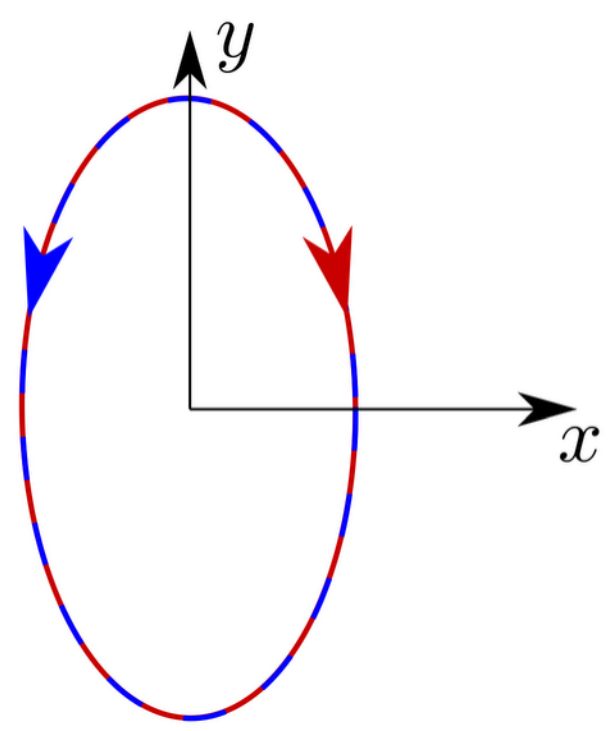

***Fig.21***: *Exaggerated schematic illustration of the polarization ellipses downstream of a waveplate oriented under $-45°$ ((i)-LCP, blue) and $+45°$ ((i)-RCP, red) with respect to the incident linear polarization direction.*

Consequently, even when the retardation deviates substantially from the ideal value of $\varphi = \frac{\pi}{2}$, the polarization ellipses corresponding to (i)-LCP and (i)-RCP light remain identical except for their opposite sense of rotation. Hence, polarization artifacts arising solely from imperfect polarization do not occur.

Despite the conceptual simplicity of this approach, several practical limitations arise when attempting to measure the CD of individual nanometer-sized objects. First, real waveplates are not perfectly homogeneous, and the retardation varies across the clear aperture. Rotating such an inhomogeneous waveplate causes different regions of the beam to experience different local retardations, resulting in position-dependent phase shifts with $\varphi_{\mathrm{r}} \neq \varphi_{\mathrm{l}}$. Second, in crystalline composite waveplates, the retardation is highly sensitive to the angle of incidence and therefore to the intrinsic divergence of the laser beam. Third, any mechanical runout or wobble of the rotation stage modulates the angle of incidence during rotation. This effect alters the downstream beam path and can generate both polarization- and intensity-related artifacts. Finally, real waveplates inevitably introduce a small beam deviation. Upon rotation, this deviation causes a displacement of the optical beam path. This effect is typically much more pronounced for achromatic and superachromatic waveplates than for zero-order waveplates.

To evaluate the feasibility of this approach, various waveplate architectures were tested, including polymer true-zero-order, composite crystalline zero-order, achromatic, and superachromatic designs. However, none of the investigated retarders allowed the total artifact contribution to be reduced below $g_{\mathrm{I}} + g_{\mathrm{P}} < 0.01$ for an LD = 1 over a spectral bandwidth of 100 nm.

### 9.2 Beam displacers and Fresnel rhomb

To eliminate the mechanical inaccuracies and systematic errors associated with rotating a waveplate, the linear polarization of the incident beam can instead be rotated by 90° before it enters a stationary retarder. An elegant implementation of this concept, which avoids any moving optical components, was proposed in [22]. Fig.22 illustrates the setup used in the present work to evaluate this approach.

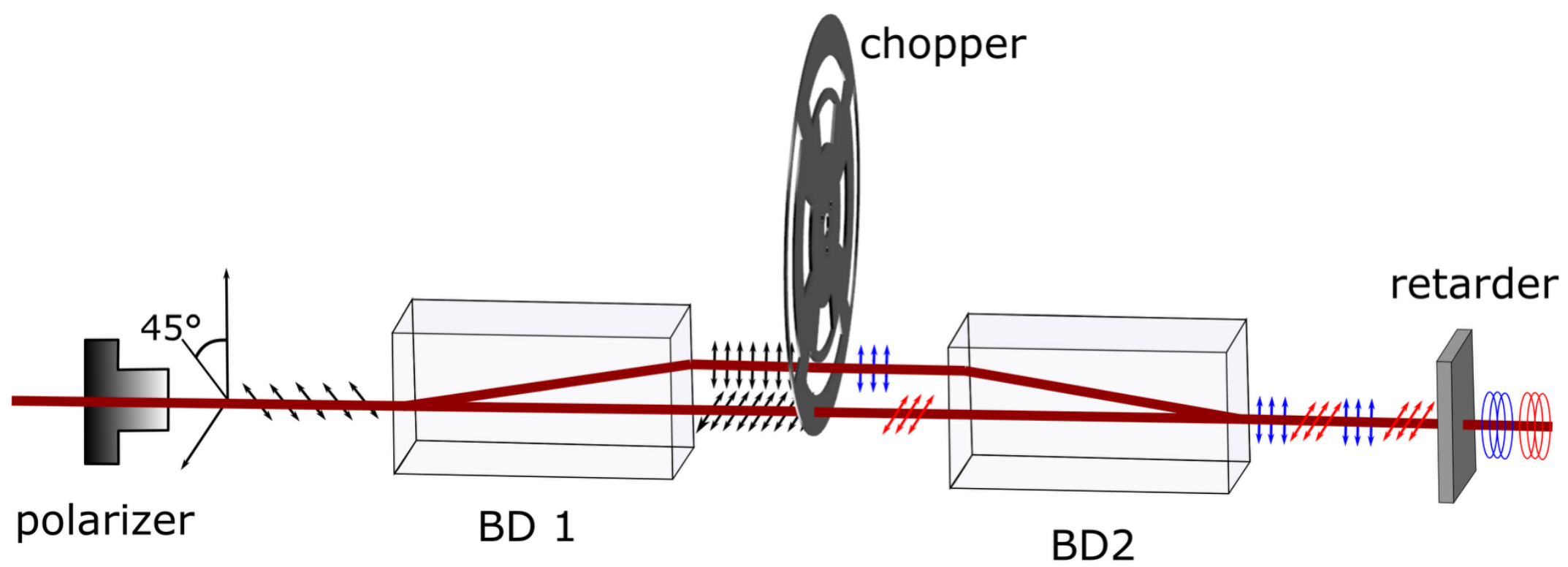

***Fig.22***: *Experimental setup used to generate alternating trains of light with mutually orthogonal linear polarization. The setup consists of a polarizer, two beam displacers (BD1 and BD2), a retarder plate, and a chopper. Adapted from[22].*

The incident beam is linearly polarized at an angle of 45° with respect to the $x$-axis of the laboratory frame. Upon passing through a birefringent calcite crystal acting as a beam displacer (BD1), the beam is split into two laterally separated components with mutually orthogonal linear polarizations. These two beams are subsequently recombined by a second, nominally identical beam displacer (BD2). A mechanical chopper positioned between the two beam displacers alternately blocks one of the two spatially separated beams. As a result, the recombined beam alternates in time between horizontal and vertical linear polarization. A stationary retarder, whose fast axis is appropriately oriented with respect to the incident polarization directions, then converts these alternating linear polarization states into LCP and RCP light.

While the retarder is still described by the Mueller matrix $M_{\mathrm{ret}}(\phi,\varphi)$, the Stokes vector of the light incident on the retarder now alternates between horizontal and vertical linear polarization. The corresponding Stokes vectors of the transmitted light are therefore given by

$$\begin{aligned}\vec{S}_{\mathrm{l}}^{(\mathrm{i})} &= M_{\mathrm{ret}}(45°,\varphi)\begin{pmatrix}1\\-1\\0\\0\end{pmatrix}=\begin{pmatrix}1\\-\cos(\varphi)\\0\\-\sin(\varphi)\end{pmatrix}\\ \vec{S}_{\mathrm{r}}^{(\mathrm{i})} &= M_{\mathrm{ret}}(45°,\varphi)\begin{pmatrix}1\\1\\0\\0\end{pmatrix}=\begin{pmatrix}1\\\cos(\varphi)\\0\\\sin(\varphi)\end{pmatrix}\end{aligned} \tag{111}$$

Comparison of eq.(41) and eq.(111)

$$\vec{S}_{\mathrm{l,r}}^{(\mathrm{i})} \overset{(41)}{\approx} \begin{pmatrix}1\\S_1^{(\mathrm{sym})} \pm S_1^{(\mathrm{asym})}\\S_2^{(\mathrm{sym})} \pm S_2^{(\mathrm{asym})}\\\mp 1\end{pmatrix} \overset{(111)}{=} \begin{pmatrix}1\\\mp\cos(\varphi)\\0\\\mp\sin(\varphi)\end{pmatrix} \tag{112}$$

yields the correspondence $S_1^{(\mathrm{asym})} = -\cos(\varphi)$. According to eq.(46b), the resulting polarization artifact is then given by

$$g_{\mathrm{P}} = 2\cdot LD\cdot S_1^{(\mathrm{asym})}\cdot cos\,2\theta = -2\cdot LD\cdot\cos(\varphi)\cdot cos\,2\theta \tag{113}$$

and the maximum value for $g_{\mathrm{P}}$ (for LD = 1, and $\theta = 0$) amounts to

$$g_{\mathrm{P}} = -2\cdot\cos(\varphi) \tag{114}$$

This result reveals the fundamental limitation of this approach. In contrast to rotating the retarder itself, modulating the linear polarization state in front of the retarder effectively interchanges the principal axes of the resulting (i)-LCP and (i)-RCP light polarization ellipses by exactly 90°, Fig.23. Consequently, any deviation of the retardation from the ideal value of $\varphi = \frac{\pi}{2}$ generates a residual linear polarization component. Because the principal axes of the two polarization ellipses no longer coincide even the slightest retardation error gives rise to polarization artifacts.

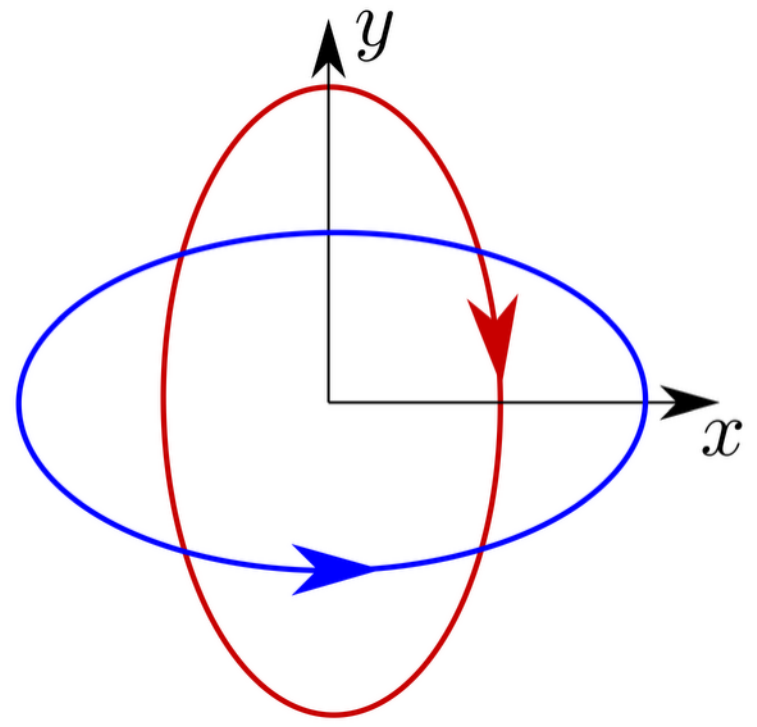

***Fig.23***: *Schematic illustration of the polarization ellipses downstream of a retarder for horizontally and vertically linearly polarized incident light. Here (i)-LCP light is indicated in blue, (i)-RCP light is indicated in red.*

The performance of this approach can be improved by replacing the waveplate with a Fresnel rhomb. Because its retardation is generated by total internal reflection rather than birefringence, the resulting phase shift is considerably less sensitive to wavelength, angle of incidence, and manufacturing tolerances. Consequently, the polarization artifacts associated with retardation errors are significantly reduced. A quarter-wave Fresnel rhomb typically employs two successive total internal reflections, Fig.24a, producing a laterally displaced output beam with a total phase retardation given by

$$\varphi = 2 \cdot 2 \cdot \arctan\left(\frac{\cos(\beta)\sqrt{\sin^2(\beta) - \left(\frac{1}{n_{\mathrm{g}}(\lambda)}\right)^2}}{\sin^2(\beta)}\right) \tag{115}$$

where $\beta$ denotes the angle of incidence at the glass-air interface, and $n_{\mathrm{g}}(\lambda)$ is the wavelength-dependent refractive index of the glass. A conventional Fresnel rhomb fabricated from N-BK7 glass typically has a cut angle of approximately 55°. According to the manufacturer's specifications (Thorlabs), both the retardation $\varphi(\lambda)$ and the corresponding ellipticity $\epsilon(\lambda)$ vary only weakly over the wavelength range from 700 nm to 780 nm, Fig.24b.

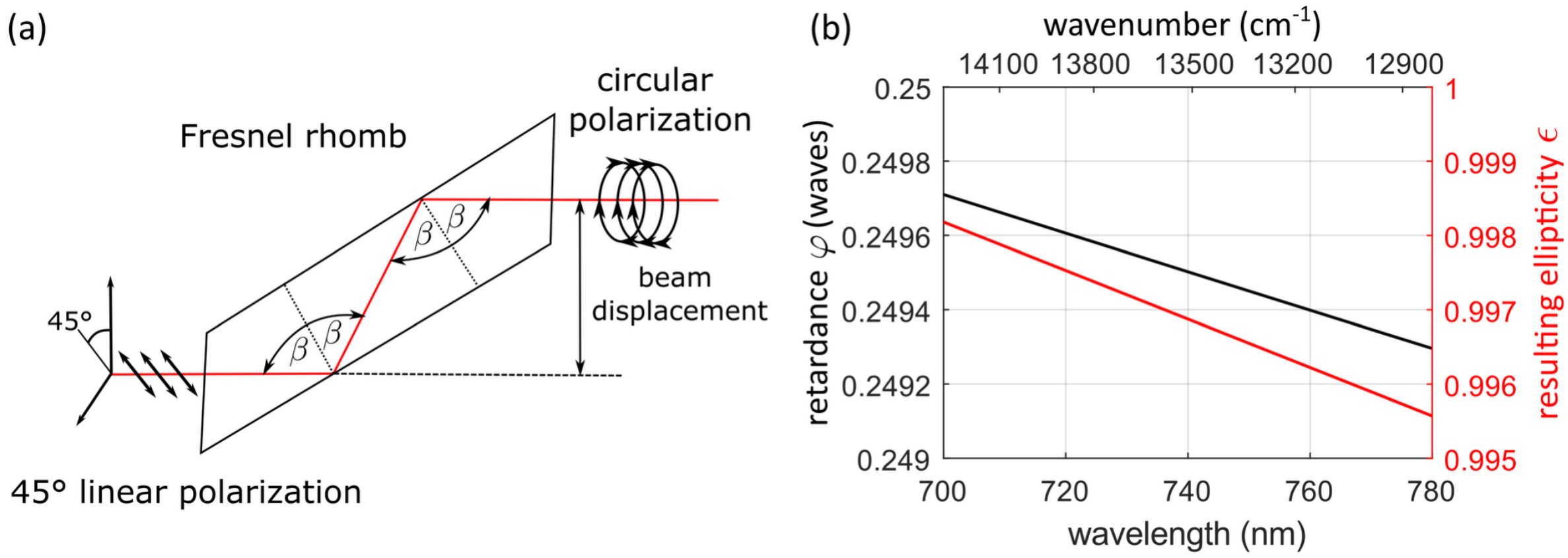

***Fig.24**: (a) Schematic illustration of the optical beam path inside a quarter-wave Fresnel rhomb. (b) Phase retardation $\varphi$ as a function of wavelength $\lambda$ (left axis, black curve) and the resulting ellipticities $\epsilon(\lambda)$ (right axis, red curve) for a commercial N-BK7 Fresnel rhomb (data adapted from[62]).*

Nevertheless, even an ideal Fresnel rhomb made of N-BK7 glass does not provide a perfectly constant quarter-wave retardation. As shown in Fig.24b, the retardation at 780 nm is $\varphi = 0.2493 \cdot 2\pi$, corresponding to a polarization artifact of approximately $|g_\mathrm{P}| \approx 0.009$ according to eq.(114). This residual error can be reduced further by slightly tilting the Fresnel rhomb with respect to the optical axis, thereby shifting the wavelength for which $\varphi = \frac{\pi}{2}$ applies closer to the center of the spectral range. In this way $g_\mathrm{P}$ can be limited to $\pm$ 0.003 across the entire wavelength interval.

In addition to these polarization-related limitations, a major practical challenge arises from the spatial recombination of the two beams. Owing to small differences in optical path length, crystal imperfections, and alignment tolerances, the recombined beams do not overlap perfectly. This imperfect spatial overlap gives rise to significant intensity artifacts. Although such artifacts can be compensated by calibration procedures in room-temperature setups, as demonstrated in[22], these calibration routines are not feasible in our cryogenic environment. As a consequence, this approach did not allow the combined artifact contribution to be reduced below $g_\mathrm{I} + g_\mathrm{P} < 0.01$ over a spectral bandwidth of 100 nm for LD = 1.

### 9.3 Flat-top beam profiles

In many experimental configurations, intensity and polarization artifacts are intrinsically coupled and measures taken to suppress intensity artifacts frequently increase polarization artifacts, and vice versa. To alleviate this coupling, one possible strategy is to replace conventional Gaussian beams with flat-top beam profiles or other tailored spatial modes. The underlying idea is that a lateral displacement of a flat-top beam should produce only negligible intensity variations at the sample position. One approach we investigated was the generation of a flat-top intensity profile through mode mixing in a square-core optical fiber. However, the multimode nature of the fiber results in a highly divergent output beam. This increased divergence leads to locally varying angles of incidence on downstream retardation optics, giving rise to pronounced spatial phase-retardation gradients across the beam profile, analogous to the field-of-view limitations encountered in Pockels cells. In addition, modal interference within the fiber generates strong laser speckle. To reduce the speckle contrast, mechanical vibration of the fiber was initially employed. However, this approach proved insufficient, as a residual speckle contrast of approximately 1 % remained. The resulting fine-scale intensity fluctuations introduced significant measurement noise due to the steep local intensity gradients within the nominally flat-top profile. Rotating diffusers were also evaluated as an alternative means of speckle suppression. Although they reduced the visibility of individual speckles, they introduced subtle physical rotations of the beam profile and failed to decrease the speckle contrast to an acceptable level.

A second approach involved strongly expanding the diameter of a conventional Gaussian beam and utilizing only its nearly uniform central region for excitation. While conceptually straightforward, this strategy requires an optical output power of the laser above several hundred mW to achieve excitation densities on the order of 10 W/cm$^2$ at the sample plane. Depending on the degree of beam expansion, the required laser power can reach impractically high levels. Such optical powers are difficult to obtain from standard tunable laser sources and may lead to thermal instabilities within the optical system. Moreover, they generate substantial autofluorescence within the microscope objective, which ultimately compromises the sensitivity of the measurement.

### 9.4 Variable retarders

Instead of relying on retarders with a static phase retardation, electro-optic modulators offer dynamic polarization control, and the following devices are widely used in commercial CD spectrometers.

### 9.4.1 Pockels cell

As described in detail in Chapter 5, Pockels cells operate reliably and provide excellent artifact suppression when precisely aligned along the optical axis, even in the worst-case scenario where LD = 1. For the standard measurements conducted in this work, a KD*P Pockels cell was utilized, as detailed above.

We note that a highly sophisticated, custom-built rubidium titanyl phosphate (RTP) Pockels cell is described in[47]. Currently, this provides state-of-the-art artifact reduction alongside fast polarization switching times below 10 µs. However, this system is not deployed for single-object CD spectroscopy but rather used to generate helicity-modulated electron beams for electron scattering experiments aimed at measuring parity violation in specific electron scattering processes.

### 9.4.2 Alternative modulators (not experimentally evaluated)

While photoelastic modulators (PEM) and liquid-crystal (LC) displays were not experimentally tested in our setup, implementations are well-documented in the literature. For example a photothermal CD configuration that combines a PEM with a Pockels cell via a dual-modulation scheme to minimize artifacts during measurements on individual gold nanorods exhibiting moderate LD values is described in[25]. Similarly, an LC device is employed as a variable retarder for CD scattering measurements on individual gammadions in[23]. These specific chiral nanostructures exhibit small intrinsic LD (LD = 0 when perfectly fabricated), which inherently relaxes the constraints on polarization artifacts.

## 9.5 Correction and averaging methods

Regardless of the specific retardation methodology selected, a critical prerequisite for establishing the credibility of single-object CD spectroscopy is the inclusion of rigorous reference measurements. Specifically, control measurements must be performed on a reference sample exhibiting high LD (oriented at both 0° and 45°) but zero intrinsic CD. Utmost caution is warranted when systematic artifacts are computationally subtracted, particularly in regimes where the magnitude of the artifact exceeds the target CD signal by orders of magnitude. Nevertheless, some artifacts can be averaged out, as discussed below.

### 9.5.1 Correction of artifacts in macroscopic samples via rotation and flipping of the sample

While this work focuses on single nano-objects, artifacts in CD spectroscopy are typically described for macroscopic media like bulk solutions or films. To understand if these error sources distort our measurements and how to average them out, it is possible to apply established continuum theory to the nanoscale.

In classical polarimetry, the light-matter interaction within an infinitesimally thin layer is described by the differential Mueller matrix $m$. Upon traveling through a macroscopic medium consisting of several non-depolarizing thin layers, the Stokes vector $\vec{S}$ of the incident light evolves according to[63]

$$\frac{d\vec{S}(z)}{dz} = m \cdot \vec{S}(z) \tag{116}$$

The differential Mueller matrix $m$ captures all optical properties of the thin layer and can be separated into an isotropic component (the absorption $a$ of unpolarized light) and an anisotropic matrix $F$, such that $m = aI + F$, where $I$ is the identity matrix and $m$ reads as[42,44,63,64]

$$m = \begin{pmatrix} a & -ld & -ld' & cd \\ -ld & a & cb & lb' \\ -ld' & -cb & a & -lb \\ cd & -lb' & lb & a \end{pmatrix} = aI + \underbrace{\begin{pmatrix} 0 & -ld & -ld' & cd \\ -ld & 0 & cb & lb' \\ -ld' & -cb & 0 & -lb \\ cd & -lb' & lb & 0 \end{pmatrix}}_{F} \tag{117}$$

The matrix $F$ contains the direction-dependent anisotropic properties of the sample. Here, the linear dichroism ($ld$) and linear birefringence ($lb$) are aligned with the primary laboratory axes, while their counterparts, $ld'$ and $lb'$, are measured at a 45° with respect to the laboratory system. The matrix also incorporates the circular dichroism ($cd$) and circular birefringence ($cb$). All properties are normalized per unit length (indicated here by lowercase letters)[42,63]. This approach is conceptually very similar to eqs.(25) and (28). However, whereas we used eqs.(25) and (28) to describe solely the linear dichroism of isolated objects via their transition dipole moments, the differential matrix model for a thin layer incorporates a wider range of optical phenomena, including linear and circular dichroism, as well as linear and circular birefringence. To derive the overall macroscopic response $M_{\text{sample}}$ of a thick, anisotropic sample, the infinitesimal thin layers are integrated over the entire optical path $z$. This yields the matrix exponential function $M_{\text{sample}} = e^{m \cdot z}$. Expanding this expression into a Taylor series[39,42] explicitly demonstrates how the macroscopic Mueller matrix $M_{\text{sample}}$ depends on the fundamental anisotropic properties of a thin layer, encoded within $F$:

$$M_{\text{sample}} = e^{m\cdot z} = e^{a\cdot z}\left[I + Fz + \frac{1}{2}F^2z^2 + \frac{1}{6}F^3z^3 + \dots\right] \tag{118}$$

This macroscopic perspective reveals the fundamental source of artifacts in optically thick samples. As the expansion progresses to higher orders in $F$, the optical parameters $ld, lb, ld', lb', cd$, and $cb$ are mixed as illustrated by the quadratic term $\frac{1}{2}F^2z^2$:

$$\frac{1}{2}F^2z^2 = \frac{z^2}{2}\begin{pmatrix} ld^2 + ld'^2 + cd^2 & ld'\cdot cb - cd\cdot lb' & cd\cdot lb - ld\cdot cb & ld'\cdot lb - ld\cdot lb' \\ cd\cdot lb' - ld'\cdot cb & ld^2 - cb^2 - lb'^2 & ld\cdot ld' + lb\cdot lb' & -ld\cdot cd - cb\cdot lb \\ ld\cdot cb - cd\cdot lb & ld\cdot ld' + lb\cdot lb' & ld'^2 - cb^2 - lb^2 & -ld'\cdot cd - cb\cdot lb' \\ ld\cdot lb' - ld'\cdot lb & -ld\cdot cd - cb\cdot lb & -ld'\cdot cd - cb\cdot lb' & cd^2 - lb'^2 - lb^2 \end{pmatrix} \tag{119}$$

This demonstrates that the matrix elements $M_{14}$ and $M_{41}$ of $M_{\text{sample}}$, which should ideally represent exclusively $cd$, have also contributions from $ld'\cdot lb - ld\cdot lb'$. Consequently, the coupling of linear dichroism and linear birefringence inside a thick sample generates an apparent CD signal, thereby creating a macroscopic artifact. Physically, this means that as light propagates through the medium, its polarization changes continuously due to the interaction with each layer. Effectively, every subsequent thin layer interacts with slightly imperfectly circularly polarized light, triggering the same problems described throughout this manuscript. As a result, through the sequential accumulation of effects like linear dichroism and linear birefringence, the sample itself generates macroscopic artifacts even under perfectly circularly polarized excitation.

Transferring this model back to the nanoscale reveals a massive experimental advantage for investigating single isolated objects. For dimensions spanning only a few tens of nanometers, the optical path $z$ approaches 0. The disruptive higher-order terms vanish, the optical effects within the matrix $M_{\text{sample}}$ decouple, and only the pure, linear first-order terms $Fz$ remain relevant. The sequential artifact accumulation characteristic of macroscopic layers loses its physical validity, thereby inherently shielding the measurements of nanoscale objects from intrinsic linear dichroism - linear birefringence interaction.

However, this advantage provides no protection against external error sources, such as residual antisymmetric linearly polarized components of the incident laser beam, as described throughout the manuscript. If imperfectly circularly polarized light interacts with a single object, this polarization error couples with the sample's linear dichroism to produce an apparent CD signal (i.e., instrumental birefringence interacts with the sample's LD). It is exactly at this point that investigating single objects becomes a massive technical disadvantage compared to macroscopic measurements. A standard procedure in macroscopic CD spectroscopy for eliminating LD-induced polarization artifacts in combination with imperfectly circularly polarized excitation involves rotating the sample around the optical axis to various azimuthal angles[38–40]. By averaging the CD signals acquired at these different orientations, the antisymmetric linear components of the incident polarization effectively cancel out. Furthermore, macroscopic sample flipping is routinely used to eliminate intrinsic cross-terms, such as the linear dichroism – linear birefringence contribution, which arises if the sample itself acts as a retarder (which is fortunately not the case for most single objects, as detailed above). While these are highly preferred and well-documented methods for macroscopic targets such as aligned films (comprehensively described in[38,39,41–43]), their application to single, isolated objects comes along with technical challenges.
For these techniques to be viable on the nanoscale, the mechanical axis of rotation must precisely intersect the nanometer-sized object. Furthermore, this approach is incompatible with wide-field illumination; spatial polarization gradients across the beam profile do not average out if the target object is significantly smaller than the focal spot. Particularly within our cryogenic setup, the mechanical tolerances required to achieve such precise, localized rotations and flips are practically hard to achieve.

### 9.5.2 Rotation of the polarization

As an alternative to rotating the sample, the polarization axis of the entire optical setup can be rotated to average out systematic errors. A particularly elegant implementation of this concept, demonstrated in[46], involves the use of "half-waveplate flips". By mechanically inserting and removing a perfect half-waveplate (a process referred to as "slow reversals") the sign of the polarization artifacts is inverted. Averaging the signals from both states effectively nullifies all artifacts originating from optical components located upstream of the half-waveplate. However, this method does pose some additional challenges. Artifacts introduced by downstream components, such as the birefringent cryostat windows, remain entirely unaffected. Additionally, any real half-waveplate inherently deviates from ideal behavior and introduces its own systematic errors, demanding rigorous alignment and calibration within the overall setup.

### 9.5.3 Numerical correction

In principle, if both the spatial orientation of the sample's LD axis and the exact polarization state of the excitation light (specifically its antisymmetric linear components) are known, LD-induced artifacts can be computationally subtracted from the measured spectra (similar to a full Mueller matrix characterization of a sample in ellipsometry). However, the accuracy of this numerical correction is fundamentally limited by the precision with which the local polarization state can be quantified. Relying on this method requires high confidence that the antisymmetric linear components of the LCP and RCP excitation light are mapped with absolute accuracy at the specific location of the structure. We deliberately avoid this approach. In our setup, measuring the local polarization state of the LCP and RCP excitation light with the required spatial precision is experimentally unfeasible, particularly within the complex environment of a cryostat. However, zeroth-moment artifacts can be corrected, as they are position-independent and can therefore be compensated more readily. After applying the correction, the residual artifact is centered around zero.

### 9.5.4 Restriction to single-point measurements

Rather than employing wide-field detection, artifacts can be suppressed by restricting the measurement to a single object positioned strictly at the center of the laser beam. By confining the excitation volume to the central optical axis, the effects of spatially varying polarization gradients are largely circumvented (as can be seen in Fig.19). When doing this, locally varying polarization gradients are easier to handle, and the system is primarily governed by a global, macroscopic antisymmetric linear component, which is inherently easier to characterize and control than local spatial variations.

## 10. Appendix

### 10.1 Design principles of the setup

For single objects, the excitonic CD strongly depends on the angle of incidence of the excitation light. For objects with a defined symmetry axis, light propagating along this axis probes the so-called face contribution to the CD spectrum, whereas excitation perpendicular to the axis reveals the fundamentally different edge contribution. Restricting the excitation to normal incidence may therefore obscure important structural information[17,34]. Consequently, the experiment was designed to enable excitation at oblique angles of incidence with respect to the substrate surface. To achieve this, a transmission geometry was employed instead of a conventional epi-fluorescence or confocal configuration. This approach minimizes the number of optical elements between the polarization-shaping optics and the sample while eliminating the need for dichroic beamsplitters, which can introduce additional polarization artifacts..

In the transmission configuration, see also Fig.2, the excitation beam is only weakly focused using a long focal length lens ($f = 100$ mm) positioned in front of the cryostat. This weak focusing offers several advantages. First, it allows the angle of incidence to be adjusted easily without requiring complex mechanical modifications inside the cryostat. Second, because all of the optical elements for the excitation remain outside the cryostat temperature- and/or pressure-induced stress birefringence is avoided. Third, artifacts associated with strong focusing are effectively suppressed. Finally, the resulting focal spot diameter of approximately 100 µm enables the simultaneous study of multiple single objects.

However, this design also introduces several challenges that require careful consideration. First, the low divergence of the weakly focused beam enhances unwanted interference effects caused by multiple reflections between the parallel surfaces of the substrate. Such effects are typically suppressed in highly divergent, strongly focused beams. In the present setup, they are minimized by employing substrates with a single-sided textured broadband antireflective surface (intensity reflection coefficient $R < 0.25$ %). Second, because the excitation lens is not rigidly connected to the sample holder and the substrate, the system becomes more susceptible to lateral vibrations and other relative movements between the sample and the excitation optics. These motions lead to fluctuations of the excitation intensity at the position of a given single object and can therefore contribute to intensity-related artifacts of the CD signal. Third, achieving an excitation intensity of about 10 W/cm$^2$ within the large excitation spot of 100 µm in diameter requires excitation powers in the range of 1 - 3 mW. Owing to the transmission geometry, the laser beam passes through the substrate with little attenuation and propagates into the detection path. Although the transmitted excitation light can largely be suppressed by optical filtering, the relatively high excitation power may induce autofluorescence in conventional commercial objectives. Since the autofluorescence originates within the objective itself, it cannot be rejected by spatial filtering, as it would be in a confocal detection scheme. To minimize this background contribution, a custom-designed low-autofluorescence objective was developed.

### 10.2 Custom-designed objective lens

As outlined in Chapter 2 and above, a custom-designed objective was employed to collect the fluorescence from a single-object inside the cryostat. This section discusses the specific requirements imposed by the experimental configuration, an overview of the design process, and the resulting performance of the objective within the setup. References[65–68] served as a practical guide for designing the custom objective lens.

#### 10.2.1 Requirements and selection of the materials

The experimental geometry and cryogenic operating conditions imposed a number of stringent requirements on the objective design. First, the limited space inside the cryostat restricted the maximum parfocal length of the objective to 30 mm. Second, the optical design had to provide aberration correction for imaging through a 1 mm quartz substrate. Third, the objective was required to be chromatically corrected over the entire emission wavelength (780 nm to 820 nm), feature an infinity-corrected design with a numerical aperture (NA) greater than 0.5, and provide a working distance of at least 0.5 mm. Fourth, cemented lenses could not be used, as optical cements tend to become opaque after repeated thermal cycling to temperatures as low as 1K. Finally, the autofluorescence of the objective, originating primarily from the lens glass materials, had to be minimized.

Correction of spherical and chromatic aberrations is typically achieved by combining low-index, low-dispersion glasses with high-index, high-dispersion materials. For the low-index component, UV fused silica (UV-FS) was selected (refractive index $n_\mathrm{d} = 1.458$, Abbe number $\nu_\mathrm{d} \approx 68$ [60]). UV-FS exhibits virtually no autofluorescence and is readily available in the form of standard commercial lens elements. The selection of a suitable high-index material proved more challenging, as highly dispersive optical glasses (e.g., N-SF5, N-SF11) exhibit strong autofluorescence, particularly for excitation wavelengths around 700 nm. After evaluating several candidate materials, SF57 ($n_\mathrm{d} = 1.847$, $\nu_\mathrm{d} \approx 24$ [55]) was identified as the best compromise between the required optical performance and sufficiently low autofluorescence.

### 10.2.2 Design strategy

Once the materials had been selected, the objective was modeled and optimized using Zemax OpticStudio (ANSYS Zemax OpticStudio, Version 24.1). Owing to the restricted parfocal length, the design was limited to two primary functional groups: a front group and a rear group.

Front Group: The first lens element, positioned closest to the substrate, is the most critical component of the objective. It largely determines the NA and working distance, while being exposed to the highest intensity of the essentially unattenuated excitation light. Consequently, it constitutes the dominant potential source of autofluorescence within the optical system. Achieving a high NA while minimizing the spherical aberrations associated with the large surface curvatures required for high-NA optics necessitates the use of a high-index material in combination with an aspheric surface profile. To meet these requirements, a commercially available aspheric lens (LightPath 355330, NA = 0.77, broadband AR-coated for 600 to 1050 nm) fabricated from D-ZLaF52LA glass ($n_\mathrm{d} = 1.810$, $\nu_\mathrm{d} \approx 41$ [69]) was selected. The high refractive index of this material enables a large NA while maintaining autofluorescence at an acceptable level for the intended application. The second element in the front group is a UV-FS meniscus lens custom-manufactured by Knight Optical (lens 2 in Table 5). This element was designed to achieve the required NA and magnification while contributing only a negligible additional autofluorescence background.

Rear Group: The rear group compensates for the residual spherical and chromatic aberrations introduced by the front group. To reduce manufacturing costs, custom lens elements were replaced wherever possible by combinations of standard, commercially available UV-FS plano-convex and plano-concave singlets. The final design employs two identical SF57 plano-concave lenses custom-manufactured by Knight Optical (lens 3 and 6 in Table 5) together with three off-the-shelf UV-FS plano-convex singlets. Collectively, these elements provide the aberration correction required to achieve the desired performance across the specified spectral range. The complete optical design of the objective is shown in Fig.25.

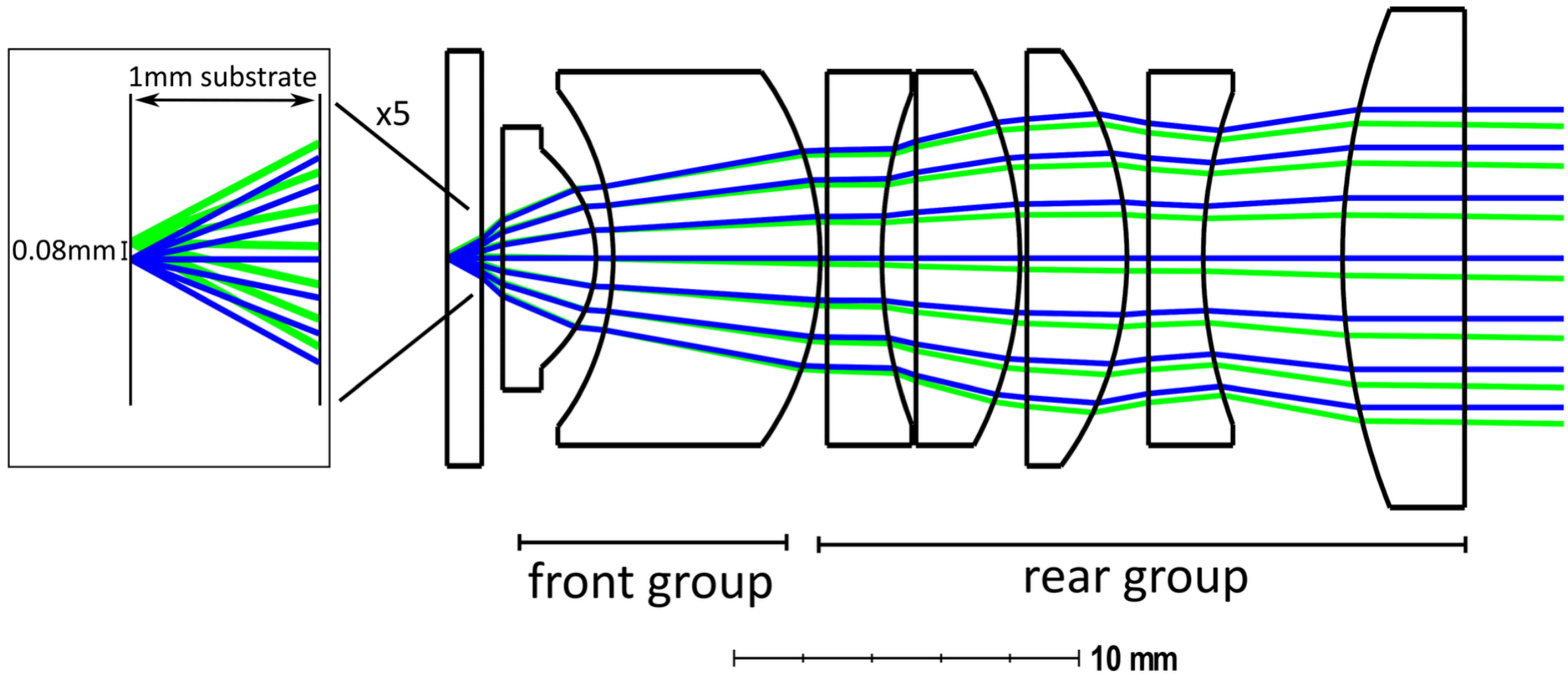

***Fig.25**: Cross-sectional view of the custom-made objective. Ray paths for on-axis and off-axis object points are shown in blue and green, respectively. The inset on the left shows the beam path within the substrate plate for a fivefold enlarged scale.*

### 10.2.3 Parameters and Performance of the Objective

The final optimized design provides a NA of 0.6, a working distance of 0.6 mm, a magnification of 33×, and a corrected field of view approximately 160 μm in diameter. The material properties and geometric parameters of the individual lens elements are listed in Table 5.

***Table 5**: Parameters of the lenses used for designing the objective in Zemax*

| Surface | Comment | Part/Vendor | Radius (mm) | Thickness (mm) | Material | Mech. Radius (mm) |
|---|---|---|---|---|---|---|
| 0 | Object | | Infinity | 0.000 | | |
| 1 | Substrate | | Infinity | 1.000 | C79-80 | 5.000 |
| 2 | Working dist. | | Infinity | 0.611 | | |
| 3 | Lens 1 (aspheric) | LightPath 355330 | Infinity | 2.706 | D-ZLAF52LA | 3.615 |
| 4 | | | -2.452 | 0.500 | | 3.615 |
| 5 | Lens 2 | Custom 1 (Knight Optical) | -5.906 | 6.000 | C79-80 | 4.500 |
| 6 | | | -6.768 | 0.200 | | 4.500 |
| 7 | Lens 3 | Custom 2 (Knight Optical) | Infinity | 1.590 | SF57 | 4.500 |
| 8 | | | 9.715 | 0.984 | | 4.500 |
| 9 | Lens 4 | Edmund #18-030 | Infinity | 3.015 | C79-80 | 4.500 |
| 10 | | | -8.250 | 0.200 | | 4.500 |
| 11 | Lens 5 | Quioptic G312256000 | Infinity | 2.900 | C79-80 | 5.000 |
| 12 | | | -7.499 | 0.625 | | 5.000 |
| 13 | Lens 6 | Custom 2 (Knight Optical) | Infinity | 1.590 | SF57 | 4.500 |
| 14 | | | 9.715 | 4.040 | | 4.500 |
| 15 | Lens 7 | Edmund #18-045 | 13.750 | 3.540 | C79-80 | 6.000 |
| 16 | | | Infinity | 0.000 | | 6.000 |
| 17 | Propagation distance | | Infinity | 150.000 | | |
| 18 | CCD tube lens | Thorlabs AC254-200-B-ML | black box parameters | | | |
| 19 | | | | | | |
| 20 | | | | | | |
| 21 | | | | | | |
| 22 | Image on CCD | | Infinity | 0.000 | | |

The theoretical imaging performance of the combined objective and tube lens was simulated using Zemax OpticStudio, see Fig.26. The on-axis simulation predicts negligible chromatic aberration for both targeted detection wavelengths (780 nm and 820 nm). For off-axis points, the simulated spot diagrams show the characteristic shape associated with coma, while remaining within the diffraction limit.

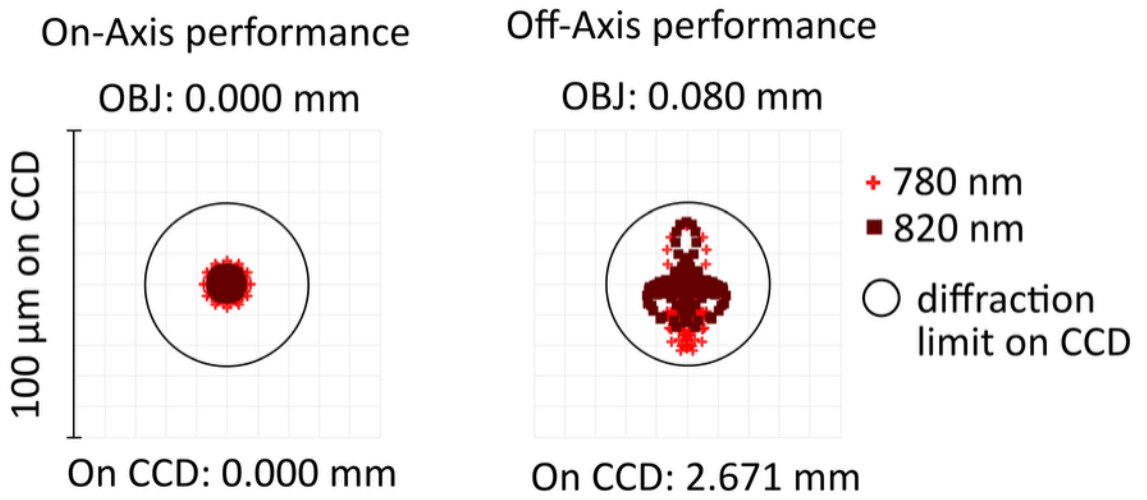

***Fig.26**: Simulated Zemax spot diagrams illustrating the spatial resolution in the EMCCD camera plane for wavelengths of 780 nm (light red crosses) and 820 nm (dark red squares). The black circle denotes the theoretical diffraction limit (Airy disk). The left panel shows the on-axis performance (the imaging of an object with no lateral displacement from the optical axis, denoted as OBJ: 0.000 mm), whereas the right panel depicts the off-axis performance at the edge of the targeted field of view (OBJ: 0.080 mm). This corresponds to a radial distance of 2.671 mm on the detector, which is equivalent to a magnification of 33.4×.*

To maintain optomechanical stability during cooling to 1 K, the objective housing and lens spacers were machined from Invar 36 (EN material 1.3912). Invar 36 features an exceptionally low coefficient of thermal expansion (CTE) ($\approx 1.3 \cdot 10^{-6}\ /K$ [70]), which approximately matches that of the fused silica lens elements (CTE $\approx 0.5 \cdot 10^{-6}\ /K$ [60]). This minimizes thermally induced mechanical stress during cooldown and prevents damage to the lenses at cryogenic temperatures.

However, this design requires a compromise with respect to the high-index materials. SF57 (CTE $\approx 8.3 \cdot 10^{-6}\ /K$ [55]) and D-ZLaF52LA (CTE $\approx 6.6 \cdot 10^{-6}\ /K$ [69]) exhibit substantially higher CTEs than the Invar housing. Consequently, these lens elements contract more strongly than the housing during cooldown, resulting in slight mechanical loosening at cryogenic temperatures. Note that all CTE values reported here are given at room temperature and decrease significantly upon cooling toward 1 K. As shown in Fig.27, images of weakly fluorescent objects (chlorosomes) reveal residual aberrations, most notably coma. These aberrations are attributed primarily to lateral decentering of the lens elements arising from the combined effects of differential thermal contraction and the finite tolerances of in-house CNC machining.

For the present application, these imaging imperfections are not critical. Since the objective is used primarily for collecting fluorescence from isolated single objects, diffraction-limited imaging performance is not required. More importantly, the objective fulfills its principal design goal of minimizing autofluorescence. Under excitation at wavelengths below 700 nm, and excitation powers of up to 3 mW, the measured autofluorescence background is only 50 to 100 photons/s/pixel. This corresponds to approximately a tenfold reduction in background signal compared with standard commercial objectives tested under identical experimental conditions.

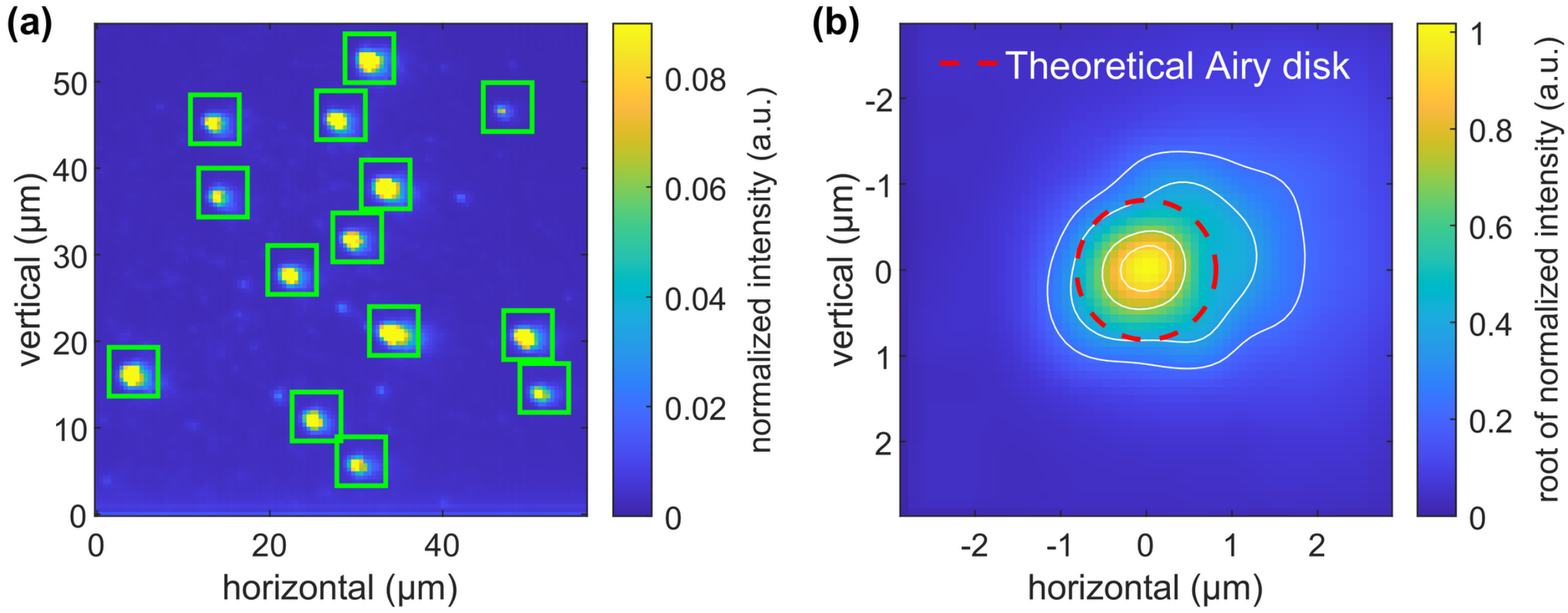

***Fig.27**: (a) Raw EMCCD image of individual chlorosomes. The green boxes highlight isolated single objects selected for reconstruction of the effective point spread function (PSF) of the objective. (b) Two-dimensional PSF of the objective obtained by aligning the centroids of the selected objects with sub-pixel accuracy and averaging the resulting images. The color scale represents the square root of the normalized intensity to enhance low-intensity peripheral features. The red dashed circle indicates the theoretical diffraction limit (Airy disk). White contours denote intensity levels of 80 %, 50 %, 13.5 %, and 5 % of the peak intensity, from the center outward.*

### 10.3 Generalized 2D intensity artifact

In Section 3.1, the intensity artifact $g_{\mathrm{I}}$ was derived for a one-dimensional Gaussian intensity profile, $I_{\mathrm{l,r}}(x) = I_{0_{\mathrm{l,r}}} \exp\left(-\frac{1}{2\sigma_{\mathrm{l,r}}^2}\left(x - \mu_{\mathrm{l,r}}\right)^2\right)$ allowing for differences between LCP and RCP intensity profiles in peak intensity, beam center position, and beam width. The resulting expression for the intensity artifact $g_{\mathrm{I}}$ is given in eq.(15a). Since the derivation for the $y$-coordinate follows analogously, it was not discussed separately. However, for a two-dimensional excitation profile, the principal axes of the Gaussian beam profiles associated with LCP and RCP light may be rotated with respect to each other. This effect is addressed in the present section.

#### 10.3.1 General 2D Gaussian intensity distribution

The intensity distribution of a general two-dimensional Gaussian beam is written in its most general form as

$$I(x,y) = I_0 \cdot \exp\left(-\frac{1}{2}\Delta r^T\, \Sigma^{-1}\, \Delta r\right)$$
$$\Delta r = r - \mu = \begin{pmatrix} x - \mu_x \\ y - \mu_y \end{pmatrix}$$
$$r = \begin{pmatrix} x \\ y \end{pmatrix} \tag{120}$$
$$\mu = \begin{pmatrix} \mu_x \\ \mu_y \end{pmatrix}$$

where $\Sigma^{-1}$ denotes the inverse of covariance matrix $\Sigma$, which is defined as

$$\Sigma(\alpha) = \begin{pmatrix} \sigma_x^2(\alpha) & \sigma_{xy}(\alpha) \\ \sigma_{xy}(\alpha) & \sigma_y^2(\alpha) \end{pmatrix} = R_\alpha \begin{pmatrix} \sigma_1^2 & 0 \\ 0 & \sigma_2^2 \end{pmatrix} R_\alpha^\top$$
$$R_\alpha = \begin{pmatrix} \cos(\alpha) & -\sin(\alpha) \\ \sin(\alpha) & \cos(\alpha) \end{pmatrix} \tag{121}$$

Here, $\sigma_1^2$ and $\sigma_2^2$ are the variances in the principal axis coordinate system of the intensity distribution. The rotation matrix $R_\alpha$ transforms these variances into the global coordinate system through a rotation by an angle $\alpha$, resulting in the variances $\sigma_x^2$, $\sigma_y^2$ and the covariance $\sigma_{xy}$. Since $R_\alpha$ is orthogonal, its inverse $R_\alpha^{-1}$ is equal to its transpose $R_\alpha^\top$. Hence, the inverse covariance matrix is then given by

$$\Sigma^{-1} = (R_\alpha^\top)^{-1} \begin{pmatrix} \frac{1}{\sigma_1^2} & 0 \\ 0 & \frac{1}{\sigma_2^2} \end{pmatrix} R_\alpha^{-1} = R_\alpha \begin{pmatrix} \frac{1}{\sigma_1^2} & 0 \\ 0 & \frac{1}{\sigma_2^2} \end{pmatrix} R_\alpha^\top \tag{122}$$

Inserting eq.(122) into eq.(120) yields

$$I(x,y) = I_0 \exp\left[-\frac{1}{2}\left(\Sigma_{11}^{-1} \cdot (x-\mu_x)^2 + \Sigma_{22}^{-1} \cdot \left(y-\mu_y\right)^2 + 2\Sigma_{12}^{-1} \cdot (x-\mu_x)\left(y-\mu_y\right)\right)\right] \tag{123}$$

with

$$\Sigma_{11}^{-1} = \frac{\cos(\alpha)^2}{\sigma_1^2} + \frac{\sin(\alpha)^2}{\sigma_2^2}$$
$$\Sigma_{22}^{-1} = \frac{\sin(\alpha)^2}{\sigma_1^2} + \frac{\cos(\alpha)^2}{\sigma_2^2} \tag{124}$$
$$\Sigma_{12}^{-1} = \cos(\alpha) \cdot \sin(\alpha) \cdot \left(\frac{1}{\sigma_1^2} - \frac{1}{\sigma_2^2}\right) = \frac{1}{2}\sin(2\alpha)\left(\frac{1}{\sigma_1^2} - \frac{1}{\sigma_2^2}\right)$$

For only small rotations of the principal axis system against the global coordinate system, the first-order approximations are $\cos(\alpha)^2 \approx 1$, $\sin(\alpha)^2 \approx 0$ and $\sin(2\alpha) \approx 2\alpha$ which yields

$$\Sigma_{11}^{-1} \approx \frac{1}{\sigma_1^2}$$
$$\Sigma_{22}^{-1} \approx \frac{1}{\sigma_2^2} \tag{125}$$
$$\Sigma_{12}^{-1} \approx \alpha\left(\frac{1}{\sigma_1^2} - \frac{1}{\sigma_2^2}\right)$$

Inserting eq.(125) into eq.(123) yields

$$I(x,y) \approx I_0 \exp\left[-\frac{1}{2}\left(\frac{(x-\mu_x)^2}{\sigma_1^2} + \frac{\left(y-\mu_y\right)^2}{\sigma_2^2} + 2\alpha\left(\frac{1}{\sigma_1^2} - \frac{1}{\sigma_2^2}\right)(x-\mu_x)\left(y-\mu_y\right)\right)\right] \tag{126}$$

Experimentally, the assumption of $\alpha$ being small requires the coordinate system of the detector (i.e., the pixel grid of the CCD camera) to be aligned with the principal axes of the incoming beam when measuring the artifact.

To express eq.(126) entirely in terms of variables measurable in the global coordinate system (camera frame), we consider the variances in eq.(121)

$$\sigma_x^2(\alpha) = \Sigma_{11}(\alpha) = \sigma_1^2 \cdot \cos(\alpha)^2 + \sigma_2^2 \cdot \sin(\alpha)^2$$
$$\sigma_y^2(\alpha) = \Sigma_{22}(\alpha) = \sigma_1^2 \cdot \sin(\alpha)^2 + \sigma_2^2 \cdot \cos(\alpha)^2 \tag{127}$$
$$\sigma_{xy}(\alpha) = \Sigma_{12}(\alpha) = \frac{1}{2}(\sigma_1^2 - \sigma_2^2)\sin(2\alpha)$$

where we find in first order for small $\alpha$

$$\sigma_x^2(\alpha) \approx \sigma_1^2$$
$$\sigma_y^2(\alpha) \approx \sigma_2^2 \tag{128}$$
$$\sigma_{xy}(\alpha) \approx (\sigma_1^2 - \sigma_2^2) \cdot \alpha$$

Inserting (128) into eq.(126) yields an expression for $I(x,y)$ for variances measured in the global coordinate system

$$I(x,y) \approx I_0 \exp\left[-\frac{1}{2}\left(\frac{(x-\mu_x)^2}{\sigma_x^2} + \frac{\left(y-\mu_y\right)^2}{\sigma_y^2} + 2\alpha\left(\frac{1}{\sigma_x^2} - \frac{1}{\sigma_y^2}\right)(x-\mu_x)\left(y-\mu_y\right)\right)\right] \tag{129}$$

This corresponds to a 2D Gaussian intensity profile whose principal axes are rotated by a small angle with respect to the global coordinate system.

### 10.3.2 Calculating the intensity artifact for a general 2D Gaussian intensity distribution

According to eq.(129), extending the one-dimensional Gaussian profile in eq.(8) to two dimensions requires the inclusion of the $y$-dependent contribution $\frac{(y-\mu_y)^2}{\sigma_y^2}$ and the cross term $2\alpha\left(\frac{1}{\sigma_x^2}-\frac{1}{\sigma_y^2}\right)(x-\mu_x)(y-\mu_y)$ as additional terms in the intensity distribution. Because the intensity profiles for LCP and RCP light can be slightly rotated with respect to each other, this is taken into account by distinguishing the angle $\alpha$ for the two circular polarizations, i.e. $\alpha_{\mathrm{l,r}} = \alpha + \delta\alpha_{\mathrm{l,r}}$ with $\delta\alpha_{\mathrm{l,r}} \ll 1$. Then the intensity distribution $I_{\mathrm{l,r}}(\delta I_{0_{\mathrm{l,r}}}, \delta\mu_{\mathrm{l,r}}, \delta\sigma_{\mathrm{l,r}}^2)$ provided in eq.(8) for LCP and RCP light depends in addition on $\delta\alpha_{\mathrm{l,r}}$. For brevity, this is written as $I_{\mathrm{l,r}}(\ldots,\ \delta\alpha_{\mathrm{l,r}})$ and the derivative

$$\left.\frac{\partial I_{\mathrm{l,r}}}{\partial\delta\alpha_{\mathrm{l,r}}}\right|_{(\ldots,0)} \cdot \delta\alpha_{\mathrm{l,r}} = -\left(\frac{1}{\sigma_x^2}-\frac{1}{\sigma_y^2}\right)(x-\mu_x)(y-\mu_y)\cdot I_{\mathrm{l,r}}(\ldots,0)\cdot\delta\alpha_{\mathrm{l,r}} \tag{130}$$

of the 2D intensity distribution with respect to variable $\alpha_{\mathrm{l,r}}$ for a small rotation of the beam profiles with respect to each other at $\alpha = 0$ needs to be introduced as well into the Taylor expansion, eq.(9). The equivalent of the intensity distribution in eq.(12) is then given as

$$I_{\mathrm{l,r}}(\ldots,\delta\alpha_{\mathrm{l,r}}) = \ldots - \left(\frac{1}{\sigma_x^2}-\frac{1}{\sigma_y^2}\right)(x-\mu_x)(y-\mu_y)\cdot I_{\mathrm{l,r}}(\ldots,0)\cdot\delta\alpha_{\mathrm{l,r}} \tag{131}$$

and, eq.(14) is extended by

$$\begin{aligned}
g_{\mathrm{I}} &= 2\frac{I_{\mathrm{l}}-I_{\mathrm{r}}}{I_{\mathrm{l}}+I_{\mathrm{r}}} \approx \frac{I_{\mathrm{l}}-I_{\mathrm{r}}}{I_{\mathrm{l,r}}(\ldots,0)}\\
&= \ldots + \frac{\left.\frac{\partial I_{\mathrm{l}}}{\partial\delta\alpha_{\mathrm{l}}}\right|_{(\ldots,0)}\cdot\delta\alpha_{\mathrm{l}} - \left.\frac{\partial I_{\mathrm{r}}}{\partial\delta\alpha_{\mathrm{r}}}\right|_{(\ldots,0)}\cdot\delta\alpha_{\mathrm{r}}}{I_{\mathrm{l,r}}(\ldots,0)}\\
&= \ldots - \left(\frac{1}{\sigma_x^2}-\frac{1}{\sigma_y^2}\right)(x-\mu_x)(y-\mu_y)\cdot(\alpha_{\mathrm{l}}-\alpha_{\mathrm{r}})\\
&= \ldots + \frac{(\sigma_x^2-\sigma_y^2)\cdot(\alpha_{\mathrm{l}}-\alpha_{\mathrm{r}})}{\sigma_x^2\sigma_y^2}(x-\mu_x)(y-\mu_y)
\end{aligned} \tag{132}$$

Using eq.(128) the following expressions are obtained

$$\begin{aligned}
\sigma_{\mathrm{l},xy} &\approx (\sigma_x^2-\sigma_y^2)\cdot\alpha_{\mathrm{l}}\\
\sigma_{\mathrm{r},xy} &\approx (\sigma_x^2-\sigma_y^2)\cdot\alpha_{\mathrm{r}}\\
\Delta\sigma_{\mathrm{lr},xy} = \sigma_{\mathrm{l},xy} - \sigma_{\mathrm{r},xy} &\approx (\sigma_x^2-\sigma_y^2)\cdot(\alpha_{\mathrm{l}}-\alpha_{\mathrm{r}})
\end{aligned} \tag{133}$$

and the additional term in eq.(132) is

$$\frac{\left(\sigma_x^2-\sigma_y^2\right)\cdot(\alpha_\mathrm{l}-\alpha_\mathrm{r})}{\sigma_x^2\sigma_y^2}(x-\mu_x)\left(y-\mu_y\right)\approx\frac{\Delta\sigma_{\mathrm{lr},xy}}{\sigma_x^2\sigma_y^2}(x-\mu_x)\left(y-\mu_y\right) \tag{134}$$

This term arises from a slight relative rotation of the intensity profiles for LCP and RCP light. Combining the one-dimensional, $x$-dependent artifact term, eq.(15a), with the corresponding contribution in the $y$ direction and the mixed term resulting from the slight relative rotation of the LCP and RCP beam profiles (eq.(134)) yields eq.(15b) in the main text.

$$\begin{aligned} g_\mathrm{I}=2\frac{I_{0_\mathrm{l}}-I_{0_\mathrm{r}}}{I_{0_\mathrm{l}}+I_{0_\mathrm{r}}}&+\underbrace{\frac{\Delta\mu_{\mathrm{lr},x}}{\sigma_x^2}\cdot(x-\mu_x)^1+\frac{\Delta\sigma_{\mathrm{lr},x}^2}{2\sigma_x^4}\cdot(x-\mu_x)^2}_{x\text{ component}} \\ +\underbrace{\frac{\Delta\mu_{\mathrm{lr},y}}{\sigma_y^2}\cdot\left(y-\mu_y\right)^1+\frac{\Delta\sigma_{\mathrm{lr},y}^2}{2\sigma_y^4}\cdot\left(y-\mu_y\right)^2}_{y\text{ component}}&+\underbrace{\frac{\Delta\sigma_{\mathrm{lr},xy}}{\sigma_x^2\sigma_y^2}(x-\mu_x)\left(y-\mu_y\right)}_{\text{slight rotation of the LCP}-\text{RCP beam profiles}} \end{aligned} \tag{135}$$

### 10.4 Useful relationships

In Chapter 3, eq.(64) quantifies how the polarization artifact $g_{\mathrm{P}_{1,2}}$ can be determined by measuring differences in the first three moments of the intensity distributions of LCP and RCP light after passing through a polarizer. For obtaining this equation, the corresponding intensity distributions for (i)-LCP and (i)-RCP light behind the polarizer were obtained from eq.(59). This section outlines briefly the derivation of eqs.(60)/(61), which expresses the resulting differences of the intensity moments based on the distributions given in eq.(59). To obtain eqs.(60)/(61), eq.(59) must be inserted into eqs.(53)-(55). For performing the required calculation the following relationships are useful:

$$\begin{aligned} \int_{-\infty}^{\infty}e^{-\frac{x^2}{2\sigma^2}}dx&=\sqrt{2\pi}\,\sigma \\ \int_{-\infty}^{\infty}x\,e^{-\frac{x^2}{2\sigma^2}}\mathrm{dx}&=0 \\ \int_{-\infty}^{\infty}x^2e^{-\frac{x^2}{2\sigma^2}}dx&=\sqrt{2\pi}\,\sigma^3 \\ \int_{-\infty}^{\infty}x^4e^{-\frac{x^2}{2\sigma^2}}dx&=3\sqrt{2\pi}\,\sigma^5 \end{aligned} \tag{136}$$

When inserting eq.(59) into eq.(53-55), uneven terms drop out. We considered all coefficients $a$, $b$, ... etc. to be small with respect to 1. Therefore, products of these coefficients were neglected. Accordingly, eq.(53) simplifies to

$$\begin{gathered} P_{\mathrm{l,r}} = TI_0\left(\sqrt{2\pi}\sigma_x\right)\left(\sqrt{2\pi}\sigma_y\right)[1+D] \\ D = \pm a \pm \frac{\sigma_x^2}{2} d \pm \frac{\sigma_y^2}{2} e \\ \frac{1}{P_{\mathrm{l,r}}} \propto \frac{1}{1+D} \approx 1 - D = 1 \mp a \mp \frac{\sigma_x^2}{2} d \mp \frac{\sigma_y^2}{2} e \end{gathered} \tag{137}$$

Eq.(54) simplifies to

$$\begin{aligned} \langle x \rangle_{\mathrm{l,r}} &= \frac{\pm\sigma_x^2 \cdot b}{1+D} \approx \pm\sigma_x^2 \cdot b \\ \langle y \rangle_{\mathrm{l,r}} &= \frac{\pm\sigma_y^2 \cdot c}{1+D} \approx \pm\sigma_y^2 \cdot c \end{aligned} \tag{138}$$

and eq.(55) simplifies to

$$\begin{aligned} \langle x^2 \rangle_{\mathrm{l,r}} &= \frac{\sigma_x^2(1 \pm a) \pm \frac{3\sigma_x^4}{2} d \pm \frac{\sigma_x^2\sigma_y^2}{2} e}{1+D} \approx \sigma_x^2(1 \pm \sigma_x^2 d) \\ \langle y^2 \rangle_{\mathrm{l,r}} &= \frac{\sigma_y^2(1 \pm a) \pm \frac{3\sigma_y^4}{2} e \pm \frac{\sigma_x^2\sigma_y^2}{2} d}{1+D} \approx \sigma_y^2\left(1 \pm \sigma_y^2 e\right) \\ \langle xy \rangle_{\mathrm{l,r}} &= \frac{\pm\sigma_x^2\sigma_y^2 f}{1+D} \approx \pm\sigma_x^2\sigma_y^2 f \end{aligned} \tag{139}$$

Note that $\langle x \rangle_{\mathrm{l,r}}^2 \propto b^2 \approx 0$ and $\langle y \rangle_{\mathrm{l,r}}^2 \propto c^2 \approx 0$ and therefore $\sigma_{\mathrm{l,r};x}^2 = \langle x^2 \rangle_{\mathrm{l,r}} - \langle x \rangle_{\mathrm{l,r}}^2 \approx \langle x^2 \rangle_{\mathrm{l,r}}$ (and analogously for $y$).

**Acknowledgement**

The whole project to develop CD spectroscopy on single objects emerged from discussions with Profs. J. Knoester and T.L.C. Jansen from the University of Groningen about the internal structure of chlorosomes from green sulfur bacteria and turned out to present us with a number of obstacles and experimental challenges. We are indebted to Chihiro Azai and Tomomi Inagaki for providing the chlorosome samples for the test experiments. Moreover, we thank Frank Neumann from the mechanical workshop at the University of Bayreuth for his assistance in designing and machining the optomechanical components of the custom objective lens. We thank Heyou Zhang for assisting us with various sample preparations. Financial support from the Deutsche Forschungsgemeinschaft (Ko 1359/27-1) and the State of Bavaria within the initiative "Solar Technologies go Hybrid" is gratefully acknowledged.